\documentclass[acmlarge]{acmart}
\renewcommand\footnotetextcopyrightpermission[1]{} % removes footnote with conference information in first column 
\makeatletter
\renewcommand\@formatdoi[1]{\ignorespaces}
\makeatother
\footnotetextcopyrightpermission

\usepackage{enumerate}
\usepackage{epsfig}
\usepackage{endnotes}
\usepackage{graphicx}
\usepackage{amsmath}
\usepackage{amssymb}
\usepackage{url}
\usepackage{subfigure}
\usepackage{xcolor,colortbl}
\usepackage{booktabs} % for table's toprule, midrule, etc.
\usepackage{tikz}
\usepackage{pifont}
\usepackage{paralist}
\usepackage[ruled]{algorithm2e} % For algorithms
\usepackage{multirow}
\usepackage{array}
\usepackage[utf8]{inputenc}
\usepackage{wrapfig}

\usepackage{blindtext,graphicx}
\usepackage[absolute]{textpos}

\usepackage{fancyhdr}
\acmJournal{IMWUT}
\acmVolume{1}
\acmNumber{3}
\acmMonth{9}
\begin{document}
\begin{textblock}{12}(2,0.2)
\noindent\small The final version of this article is published in Proceedings of the ACM on Interactive, Mobile, Wearable and Ubiquitous Technologies, Vol. 1, No. 3, Sep 2017 and is accessible on   https://dl.acm.org/citation.cfm?id=3131904
\end{textblock}

\title{Camera Based Two Factor Authentication Through Mobile and Wearable
Devices}
\author{Mozhgan Azimpourkivi}
\affiliation{
   \institution{Florida International University}
   \city{Miami}
   \state{FL}
   \country{USA}}
\author{Umut Topkara}
\affiliation{
   \institution{Bloomberg LP}
   \city{New york}
   \state{NY}
   \country{USA}}
\author{Bogdan Carbunar}
\affiliation{
   \institution{Florida International University}
   \city{Miami}
   \state{FL}
   \country{USA}}

\begin{abstract}
We introduce Pixie, a novel, camera based two factor authentication solution
for mobile and wearable devices.  A quick and familiar user action of snapping
a photo is sufficient for Pixie to simultaneously perform a graphical password
authentication and a physical token based authentication, yet it does not require any expensive, uncommon hardware.  Pixie establishes
trust based on both the knowledge and possession of an arbitrary physical
object readily
accessible to the user, called {\it trinket}. Users choose their trinkets
similar to setting a password, and authenticate by presenting the same trinket
to the camera. The fact that the object is the trinket, is secret to the
user.  Pixie extracts robust, novel features from trinket images, and leverages
a supervised learning classifier to effectively address inconsistencies between
images of the same trinket captured in different circumstances.

Pixie achieved a false accept rate below $0.09$\% in a brute force attack with
14.3 million authentication attempts, generated with 40,000 trinket images that
we captured and collected from public datasets. We identify {\it master}
images, that match multiple trinkets, and study techniques to reduce their
impact.

In a user study with 42 participants over 8 days in 3 sessions we found that
Pixie outperforms text based passwords on memorability, speed, and user
preference. Furthermore, Pixie was easily discoverable by new users and
accurate under field use. Users were able to remember their trinkets 2 and 7
days after registering them, without any practice between the 3 test dates.
\end{abstract}
%
%\begin{CCSXML}
%<ccs2012>
% <concept>
%  <concept_id>10010520.10010553.10010562</concept_id>
%  <concept_desc>Security and privacy~Authentication</concept_desc>
%  <concept_significance>500</concept_significance>
% </concept>
% <concept>
%  <concept_id>10010520.10010575.10010755</concept_id>
%  <concept_desc>Security and privacy~Usability in security and privacy</concept_desc>
%  <concept_significance>300</concept_significance>
% </concept>
%</ccs2012>  
%\end{CCSXML}
%
%\ccsdesc[500]{Security and privacy~Authentication}
%\ccsdesc[300]{Security and privacy~Usability in security and privacy}


\maketitle

\section{Introduction}

Mobile and wearable devices are popular platforms for accessing sensitive
online services such as e-mail, social networks and banking. A secure and
practical experience for user authentication in such devices is challenging, as
their small form factor, especially for wearables (e.g.,
smartwatches~\cite{smartwatch_samsung} smartglasses~\cite{smartglass_vuzix}),
complicates the input of the commonly used text based passwords, even when the
memorability of passwords already poses a significant burden for users trying
to access a multitude of services ~\cite{memorabilitymultiplepasswords}.  While
the small form factor of mobile and wearable devices makes biometric
authentication solutions seemingly ideal, their reliance on sensitive, hard to
change user information introduces important privacy and security issues
~\cite{PPJ03,OPM} of massive scale.

%In addition, the memorability of these passwords poses a
%significant burden for users who access a multitude of
%services~\cite{memorabilitymultiplepasswords}.
%\cite{textpasswordusability}. 

%\begin{figure}[t!]
%\centering
%%\vspace{5pt}
%\subfigure[]
%{\label{fig:pixie:setup}{\includegraphics[width=1.5in]{figures/pixie/rounded_corners_set.trinket.eps}}}
%\subfigure[]
%{\label{fig:pixie:login}{\includegraphics[width=1.5in]{figures/pixie/rounded_corners_test.trinket.eps}}}
%\subfigure[]
%{\label{fig:pixie:capture:2}{\includegraphics[width=1.5in]{figures/pixie/error_rounded.eps}}}
%%\subfigure[]
%%{\label{fig:match}{\includegraphics[width=2.15in,height=1in]{./figures/match/im_2_7.eps}}}
%\vspace{-5pt}
%\caption{
%Pixie:
%(a) Trinket setup. The user takes photos of the trinket placing it in the
%circle overlay. UI shows the number of photos left to take.
%(b) Login: the user snaps a photo of the trinket.
%(c) Trinket setup messages provide actionable guidance, when the image quality is low (top), or the reference images are inconsistent (bottom).
%%(d) ORB keypoint matches between trinket images. Each line connects matching
%%keypoints (small colored circles).}
%\vspace{-15pt}
%\label{fig:pixie}}
%\end{figure}

\begin{figure}[t!]
\centering
\subfigure[]
{\label{fig:pixie:setup}{\includegraphics[width=1.5in, height=2.45in]{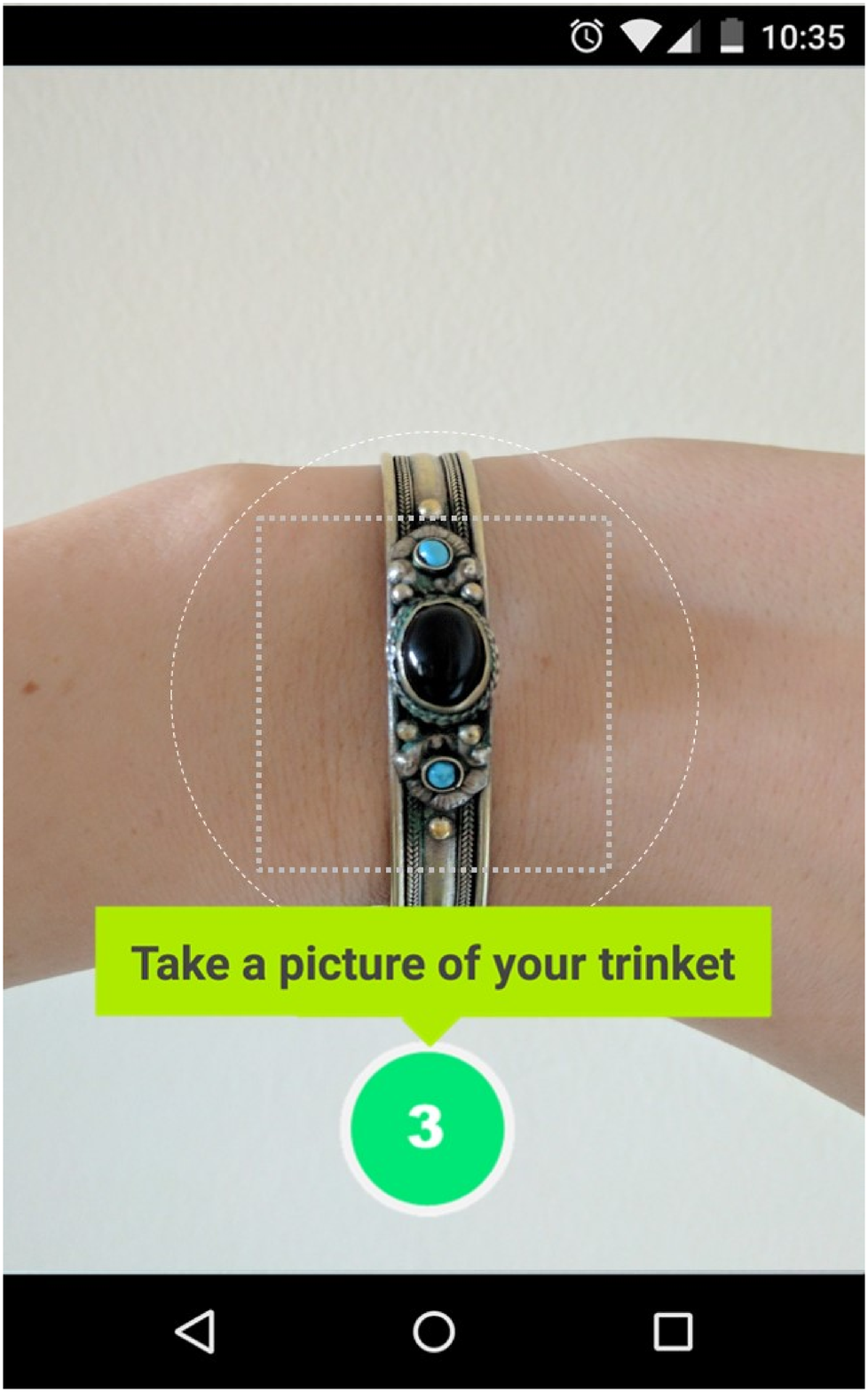}}}
\subfigure[]
{\label{fig:pixie:login}{\includegraphics[width=1.5in, height=2.45in]{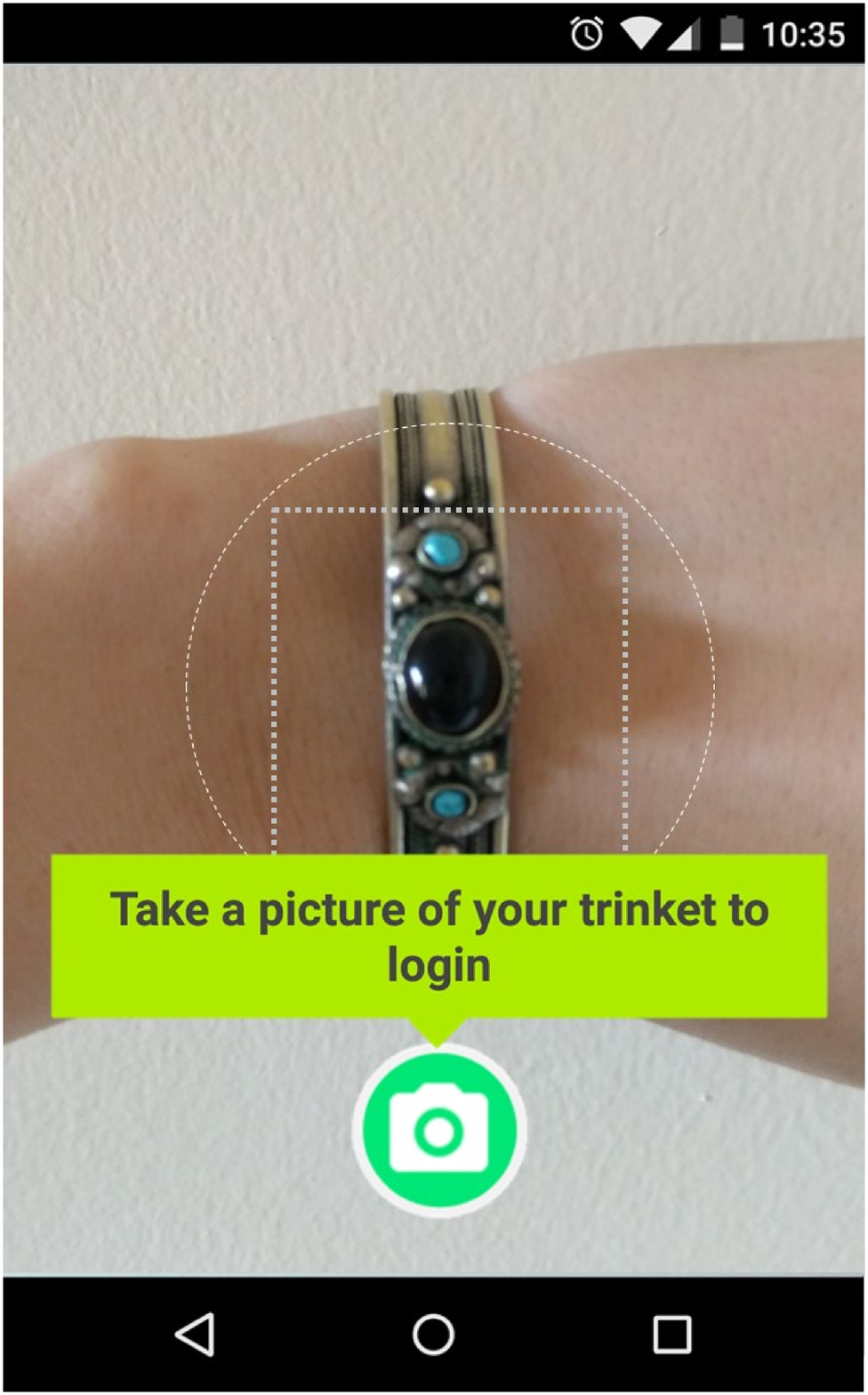}}}
\subfigure[]
{\label{fig:pixie:capture:2}{\includegraphics[width=1.5in, height=2.45in]
{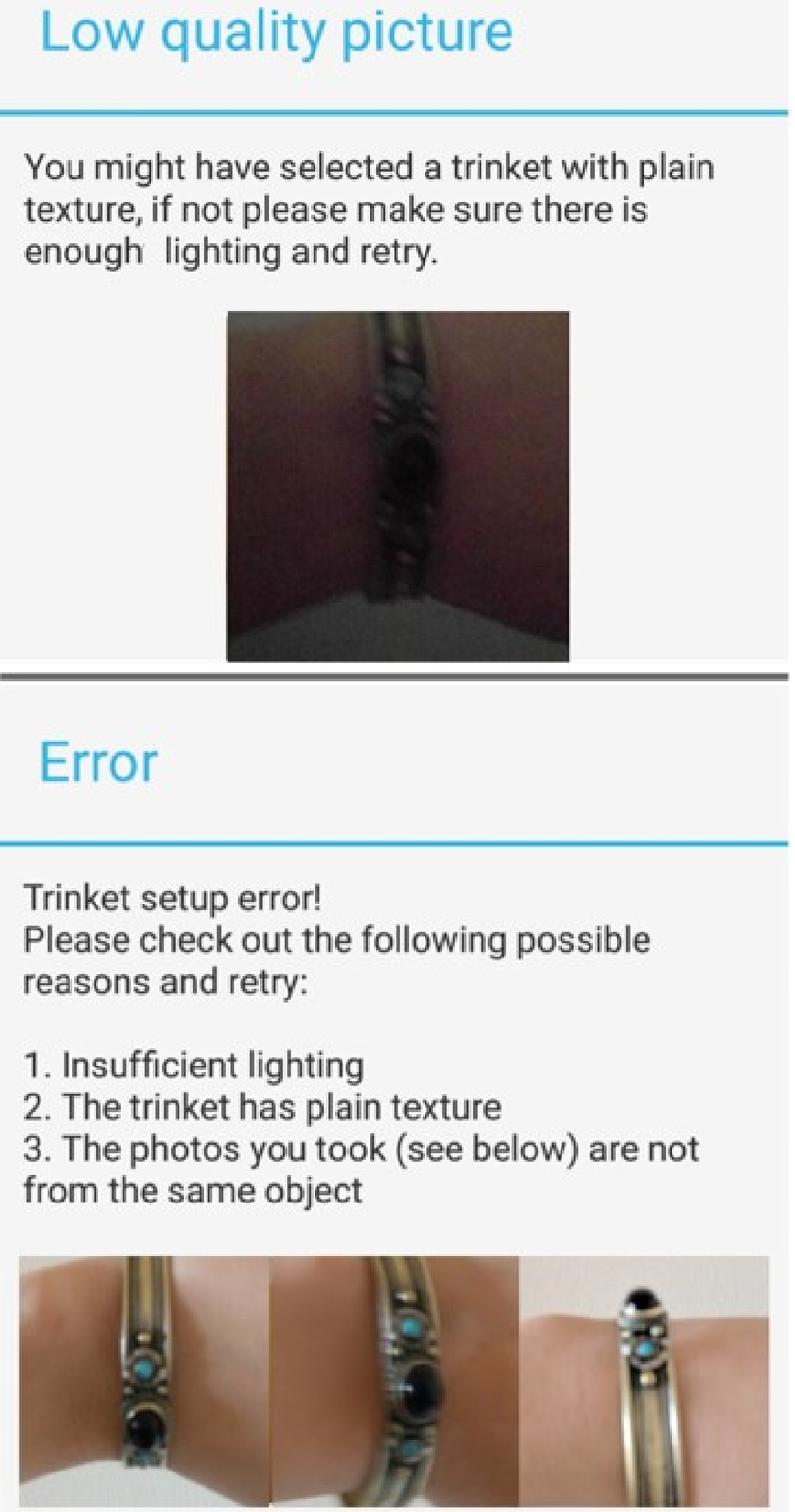}}}
%\subfigure[]
%{\label{fig:pixie:capture:errorref}{\includegraphics[width=1.5in,keepaspectratio]{figures/pixie/bracelet/error.ref.eps}}}
%\subfigure[]
%{\label{fig:pixie:capture:error}{\includegraphics[width=1.5in,keepaspectratio]{figures/pixie/bracelet/error.low.quality.eps}}}
\vspace{-5pt}
\caption{
Pixie:
(a) Trinket setup. The user takes photos of the trinket placing it in the
circle overlay. UI shows the number of photos left to take.
(b) Login: the user snaps a photo of the trinket.
(c) Trinket setup messages provide actionable guidance, when the image quality is low (top), or the reference images are inconsistent (bottom).
%(d) ORB keypoint matches between trinket images. Each line connects matching
%keypoints (small colored circles).}
%\vspace{-15pt}
\label{fig:pixie}}
\vspace{-15pt}
\end{figure}

In this paper we introduce Pixie, a camera based remote authentication solution
for mobile devices, see Figure~\ref{fig:pixie} and~\cite{Pixie.Demo} for a
short demo.  Pixie can establish trust to a remote service based on the user's
ability to present to the camera a previously agreed secret physical token.  We
call this token, the {\it trinket}.  Just like setting a password, the user
picks a readily accessible trinket of his preference, e.g., a clothing
accessory, a book, or a desk toy, then uses the device camera to snap trinket
images (a.k.a., {\it reference} images). All the user needs to do to
authenticate is to point the camera to the trinket. If the captured {\it
candidate} image matches the reference images, the authentication succeeds.

\begin{figure*}
\centering
\subfigure[]
{\label{fig:goodimage:stick}{\includegraphics[width=1.in]{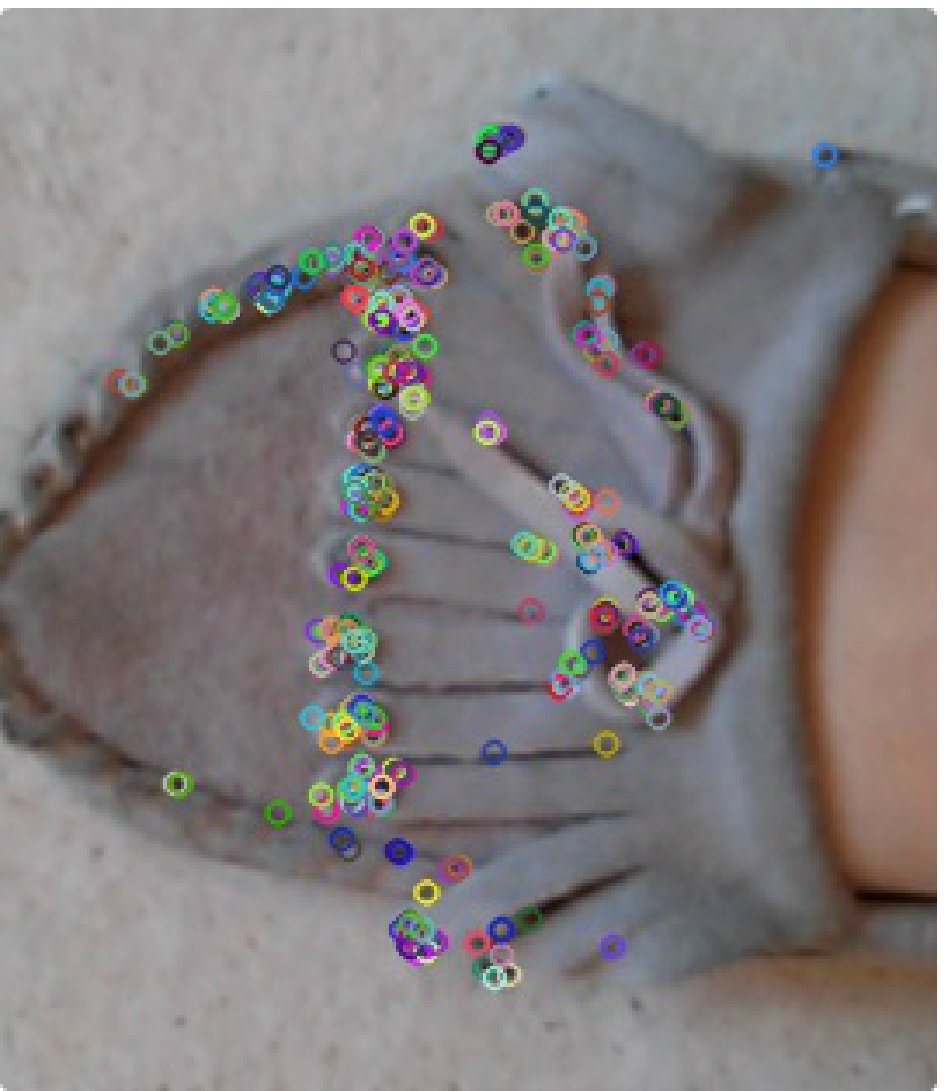}}}
\subfigure[]
{\label{fig:goodimage:santa}{\includegraphics[width=1.in]{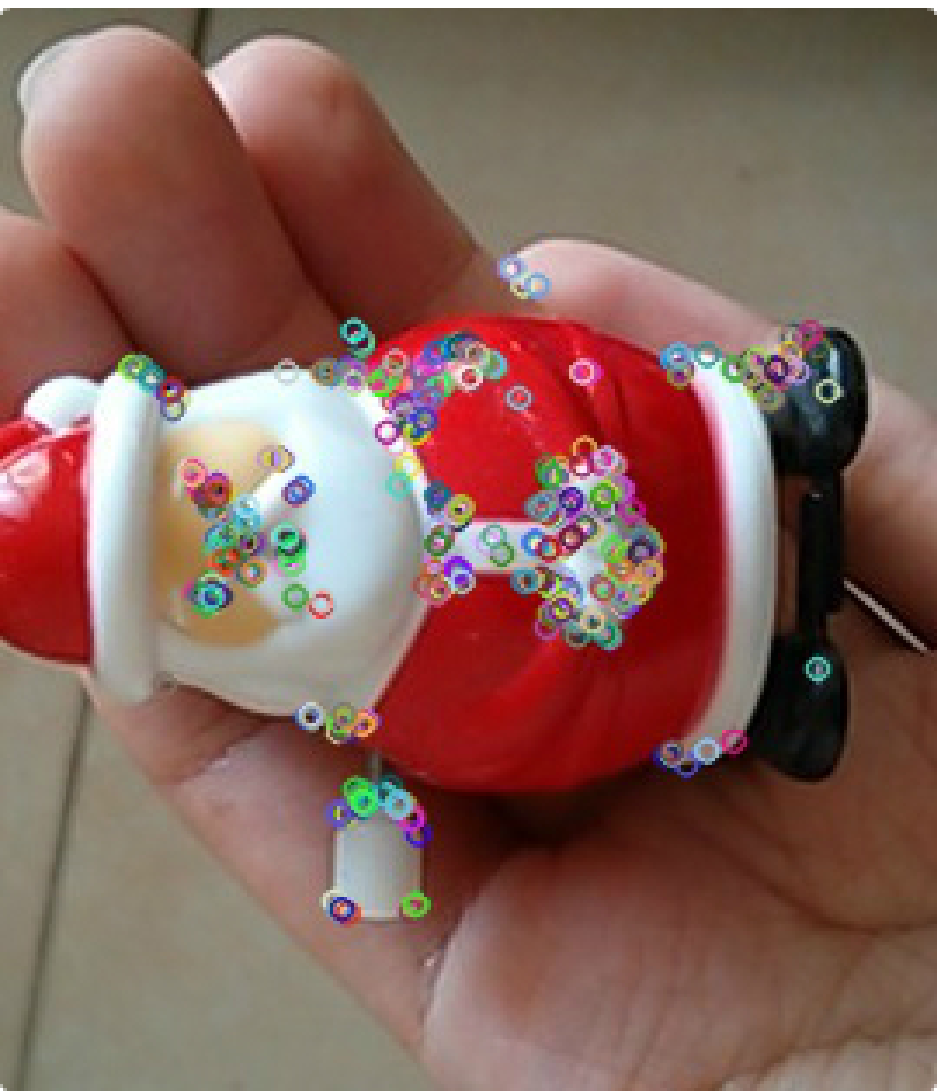}}}
\subfigure[]
{\label{fig:goodimage:bracelet}{\includegraphics[width=1.in]{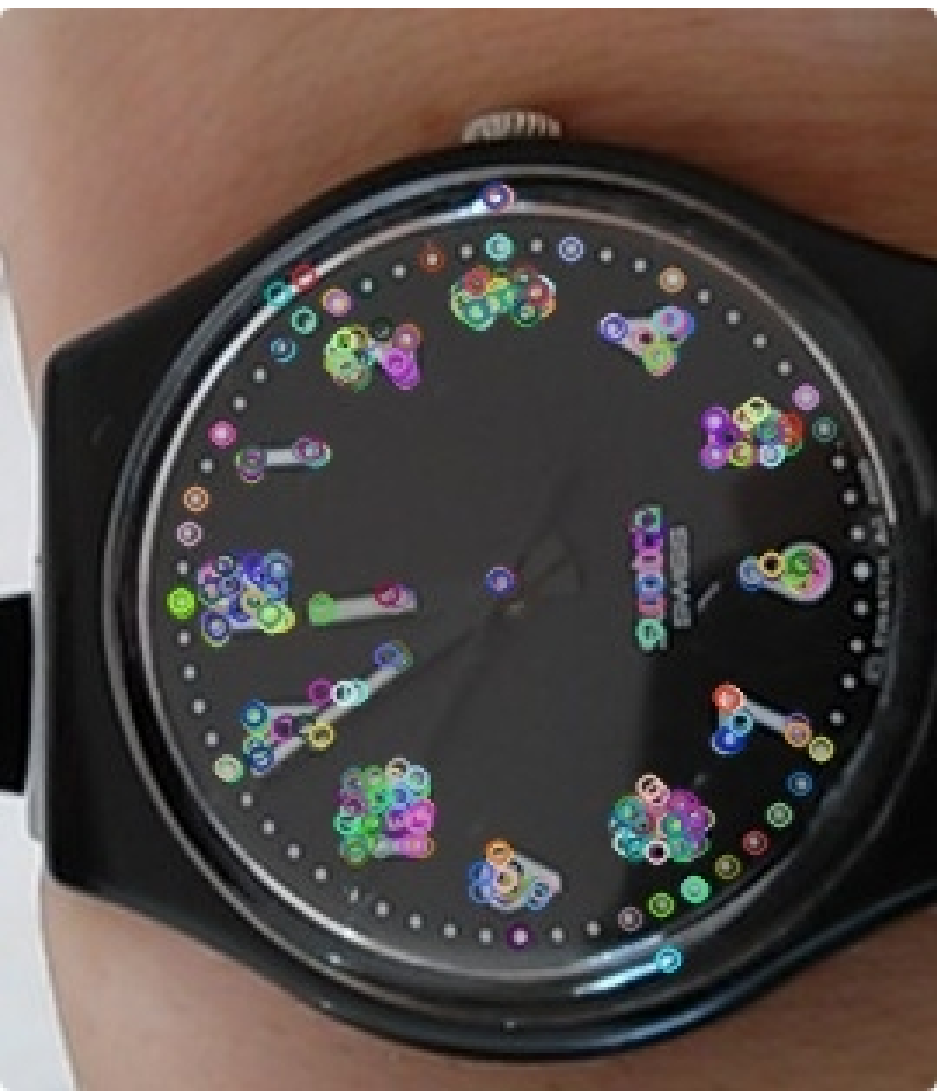}}}
\subfigure[]
{\label{fig:badimage:darkness}{\includegraphics[width=1.in]{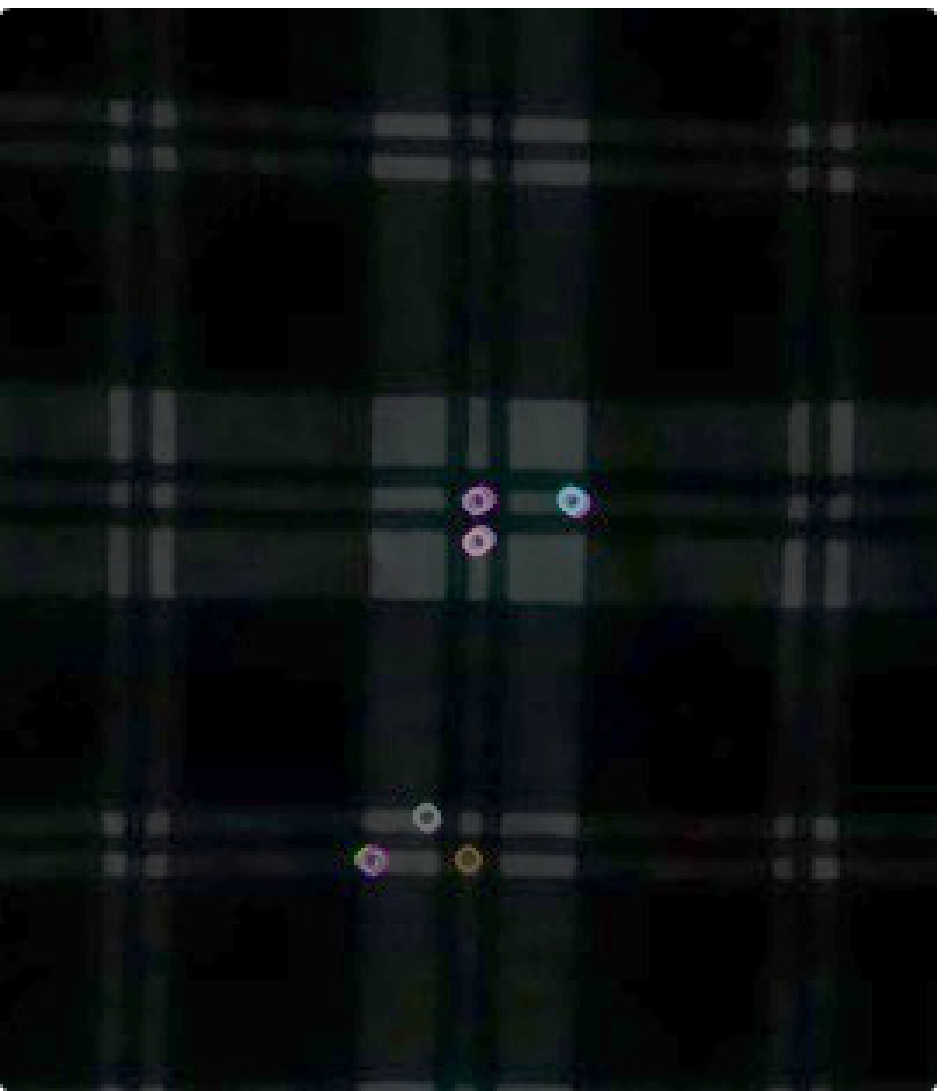}}}
\subfigure[]
{\label{fig:badimage:reflection}{\includegraphics[width=1.in]{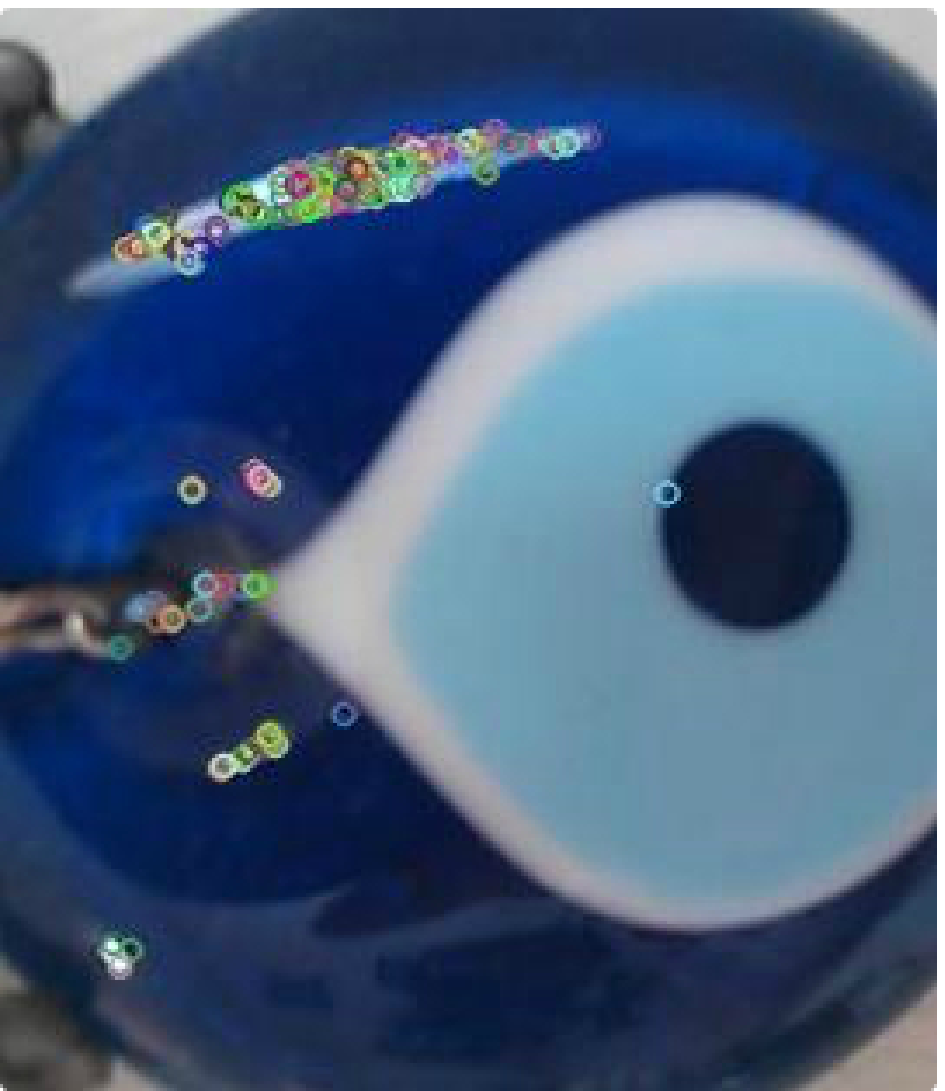}}}
\subfigure[]
{\label{fig:badimage:bluriness}{\includegraphics[width=1.in]{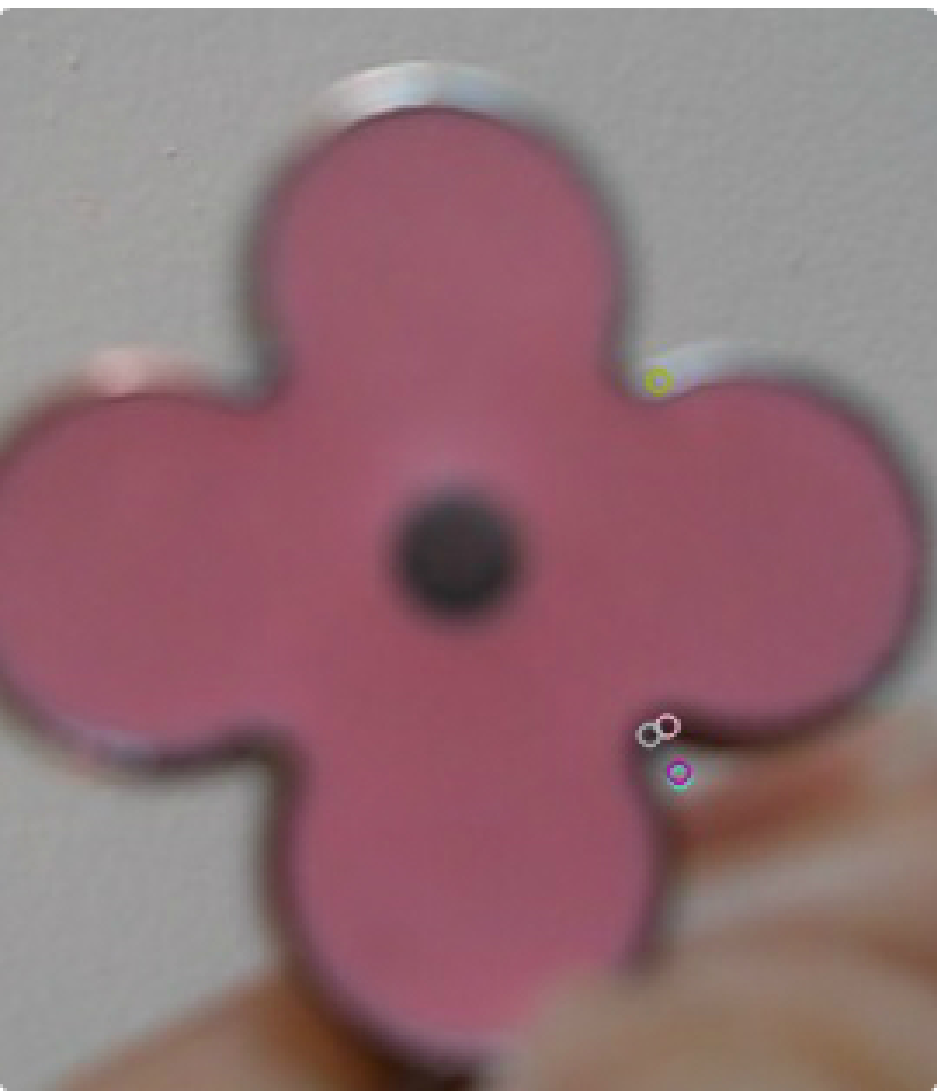}}}
%\subfigure[]
%{\label{fig:badimage:constantshade}{\includegraphics[width=1.25in]{figures/badimages/rounded_corners_plain.eps}}}
%\subfigure[]
%{\label{fig:badimage:background}{\includegraphics[width=1.25in]{figures/badimages/rounded_corners_backgr.eps}}}
\caption{
\footnotesize{
Examples of good (a-c) and low quality (d-f) trinket images.  Trinkets are
small (parts of) objects carried or worn by users, thus hard to steal and even
reproduce by adversaries.  ORB keypoints are shown as small, colored circles.
Good images have a high number of keypoints on the trinket.
%Note how the USB stick forms an ideal trinket: its cover's angle
%can form part of the secret.
Low quality images are due to
(d) insufficient light conditions on shirt section,
(e) bright light and reflection,
(f) image blur, or uniform, texture-less trinket.
%(d) uniform, texture-less trinket, and
%(e) complex background that generates most of the keypoints.
}
\label{fig:bad:image}}
\vspace{-15pt}
\end{figure*}

Pixie combines graphical password~\cite{BCV12,Passfaces,DMR04} and token based
authentication concepts~\cite{SecurID,Vasco}, into a two factor authentication
(2FA) solution based on what the user has (the trinket) and what the user knows
- the trinket, and the angle and section used to authenticate.
Figure~\ref{fig:bad:image} shows examples of trinkets.  Contrary to other 
token based authentication methods, Pixie does not require
expensive, uncommon hardware to act as the second factor; that duty is assigned
to the physical trinket, and the mobile device in Pixie is the primary device
through which the user authenticates.  Pixie only requires the authentication
device to have a camera, making authentication convenient even for wearable
devices such as smartwatches and smartglasses.

\noindent
{\bf Challenges and proposed approach}.
Building a secure and usable trinket based authentication solution is
difficult. Unlike biometrics based solutions, trinkets can be chosen from a
more diverse space than e.g., faces, thus lack the convenience of a set of well
known features.  In addition, users cannot be expected to accurately replicate
during login, the conditions (e.g. angle, distance and background) of the
trinket setup process. Thus, Pixie needs to be resilient to candidate images
captured in different circumstances than the reference images. Pixie addresses these problems in two ways: i) during the registration phase users are asked
to capture multiple trinket images, thereby revealing the variability of the trinket to Pixie, ii) to match a candidate image against these
reference images, Pixie leverages a statistical classifier using features 
which leverage robust keypoints~\cite{BTVG06,RRKB11} extracted from the 
trinket images.

In addition, in early pilot user studies, we identified new challenges for 
a successful deployment of Pixie. First, that
Pixie users may use low quality trinkets, e.g. with uniform textures, capture
inconsistent reference images with largely different viewing angles, or capture
low quality images of their trinkets, e.g., blurry, or with improper lighting
conditions, see Figure~\ref{fig:bad:image}(d)-(f).  In order to help the users
pick high quality trinkets and images thereof, we develop features that capture
the quality of reference images as defined by the likelihood of causing false
accepts or false rejects during authentication. We use these features to train
a trinket image rejection classifier that detects low quality images before
they can be used as Pixie trinkets.

Second, we found that it is crucial to give the user actionable feedback about
how to choose a better trinket when the Pixie filter rejects trinket images.
For instance, a set of reference images can be rejected because they contain
different trinkets, or because one of the images is blurry. However, most
statistical
classifiers are not easily interpretable, thus cannot indicate the nature of
the problem.  In order to provide meaningful actionable feedback, we identify
feature threshold values that pinpoint problem images and naturally translate
them into user instructions (see Table~\ref{tab:rule:filters}).

%e.g. blurry images, and inconsistent reference sets.

\noindent
{\bf Implementation and evaluation}.
We implement Pixie for Android, and show using an extensive evaluation that
Pixie is secure, fast, and usable. Pixie
achieves a False Accept Rate (FAR) of $0.02$\% and a False Reject Rate (FRR) of
$4.25$\%, when evaluated over $122,500$ authentication instances. Pixie
processes a login attempt in $0.5$s on a HTC One ($2013$ Model, $1.7$GHz CPU, $2$GB RAM).

To evaluate the security of Pixie, we introduce several image based attacks,
including an image based dictionary (or ``pictionary'') attack. Pixie achieves
a FAR below $0.09$\% on such an attack consisting of 14.3 million
authentication attempts constructed using public trinket image datasets and
images that we collected online. Similar to face based authentication, Pixie is
vulnerable to attacks where the adversary captures a picture of the trinket.
However, we show that Pixie is resilient to a shoulder surfing attack flavor
where the adversary knows or guesses the victim's trinket object type.
Specifically, on a targeted attack dataset of $7,853$ images, the average
number of ``trials until success'' exceeds $5,500$ irrespective of whether the
adversary knows the trinket type or not. In addition, we introduce and study
the concept of {\it master} images, whose diverse keypoints enable them to
match multiple trinkets.  We develop features that enable Pixie to reduce the
effectiveness of master images.

We perform a user study with 42 participants over 8 days in 3 sessions, and
show that Pixie is discoverable: {\it without prior training} and given no
external help, 86\% and 78\% of the participants were able to correctly set a
trinket then authenticate with it, respectively. Pixie's trinkets were
perceived as more memorable than text passwords, and were also easily
remembered 2 and 7 days after being set. 

%Pixie's authentication is also
%faster than well rehearsed text passwords and improves through even mild
%repetition: by the end of the study, at an average of $7.99$s per successful
%trial, users authenticated with Pixie $36$\% faster than text passwords.

Further, without any additional practice outside of the 3 sessions,
participants entered their trinket progressively faster than their text
passwords. Participants believed that Pixie is easier to use, more memorable and
faster than text passwords. We found that the preference of Pixie over text
passwords correlates positively with its preference on ease of use,
memorability and security dimensions and overall perception of trinket
memorability and willingness to adopt Pixie. In addition, 50\% of participants
reported that they preferred Pixie over text passwords.

%62\% of the participants were willing to adopt Pixie and 36\% said they would
%consider using it.

In summary, we introduce the following contributions:

\begin{compactitem}

\item
{\bf Pixie}.
We introduce Pixie, a two factor, mobile device based authentication
solution, that leverages the ubiquitous cameras of mobile devices to snap
images of trinkets carried by the users. Pixie makes mobile device based
authentication fast and convenient, and does not require expensive, uncommon
hardware. Pixie leverages a novel set of features that determine if a candidate
image contains the same token as a set of reference images
[$\S$~\ref{sec:pixie:authentication}]. We develop filters that identify low
quality images and inconsistent reference images, and provide actionable
feedback to the users [$\S$~\ref{sec:pixie:filter}].
  
\item
{\bf Security}.
We develop several image based attacks including brute force image dictionary
attacks, a shoulder surfing flavor and master image attacks. We construct more
than 14.3 million authentication instances to show that Pixie is resilient to
these attacks [$\S$~\ref{sec:evaluation:attacks}].
  
\item
{\bf User study}.
We implement Pixie in Android, and show through a user study with 42
participants that it is accurate, faster than text passwords, perceived as
such by users, and its trinkets are memorable [$\S$~\ref{sec:evaluation}].

%We study the usability and discoverability of Pixie as a novel form of
%authentication and conclude that it is faster than text passwords, perceived
%as such, and its trinkets are memorable [$\S$~\ref{sec:evaluation}].

\item
{\bf Reproducibility}.
Pixie is an open source prototype, with code and the Android installation file
available on GitHub~\cite{Pixie.github} and the Google Play
Store~\cite{Pixie.store}. We have also made our datasets, including the Pixie
attack datasets, available for download~\cite{Pixie.data}.

\end{compactitem}

%\noindent
%{\bf Organization}.
%The paper is organized as follows. We discuss related work
%($\S$~\ref{sec:related}). Then, we describe the system and adversary model
%($\S$~\ref{sec:model}). We introduce Pixie ($\S$~\ref{sec:pixie}), evaluate
%its performance, including under attack ($\S$~\ref{sec:evaluation}), and study
%its usability ($\S$~\ref{sec:user:study}).  We discuss Pixie limitations
%($\S$~\ref{sec:discussion}) and application ($\S$~\ref{sec:application}).
%Finally, we conclude this study ($\S$~\ref{sec:conclusions}).

\section{Related Work}
\label{sec:related}

Pixie is a camera based authentication solution that combines graphical
password and token based authentication concepts, into a single step 2 Factor
Authentication (2FA) solution. Pixie authentication is based on what the user 
has (the trinket) and what the user knows (the particular trinket among 
all the other objects that the user readily has access to, angle and viewpoint 
used to register the trinket). 
The unique form factor of Pixie differentiates it from existing solutions based on
typed, drawn, or spoken secrets. We briefly survey and distinguish Pixie from
existing solutions.

\subsection{Mobile Biometrics}

Biometric based mobile authentication solutions leverage unique human
characteristics, e.g., faces~\cite{DeLuca2015}, fingerprints~\cite{TouchID},
gait~\cite{gaitsmartwatch}, to authenticate users. In particular, the Pixie
form factor makes it similar to camera based biometric authentication solutions
based on face~\cite{boehm2013safe,Trewin2012,DeLuca2015} and
gaze~\cite{gazetouchpass,liu2015eyetracking}. Consequently, Pixie shares
several limitations with these solutions, that include (i) vulnerability to
shoulder surfing attacks and (i) susceptibility to inappropriate lighting
conditions, that can spoil the performance and usability of the authentication
mechanism~\cite{2015biometriciphoneandroid,majaranta2014eye}.

\begin{table*}
\centering
\resizebox{0.99\textwidth}{!}{%
\small
\textsf{
\begin{tabular}{l c c c }
\toprule
\textbf{ } & \textbf{Success rate} & \textbf{Entry Time} & \textbf{Number of trials} \\
\textbf{Solution} & \textbf{(\%)} & \textbf{(s)} & \textbf{before success} \\
\midrule
\textbf{Pixie} & \textbf{84.00} & \textbf{7.99 (Std=2.26, Mdn=8.51)} & \textbf{1.2 (Std=0.4, Mdn=1)}  \\
Text password (MyFIU) & 88.10 & 12.5 (Std=6.5, Mdn=11.5) & 1.4 (Std=1.02, Mdn=1) \\
\midrule
Text password (comp8)~\cite{shay2014can}* & 75.0-80.1 & (Mdn=13.2) & 1.3 \\
%Android PIN~\cite{harbach2016anatomy} & 96.86 & 1.96 (Std=1.67, Mdn=1.54) & 1.04 (Std=0.03, Mdn=1.03) \\
%Force-PIN~\cite{KHH16}  & 92 & 3.66 (Std=1.96) & N/A\\
\midrule
Eye tracking~\cite{liu2015eyetracking} & 77.2-91.6 & < 9.6 & 1.37 (Std=0.8, Mdn=1)-1.05 (Std=0.3, Mdn=1) \\
GazeTouchPass~\cite{gazetouchpass} & 65 & 3.13 & 1.9 (Std=1.4, Mdn=1) \\
Face biometric~\cite{Trewin2012} & 96.9 & (Mdn=5.55) & N/A \\
Face \& eyes~\cite{boehm2013safe}* & N/A & 20-40 & 1.1 \\
Face \& voice~\cite{Trewin2012} & 78.7 & (Mdn=7.63) & N/A \\
Voice biometric~\cite{Trewin2012} & 99.5 & (Mdn=5.15) & N/A \\
Gesture (stroke) biometric~\cite{Trewin2012} & 100 & (Mdn=8.10) & N/A \\
\midrule
Android pattern unlock~\cite{harbach2016anatomy} & 87.92 & 0.9 (Std=0.63, Mdn=0.74) & 1.13(Std=0.06, Mdn=1.11) \\
Passpoints~\cite{memorabilitymultiplepasswords}* & 57 & 18.1 (Mdn=15.7)& 2.2 \\
Xside~\cite{LHZMSHS14} & 88 & 3.1-4.1 &N/A \\
\bottomrule\\
{* The study device is a computer.}
\end{tabular}
}}
%\vspace{-5pt}
\caption{
Comparison of usability related metrics of Pixie's camera based two-factor
authentication approach with text, biometric and graphical password
authentication solutions. {\bf The Pixie user entry time is faster than typing
text passwords}.  The results of text-based passwords evaluated in
$\S$~\ref{sec:results} are consistent with those from previous work. Pixie's
median of login trials until success is 1, similar to other solutions.
%\bogdan{Do we have an intuition as to why Pixie has longer user entry time than
%face based authentication. Is it its novelty?} \mizhoo{yes + using a front
%camera for capturing image of face might be easier.}
}
\label{tab:comparison}
\vspace{-10pt}
\end{table*}

In contrast to biometrics, Pixie enables users to change the authenticating
physical factor, as they change accessories they wear or carry.  This 
reduces the risks from an adversary who has acquired the authentication secret
from having lifelong consequences for the victims, thereby mitigating the need
for biometric traceability and revocation~\cite{PPJ03}.

Table~\ref{tab:comparison} compares the user entry times of Pixie with various
other authentication solutions. While Pixie takes longer than
biometric authentication based on face~\cite{Trewin2012}, it is still faster
than several authentication solutions based on
gaze~\cite{boehm2013safe,liu2015eyetracking}. We note that while fingerprint
based authentication is fast and convenient ~\cite{2015biometriciphoneandroid}, 
it is only applicable to devices that invest in such
equipment.  In contrast, cameras are ubiquitously present, including on
wearable devices such as smartwatches and smartglasses.

Pixie needs to solve a harder problem than existing biometrics based
authentication solutions, due to the diversity of its trinkets: while existing
biometrics solutions focus on a single, well studied human characteristic,
Pixie's trinkets can be arbitrary objects.  We note that Pixie can be used in
conjunction with biometric authentication solutions, e.g.,~\cite{DLHBLH12}: in
touchscreen devices, one could use a touch gesture to mark the trinket, as an
additional authentication factor.

%Previous studies~\cite{2015biometriciphoneandroid,DeLuca2015} report that face
%biometrics adoption might be problematic as participants have expressed mixed
%feelings toward using them. For instance, participants in the De Luca et
%al.~\cite{DeLuca2015} user study expressed feeling awkward, as authentication
%can be perceived as taking a selfie in public.

%\bogdan{Liveness verification may be covered in the discussion section.}
%Another concern in camera based authentication solutions is to verify the
%liveness of the authentication secret to prevent spoofing attacks.  In face
%based authentication, liveness can be verified by requiring the users to blink
%or move their mouth upon capturing the image of the face
%~\cite{kollreider2008verifying}.  Boehm et al. ~\cite{boehm2013safe}
%introduced a form of challenge-response liveness verification for gaze based
%authentication where the user gaze at and follows a moving icon on the screen.
%We note that liveness verification solutions, e.g. based on consistency
%between the device motions and motion directions inferred from images captured
%by the camera~\cite{RTC16}, can bring advantages of liveness verification to
%Pixie.

\subsection{Security Tokens and 2 Factor Authentication (2FA)}

The trinket concept is similar to hardware security tokens~\cite{SecurID}, as
authentication involves access to a physical object.  Hardware tokens are
electronic devices that provide periodically changing one time passwords (OTP),
which the user needs to manually enter to the authentication device. Mare et
al.~\cite{mare2016study} found that 25\% of authentications performed in the
daily life employed physical tokens (e.g. car keys, ID badges, etc.).

%They point out that token-based authentication is an issue that requires more
%attention from the research community.

Common software token solutions such as Google's 2-step
verification~\cite{Google2Step}, send a verification code to the mobile device,
e.g. through SMS or e-mail. The user needs to retrieve the verification code
(second authentication factor) and type it into the authentication device.
This further requires the device to be reachable from the server hence
introduces new challenges, e.g. location tracing, delays in phone network, poor
network coverage.  Moreover, such solutions provide no protection when the
device is stolen.  They also impact usability, as the user needs to type both a
password and the verification code.  In contrast, the Pixie trinket combines
the user's secret and the second authentication factor. It also reduces user
interaction, by replacing the typing of two strings with snapping a photo of
the trinket.

Solutions such as~\cite{CDKWB12,SJSN14,KMSC15} treat the mobile device as a
second factor and eliminate user interaction to retrieve a token from the
mobile device to the authentication device (e.g. a desktop) by leveraging
proximity based connectivity (e.g., Bluetooth, Wi-Fi).
%
%In PhoneAuth~\cite{CDKWB12}, where the user authenticates from a browser to a
%remote server, the mobile device stores a private key of the user and uses it
%to sign server issued challenges. Shirvanian et al.~\cite{SJSN14} use the
%mobile device that is connected to the authentication device over several
%channel types to provide a secondary security token for authentication to a
%server.  In SoundProof~\cite{KMSC15}, the second factor, i.e., the proximity
%between the mobile and authentication devices, is ensured by verifying the
%consistency of the ambient noise captured by the two devices' microphone
%sensors.
%
In contrast, Pixie assigns the duty of storing the token for the second factor
to a physical object outside the mobile device. The mobile device is the sole
device that is used to access the services on remote servers. As an added
benefit, the physical factor of the trinket renders Pixie immune to the ``2FA
synchronization vulnerabilities'' introduced by Konoth et
al.~\cite{bandroidfc16}, that exploit the ongoing integration of apps among
multiple platforms. Since Pixie authentication requires a simple interaction
with the user, it is also possible to combine Pixie with a token stored on the 
mobile device. The combined Pixie and mobile device token authentication would
require the user to possess both the particular mobile device that stores the 
token and the trinket.

Pixie also differs from 2FA methods that involve visual tokens, e.g,
QR-codes~\cite{SJSN14,Dodson2010}, as the trinket is secret to the user, 
the attacker needs to discover the trinket and also take possession of it. 
We note that a Pixie variant could
be used in conjunction with the security token concept: the token displays a
pattern (e.g., a QR code, random art), which the user captures 
using Pixie.

\subsection{Wearable Device Authentication}

To address the limited input space of wearable devices, available sensors (e.g.
camera) are commonly exploited to provide alternative input techniques: Omata
and Imai~\cite{skinwatch} identify the input gesture of the user by sensing the
deformation of the skin under the smartwatch. Withana et
al.~\cite{withana2015zsense} use infrared sensors to capture the gesture input
of the user to interact with a wearable device.  Yoon et al.~\cite{YPL15}
exploit the ambient light sensor to capture the changes in light state as a
form of PIN entry for wearable devices. 

Similar to Pixie, cameras integrated in wearable devices have been used to
capture the input for authentication. Van Vlaenderen et al.
~\cite{van2015watchme} exploit the smartwatch camera to provide the device with
an input (e.g. PIN) that is drawn on a canvas, then use image processing
techniques to interpret the captured input.  Chan et al.~\cite{chan2015glass}
propose to pair and unlock smartglasses with the user smartphone by exploiting
the glass camera to scan a QR code that is displayed on the user's phone
screen.  Similarly, Khan et al.~\cite{khan2015sepia} use the smartglass camera
to scan a QR code that is displayed on a point-of-service terminals (e.g. ATM)
to connect to a cloud server for obtaining an OTP.

Wearable devices can be used as the second authentication factor, see
~\cite{bianchi2016wearable} for a survey.  Corner and Noble~\cite{ZIA} use a
wearable authentication token, which can communicate to a laptop over
short-range wireless, to provide continuous authentication to the laptop.  Lee
and Lee~\cite{lee2016implicittoken} use the smartwatch to collect and send the
motion patterns of the user for continuous authentication to a smartphone. 

As Pixie does not require uncommon sensors or hardware, but only a
camera, it is suitable for several camera equipped
wearables~\cite{smartwatch_samsung,smartglass_sony,smartglass_vuzix}. 

\subsection{Graphical Passwords}

Pixie's visual nature is similar to graphical passwords, that include recall,
recognition and cued recall systems (see~\cite{BCV12} for a survey).  Recall
based solutions such as DAS (Draw-A-Secret)~\cite{JMMRR99} and
variants~\cite{DY07,GGCWL09} ask the user to enter their password using a
stylus, mouse or finger. For instance, De Luca et al.~\cite{LHZMSHS14} proposed
to enter the stroke based password on the front or back of a double sided touch
screen device.  In recognition-based systems (e.g.,
Passfaces~\cite{Passfaces,DP00}), users create a password by selecting and
memorizing a set of images (e.g., faces), which they need to recognize from
among other images during the authentication process. 

%Cued-recall systems improve password memorability by requiring users to
%remember and target specific locations in an image~\cite{WWBBM05,WWBBMb05}.

Pixie can be viewed as a recognition based graphical password system
where the possible secret images are dynamically generated based on the
physical world around the user.  Since the user freely presents the candidate
password through a photo of the physical world, captured in different light,
background, and angle conditions, Pixie has to implement an accurate matching
of trinkets.  Trinkets can be small portions of items worn by users (e.g.,
shirt pattern, shoe section).
%
%Thus, even if the attacker is able to see and reproduce the trinket, the
%attacker does not know the required section and angle of the trinket (see
%$\S$~\ref{sec:evaluation:attacks}).
%
Pixie accurately verifies that the candidate image contains the same trinket
part as a set of previously captured reference images.  This process endows
Pixie with attack resilience properties: to fraudulently authenticate, an
adversary needs to capture both the mobile device and the trinket, then guess
the correct part of the trinket.

\subsection{Text-Based Passwords}

The usability of traditional text-based passwords has been well studied in
literature, see
e.g.,~\cite{Trewin2012,melicher2016usability,memorabilitymultiplepasswords,shay2014can}.
Trewin et al.~\cite{Trewin2012} found that face biometrics can be entered
faster than text based passwords and Table~\ref{tab:comparison} shows that
Pixie is also faster than text based passwords.  Several limitations are
associated with text passwords on memorability and usability especially when
adopted in mobile platforms.  For instance, Shay et al.~\cite{shay2014can} have
shown through a large user study of different password-composition policies,
that more than 20\% of participants had problems recalling their password and
35\% of the users reported that remembering a password is difficult. Their
reported user entry time for text passwords ranges between 11.6-16.2s (see
Table~\ref{tab:comparison}) in line with our evaluation (see
$\S$~\ref{sec:results:time}). Pixie is also perceived as more memorable than 
text passwords (see~\ref{sec:results:perception}).

Melicher et al.~\cite{melicher2016usability} found that creating and entering
passwords on mobile devices take longer than desktops and laptops. In mobile
devices, text-based passwords need to be entered on spatially limited keyboards
on which typing a single character may require multiple touches~\cite{SWKW13},
due also to typing the wrong key.  Pixie replaces typing a password with
pointing the camera to the trinket and snapping a photo of it. 

%\subsection{Summary of Pixie Differences}
%
%Pixie trinkets are the (non-electronic) authentication factors.  Instead of
%the mobile device changing the OTP, Pixie allows the user to change the
%trinket.  Pixie shares the camera input convenience with face and visual token
%(e.g., QR code) based authentication solutions. Pixie is however significantly
%more complex to implement, as Pixie tokens are drawn from a much larger space,
%and thus lack the convenience of a well defined set of features. Pixie is easy
%to use: typing two strings in conventional 2FA is replaced by pointing the
%camera at the trinket.

%Further, Pixie shares the shoulder surfing vulnerabilities and susceptibility
%to inappropriate lighting of face based autentication.

%Pixie also offers protection when both factors (the user's device and trinket)
%are captured by the adversary: the adversary also needs to correctly guess the
%trinket's section and angle with respect to the camera.

\section{System and Adversary Model}
\label{sec:model}

\subsection{System Model}
\label{sec:model:system}

\begin{figure*}
\centering
\subfigure[]
{\label{fig:system}{\includegraphics[width=0.47\textwidth, height=1.63in]{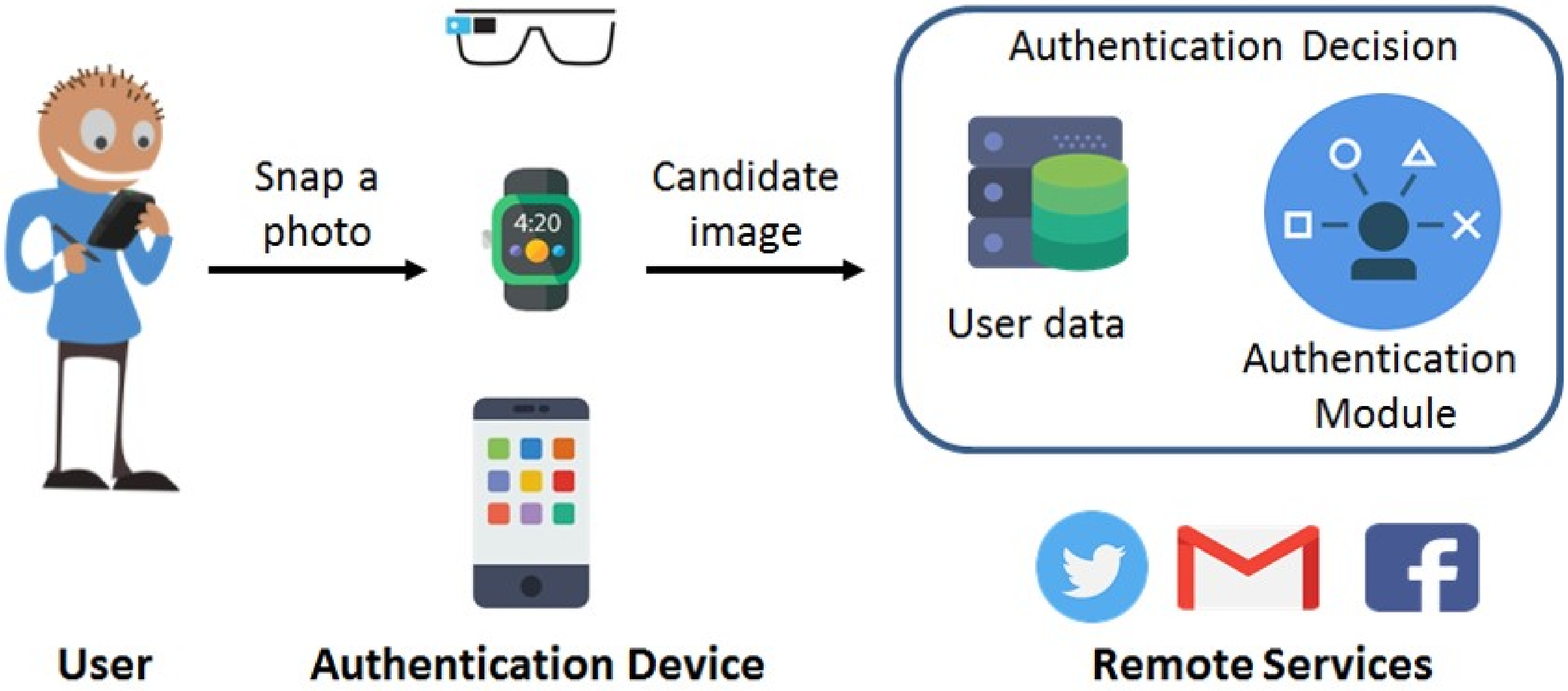}}}
\hspace{0.03\textwidth}
\subfigure[]
{\label{fig:pixie:diagram}{\includegraphics[width=0.47\textwidth, height=1.63in]{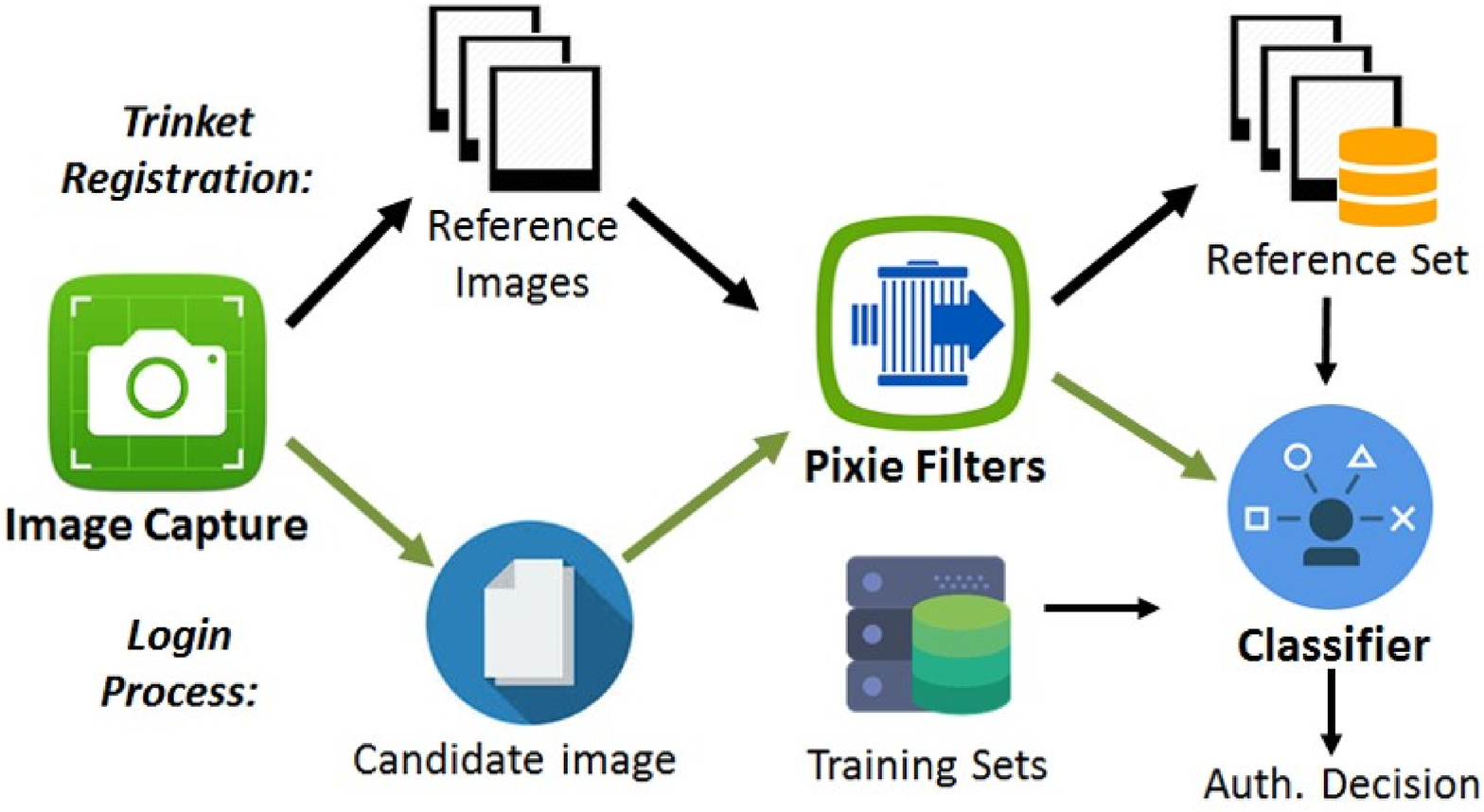}}}
\vspace{-10pt}
\caption{
(a) {\bf System model}: the user authenticates through a camera equipped device 
(smartphone, smartwatch, Google Glass, car), to a remote service, e.g., e-mail, 
bank, social network account. The remote service stores the user credentials 
and performs the authentication.
(b) {\bf Pixie registration and login workflows}: to register, the user
captures ``reference images'' of the trinket, which are filtered for quality
and consistency.  To authenticate, the user needs to capture a ``candidate
image'' of the trinket that matches the reference images.}
\vspace{-15pt}
\end{figure*}

%\begin{figure}[t!]
%\centering
%\includegraphics[width=0.47\textwidth, height=1.63in]{figures/pixie/system.three.eps}
%\vspace{-10pt}
%\caption{
%\footnotesize{
%System model: the user authenticates through a camera equipped device
%(smartphone, smartwatch, Google Glass, car), to a remote service, e.g., e-mail,
%bank, social network account. The remote service stores the user credentials
%and performs the authentication.}}
%\label{fig:system}
%\vspace{-10pt}
%\end{figure}

Figure~\ref{fig:system} illustrates the system model. The user has a camera
equipped device, called the {\it authentication device}.  Authentication
devices include smartphones, tablets, resource constrained devices such as
smartwatches and smartglasses, and complex cyber-physical systems such as cars.
The user uses the authentication device to access remote services such as
e-mail, bank and social network accounts, or cyber-physical systems, e.g., home
or child monitoring systems (see $\S$~\ref{sec:applications} for a discussion
on other related scenarios).

We assume that the user can select and easily access a physical object, the
{\it trinket}.  The user sets the authentication secret to consist of multiple
photos of the trinket, taken with the device camera. We call these
``reference'' images, or reference set. To authenticate, the user snaps a
``candidate'' image of the trinket. This image needs to match the stored,
reference set. Figure~\ref{fig:system} illustrates an approach where the remote
service stores the user's reference set and performs the image match operation.
In $\S$~\ref{sec:discussion} we compare the merits and drawbacks of this
approach to one where the authentication device performs these tasks.

Pixie can be used both as a standalone authentication solution and as a
secondary authentication solution, e.g., complementing text based passwords.

%In this paper, we focus on the case
%where the user's mobile device stores the reference images, and performs the
%image match. In addition, the device associates the reference images with the
%user's remote authentication credentials (e.g. OAuth~\cite{rfc6749}). If the
%image match succeeds, the credentials are sent to the remote service.
%\newmaterial{However, we note that the remote service can also perform
%these operations: store the reference images and perform the image
%match over candidate images captured on and sent by the mobile device.}

%While the trinket is essentially an authentication token, it is not
%(necessarily) electronic, but can be any object easily accessible to the user
%during the authentication process, e.g., wallet, watch, clothing pattern.
%Pixie also differs from biometric authentication, since the trinket can be
%changed, as users change accessories they wear and carry.

\subsection{Applications}
\label{sec:applications}

While this paper centers on a remote service authentication through a mobile
device scenario, Pixie has multiple other applications. First, authentication
in camera equipped cyber-physical systems. For instance, cars can use Pixie to
authenticate their drivers locally and to remote services ~\cite{securiCode}. 
Pixie can also
authenticate users to remote, smart house or child monitoring systems, through
their wearable devices.  Further, doorlocks, PIN pads ~\cite{securiCode,
schlage} and fingerprint readers can be replaced with a camera through which
users snap a photo of their trinket to authenticate.

Pixie can be used as an alternative to face based authentication when the users
are reluctant to provide their biometric information (e.g. in home game systems
where the user needs to authenticate to pick a profile before playing or to
unlock certain functionalities). Pixie can also be used as an automatic access
control checkpoint (e.g. for accessing privileged parts of a building). The
users can print a visual token and use it to pass Pixie access control
checkpoints.

In addition, given the large number of people who work from
home~\cite{guardiannews}, Pixie can provide an inexpensive 2FA alternative for
organizations to authenticate employees who are connecting to the private
network remotely~\cite{esecurity}: replace the hardware tokens with user chosen
Pixie trinkets.

\subsection{Adversary Model}
\label{sec:model:adversary}

We assume that the adversary can physically capture the mobile device of the
victim. We also assume that the adversary can use image datasets that he
captures and collects (see $\S$~\ref{sec:data}) to launch brute force {\bf
pictionary  attacks} against Pixie (see $\S$~\ref{sec:evaluation:attacks:pictionary}).

Similar to PIN based authentication to an ATM, Pixie users need to make sure
that onlookers are far away and cannot see the trinket and its angle. However,
we consider a {\bf shoulder surfing} attack flavor where the adversary sees or
guesses the user's trinket object type. The adversary can then use datasets of
images of similar objects to attack Pixie (see $\S$~\ref{sec:evaluation:attacks:shoulder}).

Further, we also consider an adversary that attempts to launch a {\bf master
image attack}, i.e., identify images that contain diverse features and match
many trinkets. Example master images include ``clutter'' images, with an array
of shapes, colors and shadows (see $\S$~\ref{sec:evaluation:attacks:master}).

%Similar to face based authentication, we assume that the adversary cannot
%perform a standard shoulder surfing attack, i.e., capture a picture of the
%trinket,

%The adversary cannot access the stored image password, that could be
%secured through hardware-level protection, e.g.,~\cite{TrustZone}, or through
%privacy preserving feature extraction and matching
%solutions~\cite{WHWR16,QYRCW14,HLP12}, see $\S$~\ref{sec:discussion} for a
%discussion.
%Depending on the scenario, the adversary seeks to either authenticate on the
%captured device, or to use the device to gain access to the victim's online
%service accounts.

\section{Pixie}
\label{sec:pixie}

\subsection{Pixie Requirements}
\label{sec:pixie:reqs}

In addition to being resilient against attacks (see
$\S$~\ref{sec:model:adversary}), Pixie needs to satisfy the following
requirements:

\begin{compactitem}

\item
{\bf Trinket image quality}.
Pixie needs to ensure the quality of trinkets and images.  Early pilot studies
showed that not all the trinkets that the users chose, or the photos that they
took, were suitable for authentication.

\item
{\bf Trinket match}.
Pixie needs to match images of the same trinket, even when captured with a
different background, lighting, or from a slightly different distance or angle.

\item
{\bf Discoverability}.
New users should easily discover the functionality of Pixie.

\item
{\bf Deployability}.
Pixie should be easy to integrate into existing systems.

\end{compactitem}

\noindent
Figure~\ref{fig:pixie:diagram} depicts the modular approach we use for Pixie to
address these goals. The image capture module seeks to address part of the
first requirement, by facilitating the capture of quality trinket images. The
authentication module tackles the second requirement through the use of trained
classifiers to match trinket images.  To simultaneously address the first and
third requirements, i.e., to ensure the discoverability of Pixie while guiding
new users through the capture of high quality photos and the choice of visually
complex objects as the secret, the filter module detects and eliminates low
quality images and invalid reference sets. We now detail each module.

\subsection{Image Capture \& Feedback}
\label{sec:pixie:ui}

\begin{table}
\vspace{5pt}
\centering
\resizebox{\textwidth}{!}{%
\textsf{
\begin{tabular}{l l}
\toprule
\textbf{Requirement} & \textbf{Pixie Feature} \\
\midrule
\multirow{2}{*}{\begin{tabular}[c]{@{}l@{}}
Increase the size of trinket \&\\ Reduce the background area in trinket images\end{tabular}} & 1. Disable camera zoom\\ & 2. Overlay a circle as the target area on the camera view\\
\midrule
\multirow{2}{*}{\begin{tabular}[c]{@{}l@{}}
Ensure the quality of reference and candidate images \\ Ensure consistency of reference images\end{tabular}} & \begin{tabular}[c]{@{}l@{}}1. Design prefilters for checking the quality of images\end{tabular}\\ & 2. Translate the prefilter criteria into actionable feedback to the users \\
\midrule     
\multirow{3}{*}{\begin{tabular}[c]{@{}l@{}}
Improve the discoverability of Pixie\end{tabular}} & \begin{tabular}[c]{@{}l@{}}1. Show number of remaining images to take in registration screen\end{tabular}\\ & \begin{tabular}[c]{@{}l@{}}2. Show camera capture icon for login page\end{tabular}\\ &3. Add step by step in-app instruction on how to use Pixie\\
\bottomrule
\end{tabular}
}}
\caption{Summary of user interface improvements identified during pilot studies.}
\label{tab:ui:improvements}
\vspace{-15pt}
\end{table}

We performed pilot studies to identify early problems with the Pixie user
interface. Table~\ref{tab:ui:improvements} summarizes the design improvements
we made to the Pixie UI.  For instance, during the pilot studies, some users
captured trinket photos whose background provided more features than the
trinkets. This revealed that the trinket needs to be the main object in
captured images. To simultaneously satisfy this requirement, and the trinket
quality requirement above, we design Pixie to guide the user to take larger
photos of trinkets. We achieve this by overlaying a circle on the camera image:
the user needs to fit the trinket impression inside the circle (see
Figures~\ref{fig:pixie:setup} and~\ref{fig:pixie:login}).  Since Pixie does not
allow zooming in, the user needs to bring the camera closer to the trinket,
hence take a larger photo.  Pixie crops the image, and retains only the largest
rectangle parallel to the sides of the device that fits the circle. 

In addition, we observed that the quality of trinket images captured by the
users could be low (e.g. blurry or dark), or the users may take inconsistent
trinket images in the registration phase.  To ensure the quality of trinket
images and the consistency of reference images, we identified common problems
that occur during the image capture process (e.g., insufficient light, trinket
with plain texture). Then, we mapped prefilter rejection decisions provided by
Pixie's image filter (see $\S$~\ref{sec:pixie:filter}) into informative error
messages (see Figure~\ref{fig:pixie:capture:2}). Furthermore, to facilitate the
discoverability of Pixie, we designed and included a step by step in-app
instruction guide on how to use Pixie.

\subsection{The Authentication Module}
\label{sec:pixie:authentication}

\begin{figure}
\centering
\includegraphics[width=2.9in]{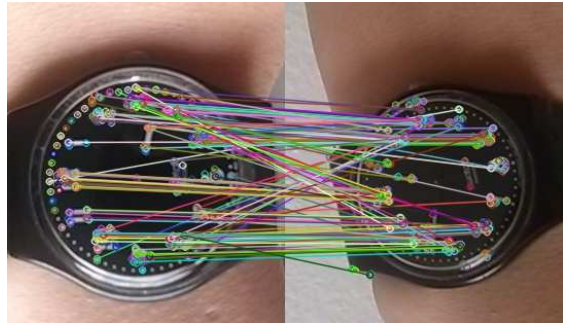}
\caption{Example ORB keypoint matches between two images of the same trinket,
taken in different conditions. {\bf Each line represents a match}: it connects
matching keypoints (shown as small colored circles) from each image.}
\label{fig:match}
\vspace{-15pt}
\end{figure}

The authentication module is responsible for addressing Pixie's second
requirement (see $\S$~\ref{sec:pixie:reqs}), of matching the candidate image
against the reference images. Pixie extracts robust keypoints from these
images, identifies a suite of features from the keypoint match process, then
uses them to train a classifier that decides if the
candidate image matches the reference set. We now detail this process. Let $C$
denote the candidate image, $\overline{R}$ be the set of reference images, and
$R$ be any of the reference images (see $\S$~\ref{sec:model:system}).

%To address the trinket
%match challenge (see $\S$~\ref{sec:pixie:overview}) Pixie extracts robust
%keypoints from these images, then computes a 1-to-1 mapping between the
%resulting keypoint sets (see Figure~\ref{fig:match} for an illustration), and
%filters out low quality matches. Pixie extracts a suite of features from the
%keypoint match process and uses supervised learning to decide if the candidate
%image matches the reference set.  We now detail this process. Let $C$ denote
%the candidate image, $\overline{R}$ be the set of reference images, and $R$
%be any of the reference images.

\noindent
{\bf Keypoint matching}.
We use SURF (Speeded Up Robust Features)~\cite{BTVG06} and
ORB~\cite{RRKB11} algorithms, to extract scale and rotation invariant image
keypoints from the candidate and reference images, e.g., shown as small colored
circles on images in Figure~\ref{fig:match} and ~\ref{fig:bad:image}. 
We also
extract the descriptors of the keypoints, which represent their characteristics.
To determine if a candidate image $C$ and a reference image $R$
contain the same trinket, we compute a 1-to-1 matching between their keypoint
descriptors (e.g., shown as lines in Figure~\ref{fig:match}). We use
brute-force matching for ORB keypoints, where each keypoint of the candidate
image is matched with the closest keypoint (in terms of Hamming distance) of
the reference image. For SURF keypoints, we use the FLANN-based
matcher~\cite{muja2009fast}.

%
% maybe for journal
%
%\begin{compactitem}
%
%\item
%{\bf Brute-Force}.
%We assume a distance function $d$ that takes as input two interest points and
%returns their distance. For ORB, that has binary string based descriptors, we
%use the Hamming distance. For each reference point $r \in KP(R)$, compute
%the distance $d(r, c)$, for all points $c \in KP(C)$. Match $r$ with the
%point $c$ ``closest'' to $r$, i.e., $d(r,c) = min \{d(r,i) | \forall i \in
%KP(C) \}$. That is, brute-force finds the closest reference point for
%each candidate point.
%
%\item
%{\bf FLANN-based Matcher (openCV)}.
%FLANN-based matcher \cite{muja2009fast} are a set of algorithms optimized for
%fast approximate nearest neighbor search in large datasets and for high
%dimensional features. Compared to the brute-force matcher, the FLANN-based
%matcher performs faster on large collections with more key points. We use the
%FLANN matcher in conjunction with SURF.
%
%\end{compactitem}

An exhaustive matching of each keypoint in the candidate image to a keypoint in
the reference image will produce low quality, {\it outlier} matches. We
experimented with several existing filters, including threshold, cross checking
and RANSAC~\cite{FB81}, to identify and remove outlier matches. The RANSAC
based filter performed the best, hence we use it implicitly in the following.

\noindent
{\bf Image similarities}.
Given two images $C$ and $R$, we define their similarity $Sim(C, R)$ to be the
ratio between the number of keypoint matches of $C$ and $R$, after the above
filter and outlier detection steps, and the number of keypoints in $C$. Given
$C$ and the set $\overline{R}$, we define the nearest neighbor similarity of
$C$ to $\overline{R}$ as $NNSim(C,\overline{R})$ = max $\{ Sim(C,R)| \forall R
\in \overline{R}\}$, and the farthest neighbor similarity,
$FNSim(C,\overline{R})$ = min $\{ Sim(C,R)| \forall R \in \overline{R}\}$.

Given a reference set $\overline{R}$, we define the average nearest neighbor
similarity value of each reference image, to the other reference images in
$\overline{R}$: $AvgRefNN(\overline{R})$ = $\frac{\Sigma_{R \in \overline{R}}
NNSim(R, \overline{R}-R)}{|\overline{R}|}$.  Similarly, we define the average
farthest neighbor similarity value of each reference image to the other images
in $\overline{R}$: $AvgRefFN(\overline{R}) = \frac{\Sigma_{R \in \overline{R}}
FNSim(R, \overline{R}-R)}{|\overline{R}|}$.

\noindent
{\bf Template image}.
Given a reference set $\overline{R}$, we define its \textit{template image},
$T(\overline{R})$, as the reference image $R$ whose value $\sum_{r \in
\overline{R}-R} Sim(r, R)$ is the maximum among all reference images in
$\overline{R}$. Intuitively, $T(\overline{R})$ is the reference image ``closest
to the center'' of the reference set. We define $AvgRefTempl(\overline{R})$ as
the average similarity of images in $\overline{R}$ to $T(\overline{R})$.

\noindent
{\bf Pixie matching features}.
We use the above concepts to extract the following features (see
Table~\ref{tab:features}, top section). We use these features to train a
supervised learning algorithm.

\noindent
$\bullet$
{\it Keypoint counts}.
The keypoint count of $C$ and $T(\overline{R})$.

\noindent
$\bullet$
{\it Match based features}.
The number of keypoints in $C$ and $T(\overline{R})$ that match, before the
RANSAC filter.  The min, max, mean and SD of the distance, size, response and
angles between the matched keypoints in $C$ and $T(\overline{R})$, after
RANSAC.

\noindent
$\bullet$
{\it Quality of the reference set}.
$AvgRefNN(\overline{R})$, $AvgRefFN(\overline{R})$ and
$AvgRefTempl(\overline{R})$.

\noindent
$\bullet$
{\it Similarity to template}.
The similarity of $C$ to $T(\overline{R})$, normalized by the
average similarity of the images in $\overline{R}$ to $T(\overline{R})$,
i.e., $\frac{Sim(C, Templ(\overline{R}))}{AvgRefTempl(\overline{R})}$.

\noindent
$\bullet$
{\it Similarity to reference set}. We define
$minSim(C, \overline{R}) =\frac{min \{ Sim(C,R)| \forall R \in \overline{R}
\}}{AvgRefFN(\overline{R})}$: the ratio of the similarity between $C$ and
``farthest'' reference image, and the average least similarity between
reference images.  Similarly,\\ $maxSim(C, \overline{R}) = \frac{max \{ Sim(C,R)|
\forall R \in \overline{R} \}}{AvgRefNN(\overline{R})}$.
%
%Similarly, compute $maxSim(C, \overline{R})$ as the ratio between the
%similarity of $C$ to the closest reference image, and the average most
%similarity between reference images.

\noindent
$\bullet$
{\it Homography}:
Output of homography between $C$ and $T(\overline{R})$: the perspective
transformation between the planes of the two images (3 features).\\

%\noindent
%We use these features to train supervised learning algorithms, using manually
%labeled data. We present experimental results in $\S$~\ref{sec:evaluation}.

%
% for journal
%
%\begin{table}
%\centering
%\textsf{
%\begin{tabular}{|c|c|c|c|}
%\hline
%Similarity Table & $R_1$ & $R_2$ & $R_3$\\
%\hline
%$R_1$ & 1 & \cellcolor{red}0.9 & 0.8\\
%\hline
%$R_2$ & \cellcolor{red}0.7 & 1 & 0.6\\
%\hline
%$R_3$ & 0.7 & \cellcolor{red}0.8 & 1\\
%\hline
%\hline
%$C$ & 0.7 & 0.5 & \cellcolor{red}0.9\\
%\hline
%\end{tabular}
%}
%\caption{Similarity table of three reference images $\overline{R} =
%\{R_1,R_2,R_3\}$ and a candidate image $C$. Red cells correspond to the nearest
%neighbor.  $R_1$ is the template image. $AvgRefNN = (0.8+0.7+0.9)/3=0.8$,
%$AvgRefFN = (0.7+0.6+0.8)/3=0.7$ and $AvgRefTempl = (0.8+0.9)/2=0.85$. Then,
%$minSim(C,\overline{R}) = 0.5/0.7$, $maxSim(C,\overline{R}) = 0.9/0.8$ and
%$TemplSim = 0.7/0.85$.
%\label{tab:similarity}}
%\end{table}

\subsection{Pixie Filters}
\label{sec:pixie:filter}

\begin{table}
\centering
\resizebox{0.75\textwidth}{!}{%
\textsf{
\begin{tabular}{l l l}
\toprule
\textbf{Solution} & \textbf{Features} & \textbf{Details}\\
\midrule
& Keypoint stats. & Statistics of ORB/SURF keypoints \\
Pixie & Keypoint nearest neighbors & Keypoint match stats. \\
& Perspective transformation & RANSAC optimal map of match keypoints \\
\midrule
& Keypoints & Count and spread of keypoints\\
Pixie filters & Edge pixels & Count and spread of edge pixels\\
& Reference quality & Reference image similarities stats.\\
\bottomrule
\end{tabular}
}}
\caption{Summary of (top) Pixie features and (bottom) Pixie filter features.}
\label{tab:features}
\vspace{-15pt}
\end{table}

Early pilot studies revealed that Pixie users can capture low quality images.
Such images, either reference or candidate, hinder the ability of the
authentication module to discern candidate images, increasing the FRR of Pixie. 
Furthermore, they impose gratuitous network latency in the
remote authentication scenario (see $\S$~\ref{sec:model:system}).

Several conditions may prevent taking high quality images
Figure~\ref{fig:bad:image}(d)-(f) shows example outcomes of such conditions,
including (i) improper lighting, (ii) unsteady hand and (iii) choice of
trinkets with constant texture.
%
%and (iv) choice of a complex background that generates irrelevant keypoints.
%
We also observed that some pilot study participants, during the reference set
registration process, took photos containing different trinkets, or different
areas of the same trinket.
%
%Such reference images may not only reduce the accuracy
%of the authentication process, but may also introduce vulnerabilities: an
%attacker may find it easier to capture a candidate image that is similar to
%one of 3 different reference images.
%
To address these issues, we introduce a set of filters (see
Figure~\ref{fig:pixie:diagram}) that reject problematic images captured by the
user. We propose the {\it two rules of filtering}, that set out the operation
space for Pixie image filters:

\noindent
$\bullet$
{\bf Filter Rule \#1}:
Pixie may not willfully fail by operating on images on which it predicts it
will fail.

\noindent
$\bullet$
{\bf Filter Rule \#2}:
Pixie may not operate in a space where it has not been trained.
\noindent

\noindent
In the following, we detail these rules and describe the resulting filters.

\vspace{-5pt}

\subsubsection{Filter Rule \#1: CBFilter and RBFilter}

We introduce CBFilter and RBFilter, filters that identify reference and
candidate images on which they predict Pixie will fail. The filters leverage
the following features, (see Table~\ref{tab:features}(bottom section) for a
summary).

\noindent
{\bf Filter features}.
First, we define KP-CNT as the keypoint count of an image. The intuition for
using this feature is that an image with a low KP-CNT (e.g.,
Figure~\ref{fig:badimage:bluriness} with only 5 keypoints) is likely to
negatively impact the accuracy of Pixie's matching process. A second feature is
based on the center, or {\it centroid} of the keypoints extracted from an
image: Let DTC-KP (distance to center of keypoints) denote the average distance
between the keypoints of the image and their centroid. DTC-KP measures the
spread of the keypoints across the image.  The intuition is that a high DTC-KP
may indicate that some keypoints do not belong to the trinket but to the
background.
Third, to detect blurry images, we use the Canny edge
detector~\cite{canny1986computational} to identify edge pixels that delimit
objects in the image. Let White-CNT denote the number of detected edge
(``white'') pixels of an image. White-CNT is an indicator of the clarity of the
image: a low White-CNT denotes a blurred image, with few trinket edges. We also
introduce DTC-White (distance to center of white pixels), the average distance
of the white pixels to their centroid. DTC-White denotes the spread of the edge
pixels, i.e., the size of the trinket.
Finally, to detect inconsistent reference images, we define $MinCrossSim(\overline{R})$,
$MaxCrossSim(\overline{R})$ and $AvgCrossSim(\overline{R})$, to be the minimum,
maximum and average similarity (see $\S$~\ref{sec:pixie:authentication}) among
all the pairs of images in $\overline{R}$. Small cross similarity values
indicate reference images of non-identical trinkets.
%We conjecture that DTC-White will identify images with complex backgrounds:
%several white (edge) pixels will belong to the background and not the trinket.

\begin{figure*}
\centering
\subfigure[]
{\label{fig:badimage:2dplot:alwayscorrect}{\includegraphics[width=1.52in]{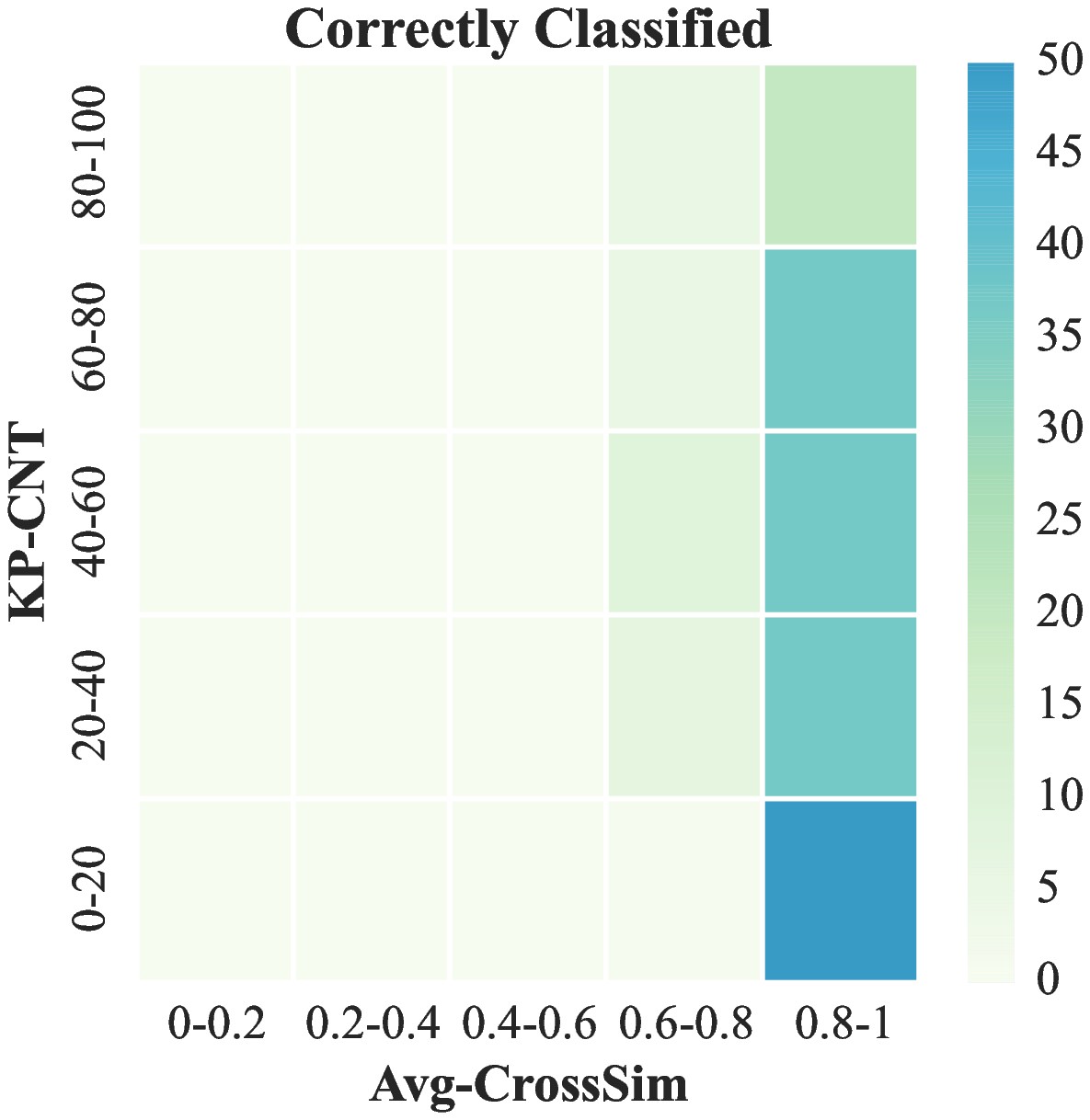}}}
\subfigure[]
{\label{fig:badimage:2dplot:falsenegative}{\includegraphics[width=1.52in]{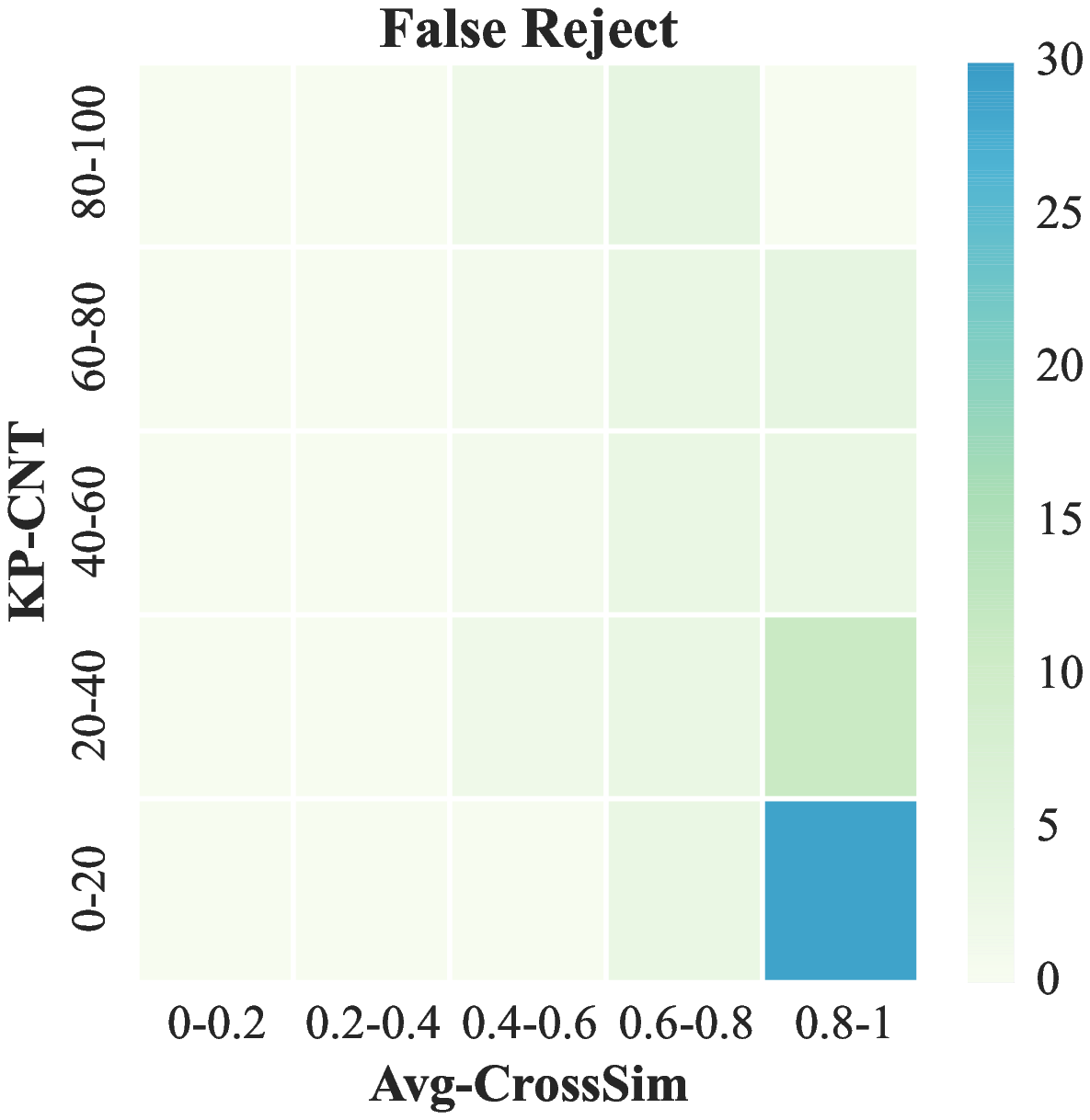}}}
\subfigure[]
{\label{fig:badimage:2dplot:flasepositive}{\includegraphics[width=1.52in]{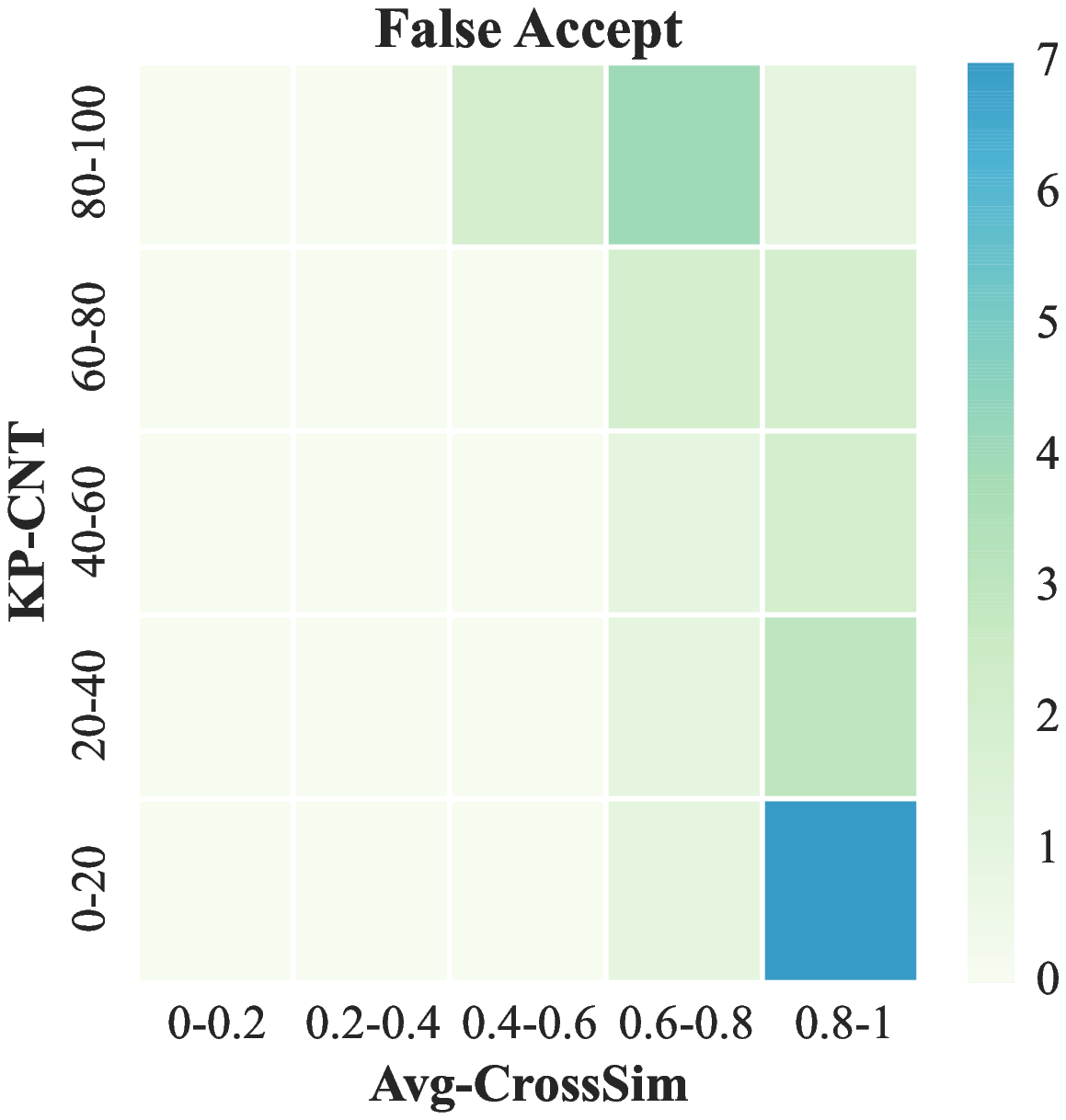}}}
\subfigure[]
{\label{fig:badimage:rulebased:avgcrosssim}{\includegraphics[width=1.52in]{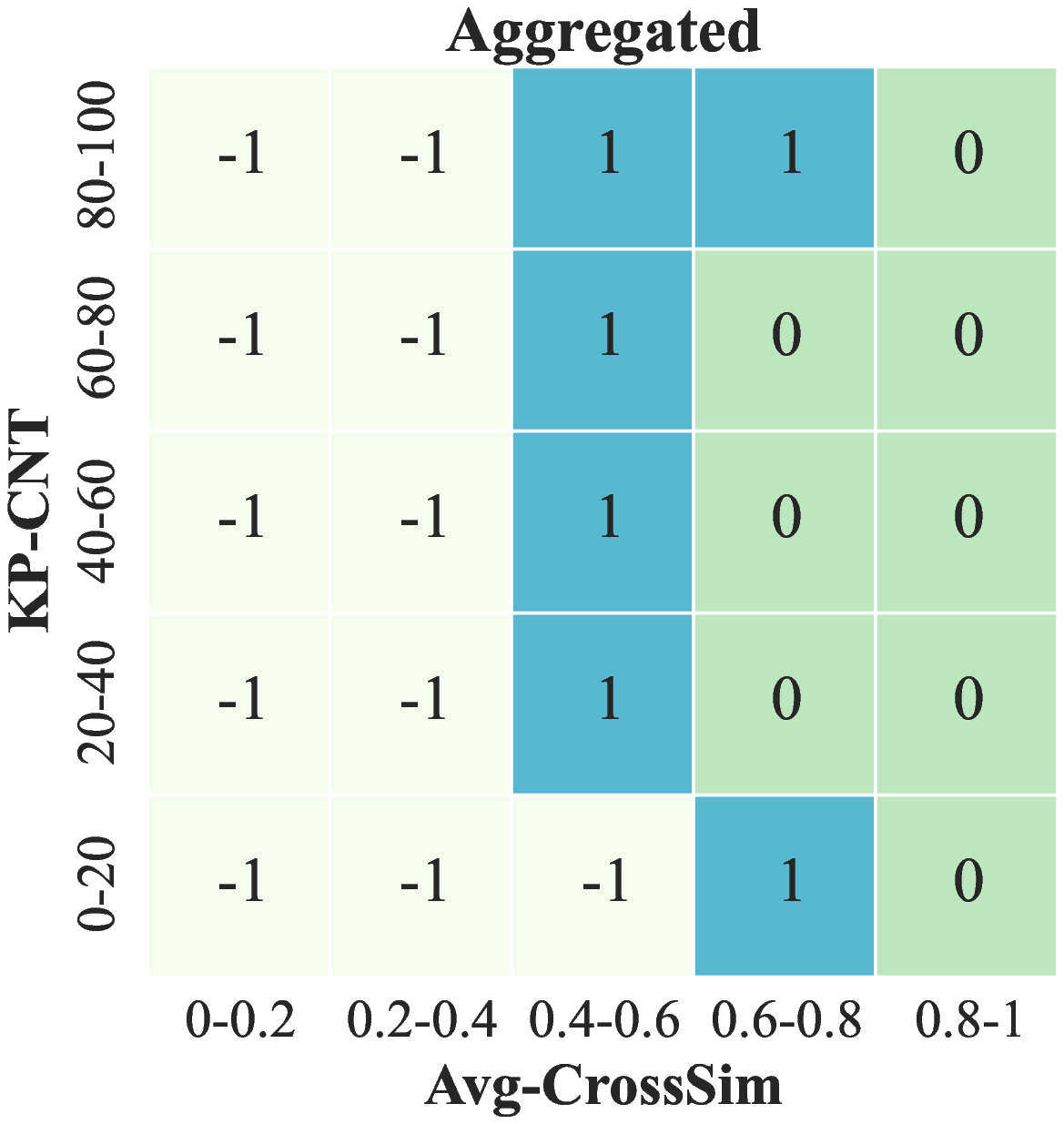}}}
\vspace{-15pt}
\caption{
\footnotesize{
Example 2D histograms of KP-CNT of template image vs.
$AvgCrossSim(\overline{R})$.
(a) Correctly classified instances.
(b) False reject instances.
(c) False accept instances. The legend in (a)-(c) shows the color code used for
the number of authentication instances.
(d) Aggregated 2D histogram. The darker regions with 1 in the center have a
greater proportion of misclassified than correctly classified instances.  The regions with -1 in the center correspond to value ranges on which we have
no template images. Conclusion: Filter out reference sets with KP-CNT $< 20$
and $AvgCrossSim(\overline{R}) < 0.6$.}
\label{fig:bad:image:2dplot:filtering}}
\vspace{-15pt}
\end{figure*}

\noindent
{\bf CBFilter: Classifier Based Filter}.
Given the reference set $\overline{R}$ and its template image $T(\overline{R})$
(see $\S$~\ref{sec:pixie:authentication}), CBFilter uses a suite of features to
train a supervised learning algorithm and determine if $\overline{R}$ is
suitable to participate in the authentication process. The features include
KP-CNT, DTC-KP, White-CNT, DTC-White of $T(\overline{R})$, the average, minimum
and maximum of KP-CNT, DTC-KP, White-CNT, DTC-White over all the images in
$\overline{R}$, and $MinCrossSim(\overline{R})$, $MaxCrossSim(\overline{R})$
and $AvgCrossSim(\overline{R})$.

\noindent
{\bf RBFilter: Rule Based Filter}.
Pilot studies demonstrated the need to give relevant feedback to users as
early as possible: early pilot study participants expressed
frustration when they discovered that the photos they took were not suitable
at the end of the registration, or worse, during the authentication process.
The output of CBFilter cannot however be used to provide meaningful feedback.

To address this limitation, we identified common problems that occur during the
image capture process, e.g., improper light, trinket with plain texture or not
identical reference images. We then developed a set of rules for these filter
features, that (i) predict if an image or image set will not perform well
during authentication, and (ii) that can be transposed to one of the problems
identified.
For instance, we found that a small KP-CNT is associated with insufficient
light, blur, or trinkets with a plain texture, while a small $AvgCrossSim$
value can indicate reference images containing non-identical trinkets.
Figure~\ref{fig:pixie}(c) illustrates the feedback provided when the user
captures a low quality trinket (top) or inconsistent reference images (bottom).

\begin{table}
\centering
\resizebox{0.59\textwidth}{!}{%
\textsf{
\begin{tabular}{l l l}
\toprule
\textbf{Image type} & \textbf{Filter Rule} & \textbf{Interpretation}\\
\midrule
Reference & KP-CNT $< 20$        & Low quality image or plain trinket \\
Reference & DTC-KP $< 30$        & Low quality image or plain trinket\\
Reference & AvgCrossSim $< 0.6$  & Non-identical trinkets in reference set\\
\midrule
Candidate & KP-CNT $< 20$        & Low quality image or plain trinket\\
%Candidate & White-CNT $< 2000$  & Blurry image\\
\midrule
Candidate & DTC-KP $> 44,600$ & Out of bounds image\\
Candidate & White-CNT $> 22,400$ & Out of bounds image\\
Candidate & DTC-White $> 160$ & Out of bounds image\\
\bottomrule
\end{tabular}
}}
\caption{
RBFilter and UBounds filter rules for reference and candidate images,
and their real world interpretation. RBFilter (top 2 sections) filters images on
which it predicts Pixie will fail. UBounds (bottom section) filters images
outside the space seen by Pixie during training.}
\label{tab:rule:filters}
\vspace{-15pt}
\end{table}

To identify such rules, we run Pixie on the Pixie dataset,
and investigate reference sets and candidate images
that contributed to misclassified instances as follows.
%To identify filter rules that detect low quality reference sets, we perform the
%following process. 
For each pair of the above filter features, we plot the 2D
histogram of instances that were correctly classified, and that contributed to
false accepts (FA) and false rejects (FR).
Figures~\ref{fig:bad:image:2dplot:filtering}(a)-(c) illustrates this process
for the KP-CNT of template images $T(\overline{R})$ vs. 
$AvgCrossSim(\overline{R})$ pair of features. 
Then, we aggregate the results for the three 2D histograms, see
Figure~\ref{fig:bad:image:2dplot:filtering}(d), by calculating the contribution
of each type of classification result (i.e., FA, FR, True Accept (TA) and True
Reject (TR)) in a cell of the 2D histogram. The dark regions have a larger
proportion of misclassified than correctly classified instances.  This enables
us to identify ``problem'' regions, where the contribution of misclassified
instances (FA and FR) is larger than that of correctly classified instances (TA
and TR). We then define rules, i.e., threshold values, that avoid clusters of
problem regions.
For instance, based on the bottom area of
Figure~\ref{fig:badimage:rulebased:avgcrosssim}, we reject reference
sets whose template has KP-CNT $< 20$. Similarly, we reject reference sets with
$AvgCrossSim(\overline{R}) < 0.6$, as we have none with 
$AvgCrossSim(\overline{R}) < 0.4$ (cells with $-1$), and those in $[0.4,
0.6]$ are frequently misclassified.

%\begin{figure}
%\centering
%\subfigure[]
%{\label{fig:badimage:rulebased:dtckp}{\includegraphics[width=2.5in]{figures/rulebased.filter/ORB.Reference/DTC-KPs.eps}}}
%\vspace{-10pt}
%\caption{Aggregated 2D histogram for KP-CNT vs. DTC-KP of template images.
%Conclusion: Filter out images with DTC-KP \textless 30 and images with KP-CNT
%\textless 20 (duplicate rule). \bogdan{I don't buy this, as 2 out of 5 cells
%are 0, and only 3 are 1.}
%\label{fig:bad:image:rulebased:filtering}}
%\vspace{-15pt}
%\end{figure}

%Similarly, Figure~\ref{fig:bad:image:rulebased:filtering} plots the
%aggregate 2D histogram for the KP-CNT vs. DTC-KP values of $T(\overline{R})$.
%It confirms that we can filter reference sets whose template image has KP-CNT $< 20$.
%It also shows that we can filter template images with DTC-KP $< 30$.

%We do not have any instances of reference sets with $AvgCrossSim(\overline{R})
%< 0.4$. Therefore, we can extend the rule to these regions, as a small value of
%$AvgCrossSim(\overline{R})$ can indicate non-identical trinkets in the
%reference images.

Through a similar process, we have identified several other filtering rules
for reference sets and candidate images, and their real world interpretation,
%a rule that rejects reference
%sets whose template images have DTC-KP $< 30$, as well as a rule that rejects
%candidate images with KP-CNT $< 20$.
see Table~\ref{tab:rule:filters} (top 2 sections).  RBFilter uses these rules
to reject low quality reference and candidate images, and extend Pixie with
informative error messages that guide users to improve the quality of captured
images.

\vspace{-5pt}

\subsubsection{Filter Rule \#2: UBounds}

We train Pixie on a dataset of images that do not cover the entire value space
of the filter features. Pixie cannot make informed decisions on
candidate images whose features take values in sub-areas not
seen during training. We have identified several such sub-areas for the Pixie
dataset. The UBounds filter consists of the ``universe boundary'' rules listed
in Table~\ref{tab:rule:filters} (bottom section), that define these sub-areas.

\section{Evaluation}
\label{sec:evaluation}

%\begin{figure*}
%\centering
%\subfigure[]
%{\label{fig:pixie:keypoint}{\includegraphics[width=2.1in]{graphs/speed/keypoint/pixie.key.point_ggplot.eps}}}
%\subfigure[]
%{\label{fig:pixie:match}{\includegraphics[width=2.1in]{graphs/speed/match/pixie.match_ggplot.eps}}}
%\subfigure[]
%{\label{fig:pixie:classify}{\includegraphics[width=2.1in]{graphs/speed/keypoint/pixie.key.point_ggplot.eps}}}
%\caption{Pixie performance time on laptop and Nexus 5 devices.
%(a) Average time to extract keypoints from images by ORB and SURF, when run
%on laptop and on Nexus 5. Even SURF takes 10ms on the Nexus 5.
%(b) Average time to match and filter the key points of
%two images, when ORB and SURF are used, on a laptop and a Nexus 5
%device. Even on ``large'' images (with an average of 447 key points
%for ORB and xxx for SURF \bogdan{Mozhgan, please update this.})
%ORB takes 250ms and SURF ... \bogdan{What is Figure (c)? Should we reference it
%in the text?}
%\label{fig:pixie:speed}}
%\end{figure*}

We have implemented Pixie using Android 3.2, OpenCV 2.4.10 and
Weka~\cite{Weka}.  In the following we first evaluate the performance of the
Pixie features, and parameters, under several supervised learning algorithms.
We then evaluate the performance of Pixie's optimal configuration under the
attacks introduced in $\S$~\ref{sec:model:adversary}.
We report the performance of Pixie through its False Accept Rate (FAR), False
Reject Rate (FRR), Equal Error Rate (EER), the rate at which $FAR=FRR$, 
F-measure, and Failure to Enroll (FTE).
For our experiments, we have used a Mac OS X (2.9 GHz Intel Core i7 CPU, and
8GB DDR3 RAM) and a Nexus 4 smartphone (Quad-core 1.5 GHz Krait, and 2GB RAM).
%, the harmonic mean of precision and recall.

%Throughout this section, we have applied 10-fold cross-validation
%tests~\cite{Kohavi} to assess how the results of the statistical analysis will
%generalize to an independent data set. The advantage of this method is that
%all observations are used for both training and validation, and each
%observation is used for validation exactly once.  We have used the Weka
%version 3.7.9 data mining suite~\cite{Weka} to perform the experiments, with
%default settings: For the backpropagation algorithm of the MLP classifier, we
%set the learning rate to 0.3 and the momentum rate to 0.2.

%We measure the False Acceptance Rate (FAR), False Rejection Rate (FRR) and
%Equal Error Rate (EER) of the authentication schema to evaluate its
%performance. FAR can be defined as a the ratio of number of time an
%authentication system allows an unauthorized access divided by the number of
%identification attempts. FPR can be defined as the ratio of the number of
%times an authentication system incorrectly rejects the access of an authorized
%user divided by the total number of identification attempts. EER defined the
%rate at which both acceptance and rejection are equal.

\subsection{Data}
\label{sec:data}

We have collected and generated the following datasets:

\noindent
{\bf Nexus image dataset}.
We used a Nexus 4 device to capture $1,400$ photos of $350$ unique
trinkets, belonging to $33$ object categories. We selected only objects that
can be easily carried by users and are thus ideal candidates for image-based
trinkets, e.g., watches, shoes, jewelery, shirt patterns, credit cards and logos.
We have captured 4 images for each trinket, that differ in background and
lighting conditions, i.e., either indoors using artificial light or outdoors in
daylight conditions.

\noindent
{\bf Pixie dataset}.
To evaluate Pixie, we generate authentication instances that consist of one
candidate image and 3 reference images. To prevent ``tainting'', we need to
ensure that instances used for testing do not contain reference images that
have appeared in a training instance. For this, we use the $1,400$ images of
the $350$ trinkets, to generate 10 Pixie subsets, each containing 10 folds, as
follows. To generate one of the 10 folds of one of the 10 subsets, we first
randomly split the $350$ trinkets into 10 sets of $35$ trinkets each. For each
trinket in a set, we randomly select one of its $4$ images as candidate; the
remaining $3$ images are reference images.  The trinket then contributes to the
fold by one genuine instance (its candidate + its 3 reference images) and 34
``fraud'' instances.  Each fraud instance combines the trinket's candidate
image with the 3 reference images of one of the other 34 trinkets in the
subset. Thus, each fold consists of $35$ authentic and $1190 = 35 \times 34$
fraud instances. Then, one of the 10 Pixie subsets contains $12,250$
authentication instances. Thus, the Pixie dataset has a total of
$122,500$ authentication instances.

%To evaluate Pixie against the attacks of $\S$~\ref{sec:model:adversary}, we
%use the following datasets. To emulate attack scenarios, we do not
%use these datasets to train the Pixie classifier.

\noindent
{\bf Google Image dataset}.
We used Google's image search site to retrieve at least $200$ images from
each of the $33$ object categories of the Nexus image dataset, for a total of
$7,853$ images.  This dataset forms the basis of a shoulder surfing attack
(see $\S$~\ref{sec:evaluation:attacks}).

\noindent
{\bf ALOI dataset}.
We use the illumination subset of the Amsterdam Library of Object Images
(ALOI)~\cite{GBS05} dataset, that contains $24$ different images for $1000$
small objects (i.e., natural trinket choices) captured under various
illumination conditions. We cropped these images to the size of the Nexus
images ($270 \times 312$ pixels), while keeping their object centered.

\noindent
{\bf Caltech101 dataset}.
Caltech101~\cite{FFP07} is a collection of $9,145$ images
% of small and large objects,
from $101$ object categories.

\subsection{Parameter Choice for Pixie}
\label{sec:evaluation:parameters}

\noindent
We first identify the parameters where Pixie performs best.

\begin{table}
\centering
\resizebox{0.55\textwidth}{!}{%
\textsf{
\begin{tabular}{l c c c c}
\toprule
\textbf{Keypoint Detector} & \textbf{FAR(\%)} & \textbf{FRR(\%)} & \textbf{F-measure(\%)} & \textbf{EER(\%)}\\
\midrule
ORB & 0.10 & 9.83 & 93.08 & 4.87\\
\midrule
SURF & 0.07 & 4.80 & 96.40 & 2.80\\
\bottomrule
\end{tabular}
}}
\caption{ORB vs. SURF based Pixie (MLP classifier, no filter) performance, on
the Pixie dataset. SURF has lower FAR and FRR compared to ORB.}
\label{tab:pixie:orb:surf}
\vspace{-15pt}
\end{table}

\begin{table}
\centering
\resizebox{0.55\textwidth}{!}{%
\textsf{
\begin{tabular}{l c c c c}
\toprule
\textbf{Pixie Classifier} & \textbf{FAR(\%)} & \textbf{FRR(\%)} & \textbf{F-measure(\%)} & \textbf{EER(\%)}\\
\midrule
MLP & 0.10 & 9.83 & 93.08 & 4.87\\
\midrule
RF & 0.02 & 10.74 & 93.90 & 3.82\\
\midrule
SVM & 0.00 & 12.57 & 93.04 & 10.74\\
\midrule
Decision Tree (C4.5) & 0.17 & 11.54 & 91.01 & 7.66\\
\bottomrule
\end{tabular}
}}
\caption{Classifier performance on Pixie dataset using ORB keypoint extractor
and no filter. {\bf Random Forest and MLP achieve the lowest EER}, thus we only
use them in the following.
\label{tab:pixie:classifiers:orb}}
\vspace{-15pt}
\end{table}

\noindent
{\bf ORB vs. SURF}.
We compare the performance of Pixie when using two popular keypoint
extraction algorithms, ORB~\cite{RRKB11} and SURF~\cite{BTVG06}. We use 
Multilayer Perceptron (MLP) 
for the Pixie classifier, and no filter. We perform the evaluation through
10-fold cross validation on each of the 10 subsets of the Pixie dataset (see
$\S$~\ref{sec:data}).  Table~\ref{tab:pixie:orb:surf} reports the 
performance of ORB and SURF: SURF has lower FAR and FRR, leading
to an EER that is smaller by $2$\% than that of ORB.

However, ORB is faster than SURF: In an experiment on $100$ Nexus
images, ORB took an average $0.15$s to extract keypoints on the Nexus 4, while
SURF took $2.5$s on the Mac and almost $5$s on the Nexus 4. ORB's keypoint
match is also faster: In an experiment over $10,000$ image pairs, on the Nexus
4, SURF's keypoint match took an average of $2.72$s, while ORB took
$0.66$s.

Given the trade-off between speed and accuracy, SURF is more suitable
when the image processing and matching tasks can be performed on a
server. The faster ORB should be preferred in the mobile authentication
scenario, when these tasks have to be performed by a mobile device. In the
following experiments, we set Pixie's keypoint extraction algorithm to be ORB.

\noindent
{\bf Classifier Choice}.
We use the Pixie dataset to identify the best performing classifier for
Pixie's authentication module.  Table~\ref{tab:pixie:classifiers:orb} shows the
results: Random Forest (RF) and MLP outperform 
Support Vector Machine (SVM) and Decision Tree (DT) through lower FAR and FRR. 
In the following, we use only RF and MLP as Pixie's classifiers.

\begin{table}
\centering
\resizebox{0.55\textwidth}{!}{%
\textsf{
\begin{tabular}{l c c c c c}
\toprule
\textbf{Images} & \textbf{Filtering Rule} & \textbf{FAR(\%)} & \textbf{FRR(\%)} & \textbf{F-measure(\%)}\\
\midrule
Reference & KP-CNT \textless 20 & 0.09 & 6.60 & 95.06\\
\midrule
Reference & DTC-KP \textless 30 & 0.10 & 9.12 & 93.46 \\
\midrule
Reference & AvgCrossSim \textless 0.6 & 0.07 & 8.10 & 94.53 \\
\midrule
Reference & All 3 Filters & 0.06 & 4.46 & 96.75\\
\midrule
%Candidate & KP-CNT \textless 20 & 0.27 & 12.39 & 89.33 & 6.15\\
%\midrule
%Candidate & White-CNT \textless 2000 & 0.25 & 10.55 & 90.61& 5.16\\
%\midrule
%Candidate & Both Filters & 0.25 & 9.86 & 91.04& 4.69\\
%\midrule
Ref. \& cand. & All RBFilter Rules & 0.04 & 5.25 & 96.58 \\
\bottomrule
\end{tabular}
}}
\caption{
Performance of Pixie MLP classifier with RBFilter on the Pixie dataset. The
disjunction of all the RBFilters on the reference images reduced the FAR and
FRR by more than $40$\%.}
\label{tab:rfixie:orb}
\vspace{-15pt}
\end{table}

%\begin{table}
%\centering
%\resizebox{0.47\textwidth}{!}{%
%\textsf{
%\begin{tabular}{l c c c c c c}
%\toprule
%\textbf{Images} & \textbf{Filtering Rule} & \textbf{FAR(\%)} & \textbf{FRR(\%)} & \textbf{F-measure(\%)} & \textbf{EER(\%)}\\
%\midrule
%Reference & KP-CNT \textless 20 & 0.28 & 11.01 & 89.94 & 5.13\\
%\midrule
%Reference &DTC-KP \textless 30 & 0.27 & 13.53 & 88.52& 6.73\\
%\midrule
%Reference &Avg-CrossSim \textless 0.6 & 0.25 & 12.46 & 89.63& 6.82\\
%\midrule
%Reference & All 3 Filters & 0.26 & 8.63 & 91.78& 4.56\\
%\midrule
%Candidate & KP-CNT \textless 20 & 0.27 & 12.39 & 89.33 & 6.15\\
%\midrule
%Candidate & DTC-KP \textless 30 & 0.25 & 10.55 & 90.61& 5.16\\
%\midrule
%Candidate & Both Filters & 0.25 & 9.86 & 91.04& 4.69\\
%\midrule
%Ref. \& cand. & All RFixie Rules & 0.25 & 6.73 & 93.32& 3.53\\
%\bottomrule
%\end{tabular}
%}}
%\caption{Pixie + RFixie performance, for the rules of
%Table~\ref{tab:rule:filters}, on the Pixie dataset. The combination of all the
%RFixie filters performs best, with an EER reduction of $49$\% (from $7.04$\% to
%$3.53$\%).}
%\label{tab:rfixie:orb}
%\vspace{-15pt}
%\end{table}

\noindent
{\bf Pixie with RBFilter}.
We evaluate the effects of the RBFilter rules of
Table~\ref{tab:rule:filters} on the performance of Pixie. For this, in each of
the 100 classification experiments (10 folds cross validation over each of the
10 subsets of the Pixie dataset), we remove from the Pixie test fold
all the authentication instances that satisfy the rules.
%
%Specifically, we have removed the
%instances whose (i) reference images have fewer than 20 keypoints, a distance
%to the centroid of keypoints of less than 30, or an average cross similarity of
%under 0.6, and (ii) whose candidate image has fewer than 20 keypoints.
%
We then run Pixie (with MLP) on this filtered dataset.

Table~\ref{tab:rfixie:orb} shows that all the rules are effective: each
increases Pixie's F-measure. The disjunction of all the reference set
filter rules is the most effective, for an F-measure of $96.75$\% ($3.8$\%
improvement from the unfiltered $93.08$\% of Table~\ref{tab:pixie:orb:surf}).
The 3 reference set filter rules remove an average of $6.68$ reference sets
from a testing fold. When also using the candidate image filter, that removes
an average of $82.23$ authentication instances per testing fold, Pixie's
F-measure drops to $96.58$\%. This is because we count the ``valid''
instances removed by the candidate filter as part of FRR, even though they are
likely of low quality and can mislead Pixie.
%Thus, the FRR of $5.25$\%
%achieved by Pixie when all the rules of RBFilter are applied, is an upper bound
%on the real value.

%\begin{figure}[t!]
%\centering
%\includegraphics[width=2.9in]{./figures/sfixie.filter/sfixie.filter.eps}
%\caption{CBFilter test methodology: create a large training set (RSB)
%that does not contain reference images from the fold on which we later run
%Pixie ($F_1$). Run CBFilter on the reference sets from fold $F_1$, filter
%the reference sets that fail, then run Pixie on the filtered $F_1$.}
%\label{fig:sfixie:test}
%\vspace{-5pt}
%\end{figure}

\noindent
{\bf Pixie with CBFilter}.
%Figure~\ref{fig:sfixie:test} illustrates the process we employed to evaluate
%the impact of CBFilter on the performance of Pixie.
To provide a large training set for CBFilter, we first build a
Reference Set Bank (RSB), that contains all the reference sets that appear in
the $10$ subsets of the Pixie dataset.  For each such reference set, the RSB
also stores its ``class'', according to the outcome of Pixie:
%
%(see step 1 of Figure~\ref{fig:sfixie:test})
%
if the reference set has been part of any
authentication instance (in the Pixie dataset) that was incorrectly classified
by Pixie (i.e., either as FR or FA), its class is $1$, otherwise it is $0$.

\begin{table}[t]
\centering
\resizebox{0.55\textwidth}{!}{%
\textsf{
\begin{tabular}{l l c c c c}
\toprule
\textbf{Pixie} & \textbf{CBFilter} & \textbf{FAR(\%)} & \textbf{FRR(\%)} & \textbf{F-measure(\%)} & \textbf{EER(\%)}\\
\midrule
MLP & MLP & 0.07 & 6.34 & 95.54 & 2.97\\
\midrule
MLP & RF & 0.06 & 4.70 & 96.52 & 1.87\\
\midrule
MLP & C4.5 & 0.02 & 7.19 & 95.92 & 2.35\\
\midrule
MLP & SVM & 0.10 & 9.83 & 93.08 & 4.87\\
\midrule
RF & MLP & 0.02 & 7.64 & 95.63 & 2.72\\
\midrule
RF & RF & 0.01 & 5.74 & 96.77 & 1.96\\
\midrule
RF & C4.5 & 0.02 & 7.19 & 95.92 & 2.35\\
\midrule
RF & SVM & 0.02 & 10.74 & 93.90 & 3.82\\
\bottomrule
\end{tabular}
}}
\caption{
Pixie + CBFilter performance, for various combinations of supervised learning
algorithms. {\bf CBFilter is effective: when using RF, it reduces
the EER of Pixie (with MLP) to $1.87$\%.}}
\label{tab:pixie:sfixie:classifier}
\vspace{-15pt}
\end{table}

We use the RSB set for the following evaluation process, performed separately for
each subset of the Pixie dataset. Each of the subset's 10 folds,
%
%(step 2)
%
is used once for testing. Given one such fold, e.g., $F_1$,
%
%($F_1$ in
%Figure~\ref{fig:sfixie:test})
%
we extract its reference sets. We train CBFilter on all the reference
sets of RSB, that are different from the reference sets of fold $F_1$, then
test CBFilter on the reference sets of $F_1$.
%
%(step 4).
%
We filter from $F_1$ all the reference sets that are labeled as 1 by
CBFilter.
%
%i.e., predicted to be a likely culprit of a future false rejection or
%false acceptance.
%
Finally, we train Pixie on the 9 other folds ($F_2 ..  F_{10}$) and test it on
the filtered $F_1$. We repeat this process $100$ times (for the $10$ folds
of each of the $10$ subsets of the Pixie dataset).

Table~\ref{tab:pixie:sfixie:classifier} compares the performance of various
classifiers for both Pixie and CBFilter. It shows that
CBFilter is effective: when using Random Forest, it reduces the EER of Pixie to
$1.87$\% (from $4.87$\%), and removes $3.45$ reference sets on average from a
testing fold.

%SFixie using Random Forest also reduces the EER of Pixie with RF,
%to $1.96$\% by removing $3.65$ reference sets on average.

\begin{table}
\centering
\resizebox{0.55\textwidth}{!}{%
\textsf{
\begin{tabular}{l c c c c}
\toprule
\textbf{Algo} & \textbf{FAR(\%)} & \textbf{FRR(\%)} & \textbf{F-measure(\%)}\\
\midrule
Pixie & 0.10 & 9.83 & 93.08\\
\midrule
Pixie \& RBFilter & 0.04 & 5.25 & 96.58\\
\midrule
Pixie \& CBFilter & 0.06 & 4.70 & 96.52\\
\midrule
Pixie \& RBFilter \& CBFilter & 0.02 & 4.25 &97.52\\
\bottomrule
\end{tabular}
}}
\caption{
Filters effects on Pixie performance. The combination of RBFilter
and CBFilter (RF) has the best performance.}
\label{tab:pixie:rfixie:sfixie}
\vspace{-15pt}
\end{table}

\noindent
{\bf Pixie, RBFilter and CBFilter}.
When used in combination with RBFilter, CBFilter removes an additional $0.9$
reference sets on average from a testing fold. RBFilter's candidate rule also
removes $79.59$ instances.  Table~\ref{tab:pixie:rfixie:sfixie} compares the
performance of the combined Pixie, RBFilter and CBFilter against the
performance of the unfiltered Pixie, as well as Pixie's combination with only
one of the filters. When used together, the filters reduce the FAR of the basic
Pixie by $80$\% and its FRR by $56$\%.

\noindent
{\bf Comparison to other authentication methods}.
The performance of Pixie (EER=$1.87$) compares favorably with the performance
of other biometric based authentication solutions. For instance, Meng et
al.~\cite{meng2015surveying} report EERs of 2-4\% and 2-6\% for authentication
solutions based on face and fingerprint. Samangouei et al.~\cite{faceatt}
report EERs of 13-30\% for attribute based face authentication, and Taigman et
al.~\cite{deepface} report an EER of 8.6\% for face recognition using features
extracted by deep neural networks. The gaze-challenge authentication solution
of Sluganovic et al.~\cite{SRRM16} has an EER of 6.3\%, while Zhao et
al.~\cite{zhao2014mobile} report EERs between 4.1-9.6\% for touch gesture based
authentication.

\subsection{Pixie Under Attack}
\label{sec:evaluation:attacks}

We investigate the performance of Pixie, trained on one of the $10$ Pixie
subsets, under the attacks of $\S$~\ref{sec:model:adversary}. We use
the previously identified parameters: the ORB keypoint extractor, MLP for the
Pixie classifier, RF for the CBFilter classifier, and all the rules
for RBFilter. Using UBounds filter we obtain a conservative
performance of Pixie: with UBounds, Pixie would easily reject out of bounds
images, artificially boosting its accuracy.

%Table~\ref{tab:pixie:pictionary:aloi} summarizes the performance of Pixie under
%the ALOI attack dataset. The FRR is $0$ and not included.

\noindent
{\bf Attack datasets}.
We use the Nexus dataset ($\S$~\ref{sec:data}) to build 3 {\it authentication
attack datasets} based on the ALOI, Google Image and Caltech101 sets. We
use the Google Image based attack dataset for a shoulder surfing attack,
and, along with the ALOI and Caltech101 datasets, to evaluate brute force
pictionary attacks.

We generate the authentication attack instances for each attack dataset, and
group them into $10$ folds, as follows. We randomly split the $350$ unique
trinkets of the Nexus dataset into $10$ subsets of $35$
trinkets each. For each trinket in a subset, we randomly select $3$ out of its
$4$ images, to form a reference set. We then combine this set with each of the
images from ALOI, Google Image, and Caltech101 datasets, respectively. We
repeat this process for all the $35$ reference sets in a fold. Thus, in the
ALOI attack dataset, a fold contains $840$K = $35 \times 24$K attack instances,
for a total of $8.4$M ALOI based attack instances.  Similarly, the Google Image
attack dataset contains $2.7$M+ attack instances, while the Caltech101 attack
dataset contains $3.2$M+ instances.

\begin{table}
\centering
\resizebox{0.25\textwidth}{!}{%
\textsf{
\begin{tabular}{l c c c}
\toprule
\textbf{Attack Dataset} & \textbf{FAR(\%)} \\%& \textbf{Avg. \# filtered candidates} & \textbf{Avg. \# filtered reference sets}\\
\midrule
Google  &  0.054 \\ %& 82 & 13 \\
\midrule
ALOI & 0.087  \\ %& 1,449 &  10 \\
\midrule
Caltech101 & 0.042  \\ %& 5 & 11 \\
\bottomrule
\end{tabular}
}}
\caption{
Performance of Pixie (with RBFilter and CBFilter)
on the ALOI, Caltech101 and Google attack datasets: {\bf On more than
14M attack authentication samples, the FRR of Pixie is less than 0.09\%.}
\label{tab:attack:pictionary}}
\vspace{-15pt}
\end{table}

%\begin{table}
%\centering
%\resizebox{0.47\textwidth}{!}{%
%\textsf{
%\begin{tabular}{l c c c}
%\toprule
%\textbf{Algo} & \textbf{FAR(\%)} & \textbf{Avg. \# filt. cand.} & \textbf{Avg. \# filt. refs.}\\
%\midrule
%Pixie & 0.504  & 0 &  0 \\
%\midrule
%Pixie \& RBFilter & 0.136  & 1,449 & 9 \\
%\midrule
%Pixie \& RBFilter \& CBFilter  &  0.144 & 1,449 & 10.5 \\
%\midrule
%Pixie \& RBFilter \& UBounds  &  0.505 & 36 & 0 \\
%\midrule
%Pixie \& RBFilter \& UBounds \& CBFilter & 0.144  & 1,483 & 10.5\\
%\bottomrule
%\end{tabular}
%}}
%\caption{Performance of Pixie under the ALOI based attack dataset. FRR is
%naturally $0$. After applying RBFilter and CBFilter, Pixie's FAR of $0.504$\%
%decreases to $0.144$\% ($1$ in $690$ trials). UBounds does not reduces the FAR.
%\label{tab:pixie:pictionary:aloi}}
%\vspace{-5pt}
%\end{table}

%Table~\ref{tab:pixie:pictionary:aloi} summarizes the performance of Pixie under
%the ALOI attack dataset. The FRR is $0$ and not included.

\subsubsection{Pictionary Attack}
\label{sec:evaluation:attacks:pictionary}

Under the Google Image attack dataset, %when using both RBFilter and CBFilter,
Pixie achieved a FAR of $0.054$\%, see Table~\ref{tab:attack:pictionary}.
%
%RBFilter removes about $82$ out of $7,853$ candidates,
%and $8.8$ out of $35$ reference sets per fold.  CBFilter removes an
%additional $4.3$ reference set on average per fold.
%
$216$ of the $350$ trinkets were unbreakable. However, we counted each such
trinket as success at $7,853$ trials. Then, the average number of Google
dataset based ``trials until success'', over the $350$ trinkets is $5,766.12$.
For the ALOI based attack, when using both RBFilter and CBFilter, Pixie
achieved a FAR of $0.087$\%.
%
%On average per fold, RBFilter removes $6.7$ out of
%$35$ reference sets, and $1.4$K out of $24$K candidate images.  On average,
%CBFilter removes $3.6$ additional reference sets per fold.
%
Under the Caltech101 attack, Pixie's FAR is $0.042$\%.
%
%On average per fold, RBFilter removes
%about $5$ out of $9,145$ candidates, and $6.4$ out of $35$ reference sets.
%CBFilter removes an average of $4.6$ additional reference sets per fold.
%
The higher FAR of the ALOI pictionary attack dataset may be due
to the similarity of its images of small objects to images in the Pixie datase.
Pixie filters about 10 reference sets from each attack dataset. 
In addition, it filters a small number of candidate images (82 and 5) from the Google
and Caltech101 datasets, but 1,449 candidate images from the ALOI dataset.

\subsubsection{Restricted Shoulder Surfing Attack}
\label{sec:evaluation:attacks:shoulder}
We use the Pixie and Google Image datasets to evaluate the ``guessing
entropy''~\cite{DMR04} of the restricted shoulder surfing attack. The attack proceeds as
follows: For each reference set of a Pixie dataset trinket, we re-order the
Google dataset images to start the brute force attack with images of the same
type as the trinket. We then use as candidate, each image in the re-ordered
Google dataset, and count the number of trials before a match (false accept)
occurs. Thus, this experiment evaluates the scenario where the adversary
exploits his knowledge of the trinket type.

As in the pictionary attack above, we counted each of the $216$ unbreakable
trinkets  as ``success'' at $7,853$ (the size of the attack dataset) trials.
Then, the average number of ``trials until success'', over the $350$ Pixie
dataset trinkets was $5,639.53$. This result is similar to the above pictionary
attack: In fact, an unpaired t-test did not find a statistically
significant difference in the number of trials to break a reference set between
the two scenarios ($p-value=0.44$, for $\alpha=0.05$). Thus, in our
experiments, knowledge of the trinket type does not provide the adversary with
a significant guessing advantage.

%This confirms that this attack has a similar success rate compared to a brute
%force approach.  Even when correctly guessing the trinket type, the adversary
%needs on average $1851$ trials to authenticate fraudulently.

%\begin{figure}
%\centering
%\vspace{-5pt}
%\subfigure[]
%{\label{fig:master:aloi}{\includegraphics[width=1.53in,height=1.3in]{graphs/hist.master.candidates/hist_aloi_matched_refsets.eps}}}
%\subfigure[]
%{\label{fig:master:google}{\includegraphics[width=1.53in,height=1.3in]{graphs/hist.master.candidates/hist_google_matched_refsets.eps}}}
%\vspace{-10pt}
%\caption{Histograms of the number of matched reference sets for candidate
%images with more than one match for
%(a) attack dataset based on ALOI dataset,
%(b) attack dataset based on Google image dataset.
%\label{fig:master:attack}}
%\vspace{-15pt}
%\end{figure}

\begin{figure*}
\centering
\includegraphics[width=0.69\textwidth]{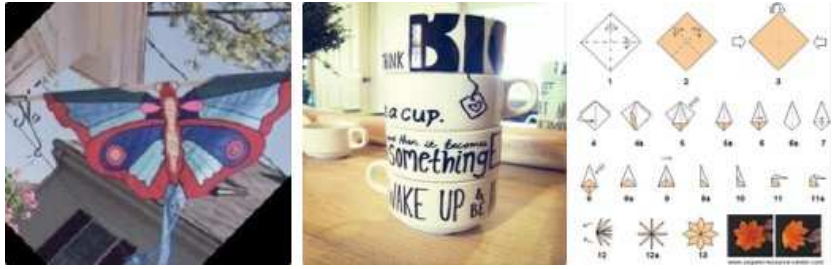}
\vspace{-5pt}
\caption{
Example master images for Pixie: each of these images matches multiple
reference sets of the Pixie dataset. Master images tend to have a rich
combination of shapes, shadows, colors and letters.}
\label{fig:master:image}
\vspace{-5pt}
\end{figure*}

\subsubsection{The Master Image Attack and Defense}
\label{sec:evaluation:attacks:master}

We identified $788$ master images in the ALOI dataset, $75$ in the Caltech
Image dataset, and $127$ in the Google dataset. Master images match multiple
Pixie reference sets.  Upon manual inspection, we observed that master images
are not of the same type of trinket as the reference set that they match.
Instead, they contain an array of shapes, shadows, letters and colors, that
translate into a diverse sets of keypoints, see Figure~\ref{fig:master:image}
for examples.
%
%Figure~\ref{fig:master:attack} shows
%the histograms of the number of reference sets matched by master images
%from the ALOI and Google Image datasets.
%
%224, 34, and 30 master images in ALOI, Caltech101 and Google datasets match 5 or more reference sets respectively.
Less than half of the master images in the ALOI (224), Caltech101 (34) and
Google (30) datasets match at least 5 reference sets. $1$ master image in the
Caltech101 dataset matches 51 reference sets.

\noindent
{\bf Defense}.
The shape formed by the matched keypoints in a master image is likely to be
inconsistent with that of the ``victim''
reference set. We leverage this observation to introduce several new features:
the distance to the centroid of the matched keypoints (DTC-MKP) in the
candidate and template images, and the min, max and mean of the DTC-MKP over
all pairs of candidate and reference images. We train the Pixie
classifier using this enhanced feature set, and test it on the ALOI
attack dataset. The enhanced Pixie reduces the number of effective ALOI master
images (matching at least $5$ reference sets) by $60$\%, i.e., from
$224$ to $88$.

To evaluate the effect of the new features on the FRR, we run Pixie with both
RBFilters and CBFilters on the 10 Pixie data subsets (see $\S$~\ref{sec:data})
in a 10-fold cross validation experiment similar to that of
$\S$~\ref{sec:evaluation:parameters}. We observed that when new features are
included in the classification task, the FRR of Pixie decreases slightly from
$4.25\%$ (last row in Table~\ref{tab:pixie:rfixie:sfixie}) to $4.01\%$, while
its FAR remained unchanged ($0.02\%$). We conclude that the newly added
features do not increase Pixie's FRR.

%We have used the $40$ most effective master images (each matched at least $5$
%reference sets) on Pixie enhanced with these features.  Specifically, we have
%created an attack dataset, that combines $315$ reference sets of the Pixie
%dataset with all the 40 master images, as candidates ($12,600$ authentication
%instances).  The enhanced Pixie reduced to 16 the number of master images,
%where only 10 master images matched more than 1 reference set.

%An alternative approach to identify the features of master images is to
%exhaustively try random combinations of features on the trained Pixie.

%The identified master images and the reference sets they match can be used to
%develop a new filter (e.g., based on supervised learning), that detects
%vulnerable reference sets, likely to be matched by master images. We leave this
%investigation for future work.

\section{User Study}
\label{sec:user:study}

We have used a lab study to evaluate the usability of Pixie's trinket based
authentication and compared it against text-based passwords. In this section,
we describe the methodology and results.

\subsection{Design and Procedure}

We performed a within-subjects study, where all the participants were exposed
to every conditions considered. Specifically, the conditions were to
authenticate from a smartphone to the Florida International University Portal Website
(MyFIU), using (i) their username and text-based password and (ii) their Pixie
trinket. MyFIU is a site that provides students with information about class
schedules and administrative functionality.

We have recruited participants from the university campus over e-mail lists,
bulletin boards and personal communications. All the participants were students
enrolled at the university. The reason for selecting students for the study was
to ensure a consistent and familiar login procedure to remote services. The
participants in our study achieved text password authentication times on par
with previously reported results (see $\S$~\ref{sec:results:time}).
Considering the ubiquity of mobile devices, we believe that the participants
had no unfair advantage when compared to other social groups of similar age,
with respect to their ability to perform the basic action of snapping a picture
with a smartphone.

In the following, we first presents some demographic information about the
(n=42) participants in this study, then describe the procedure we used to
perform the user study.

\begin{table}
\resizebox{0.59\textwidth}{!}{%
\textsf{
\begin{tabular}{l c c}
\toprule
\textbf{Demographic} & \textbf{Number} & \textbf{Proportion (\%)}\\
\midrule
\textbf{Gender} & & \\
Female & 11 &  26\\
Male & 31 & 74  \\
\midrule
\textbf{Age} & &\\
Min & 18 &\\
Max & 50 &\\
Median & 28 &\\
\midrule
Android & 20 & 48 \\
iPhone & 21 & 50\\
Windows phone & 1 & 2 \\
\midrule
Undergraduate & 16 & 38 \\
Graduate & 26 & 62\\
\midrule
CS/IT & 38 & 90 \\
Other majors & 4 & 10\\
\midrule
Use phone to login to remote services?  & 41 & 98\\
\bottomrule
\end{tabular}
}}
\centering
\caption{Participant demographics. We chose only students
in order to have a consistent experience for remote authentication (on
the university portal website, MyFIU).}
\label{tab:demographics}
\vspace{-15pt}
\end{table}

\noindent
{\bf Demographics.}
We have recruited 42 participants for our lab study.
Table~\ref{tab:demographics} shows the demographics of the participants,
obtained through the study questionnaires. 
In addition, 41 ($98$\%) participants said they use their phones to
login to their online accounts.

%\begin{table}
%\centering
%%\resizebox{0.45\textwidth}{!}{%
%\textsf{
%\begin{tabular}{l c c c}
%\toprule
%\textbf{Category} &\multicolumn{3}{c}{\textbf{Ease of Remembering}}\\
%\textbf{ } & \textbf{Mean} & \textbf{STD}& \textbf{Median}\\
%\midrule
%%Place & 4.2 & 0.9  &4\\
%Face & 3.0 & 1.9  &3\\
%Photo & 2.9 & 1.9 &3\\
%%Object & 2.4 & 1.9 &2\\
%Text & 2.2 & 1.5 & 2\\
%\bottomrule
%\end{tabular}
%}
%\caption{Participant reported scores on the ease of remembering faces, photos
%and text.  Score range from 1 (``Hard to remember'') to 5 (``Easy to
%remember''). It shows that
%participants perceive photos as being more memorable than text, and almost as
%memorable as faces.
%%While on average participants think it is difficult to remember text, we did
%%not find a statistically significant difference between the means of the
%%scores corresponding to different text, object, photo and face. However there
%%is a significant difference between the ease of remembering for places
%%compared to other items.
%}
%\label{tab:demographics:easeofremembering}
%\vspace{-15pt}
%\end{table}
%A Kruskal-Wallis
%test did not reveal any significant difference among the medians of the scores
%give to text, photo or face. 

\begin{figure}
\centering
\includegraphics[width=0.69\textwidth, keepaspectratio]{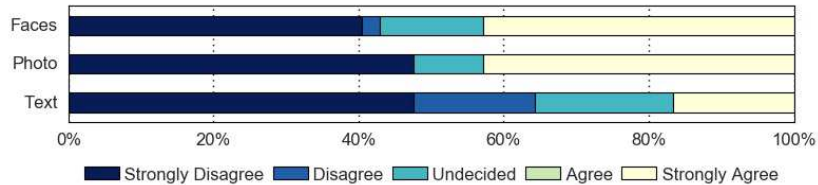}
\caption{The level of agreement of the participants with ease of remembering
faces, photos and text. 42\% of the participants strongly agree to their ease
of remembering photos and faces vs. only 16\% who agreed it is easy for them to
remember text.
%A pairwise Wilcoxon-Mann-Whitney test revealed no significant in perceived
%memorability between different  pairs.
}
\label{fig:prestudy:easeofmemorizing}
\vspace{-5pt}
\end{figure}

We also asked the participants to express their level of agreement using a
5-point Likert scale (from 1-strongly disagree to 5-strongly agree) with how easy
it is for them to remember text, photos and faces.
Figure~\ref{fig:prestudy:easeofmemorizing} shows the summary of the
participants responses. More than 42\% of the
participants strongly agreed that it is easy for them to remember faces and
photos. However, only 16\% of the participants strongly agreed it is easy for
them to remember text.  While 64.29\% of the participants said it is not easy
for them to remember text, a lower 47.62\% and 42.86\% of the participants said
it is not easy for them to remember photos and faces respectively.
A pairwise non-parametric Wilcoxon-Mann-Whitney test revealed no
significant difference between the perceived memorability for different items.
Based on this analysis and given the picture superiority
effect~\cite{pictorialSuperiority},  we posit that memorizing trinkets and
their secret angles could be perceived to be as memorable as faces and text.
We compare the perceived memorability of trinkets and text passwords in
$\S$~\ref{sec:results:memorability}.

\begin{figure*}[t!]
\centering
\subfigure[]
{\label{fig:pixie:1}{\includegraphics[height=2.5in,keepaspectratio]{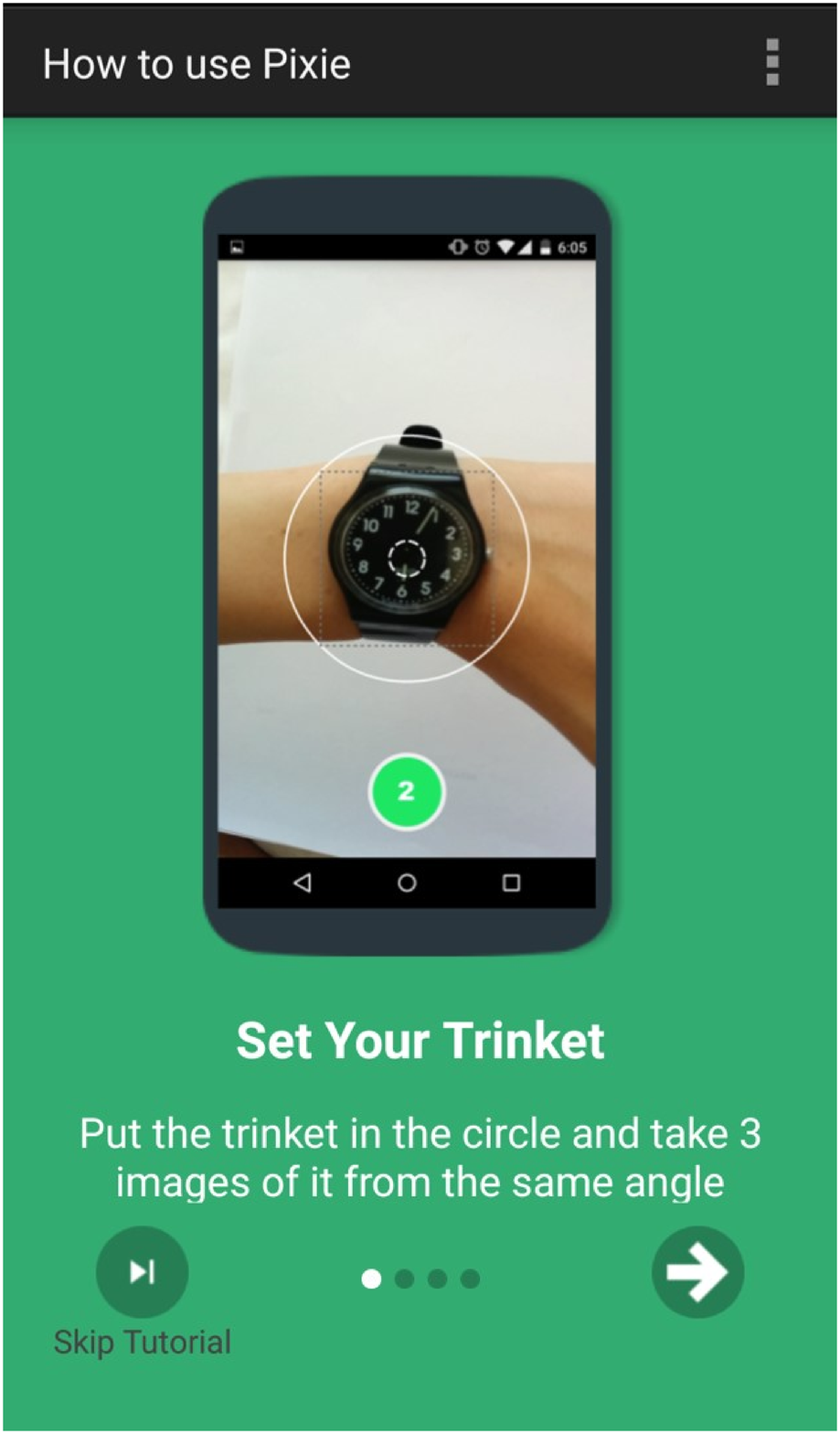}}}
\subfigure[]
{\label{fig:pixie:2}{\includegraphics[height=2.5in,keepaspectratio]{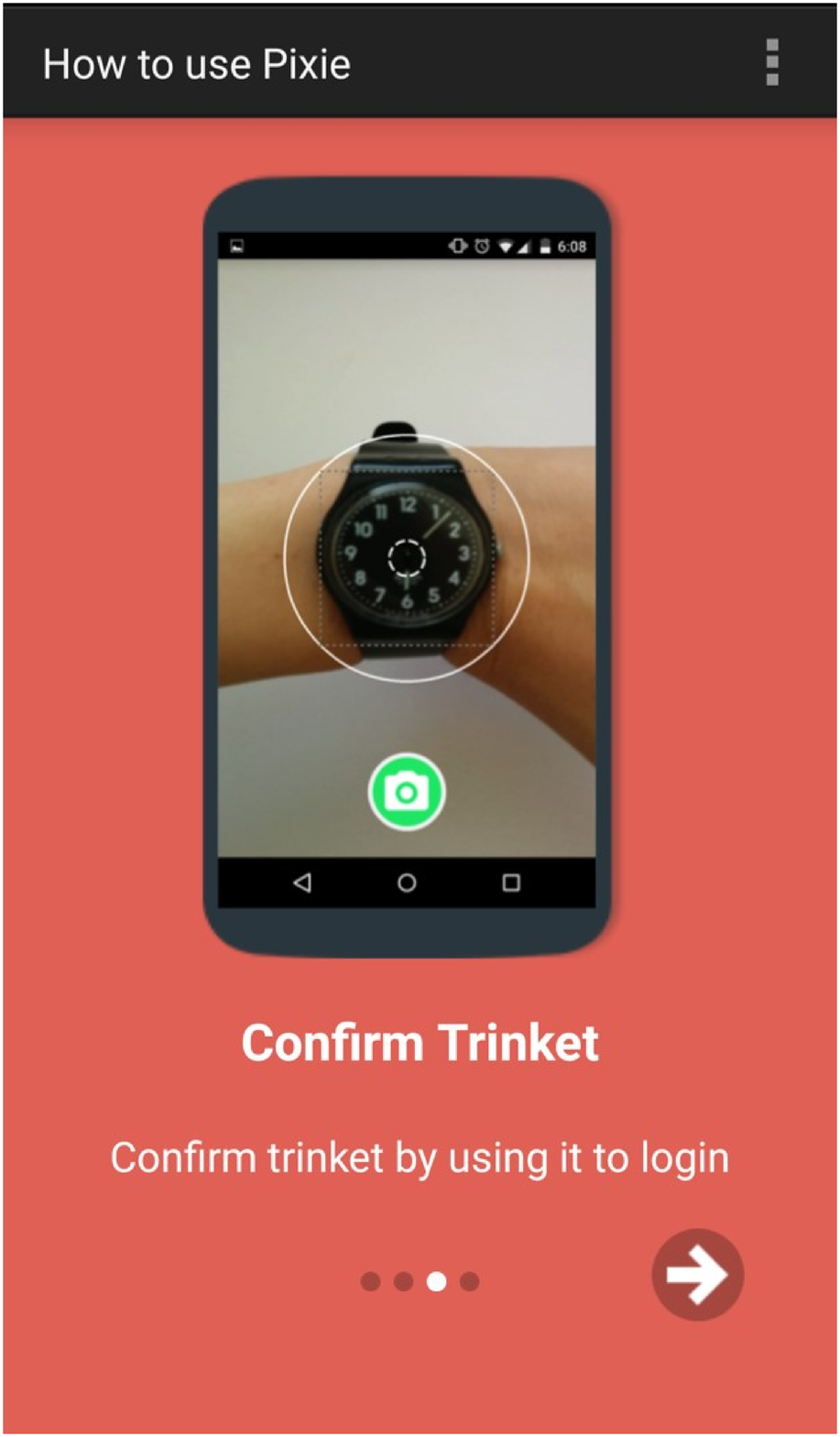}}}
\subfigure[]
{\label{fig:pixie:3}{\includegraphics[height=2.5in,keepaspectratio]{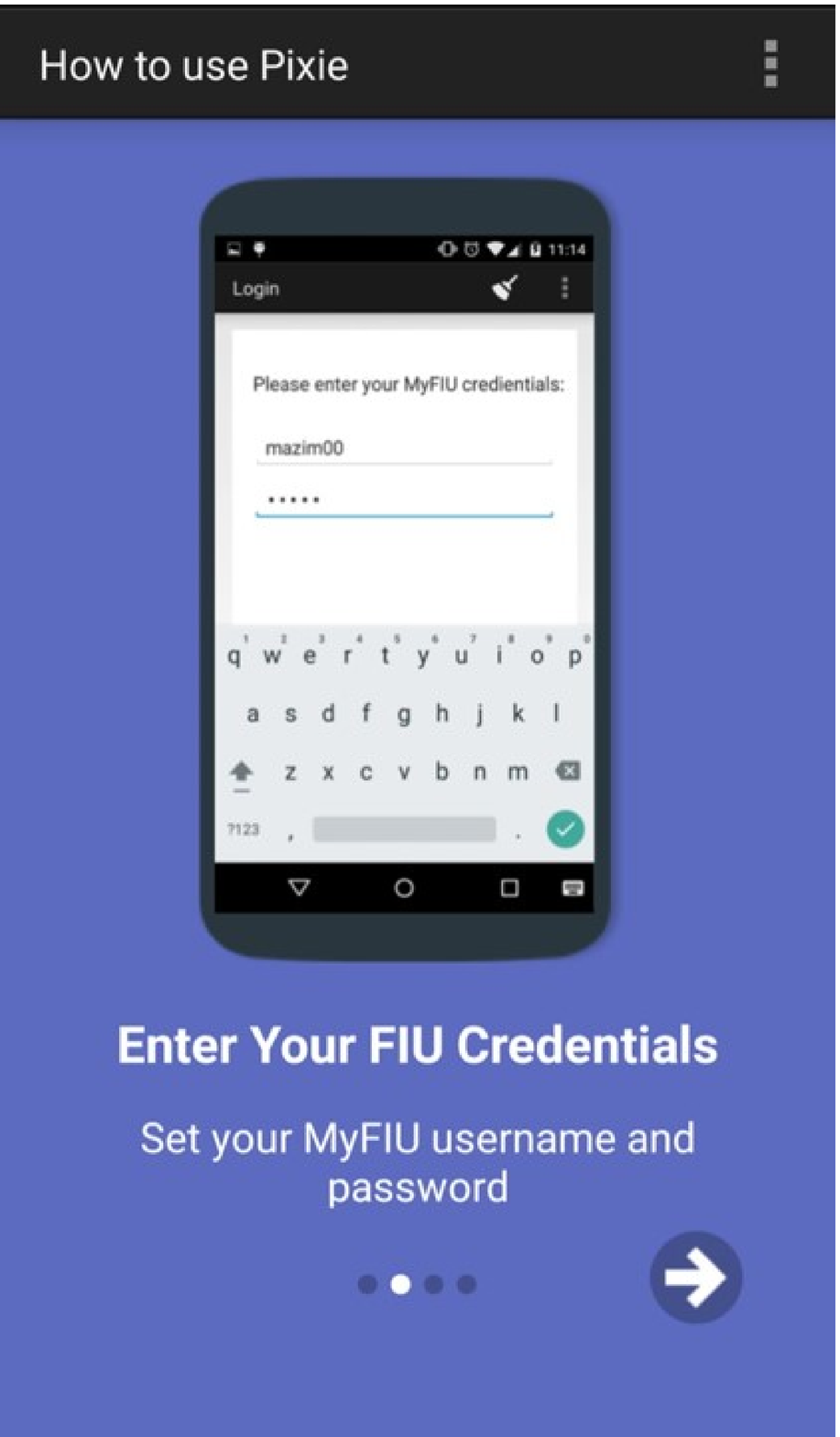}}}
\subfigure[]
{\label{fig:pixie:4}{\includegraphics[height=2.5in,keepaspectratio]{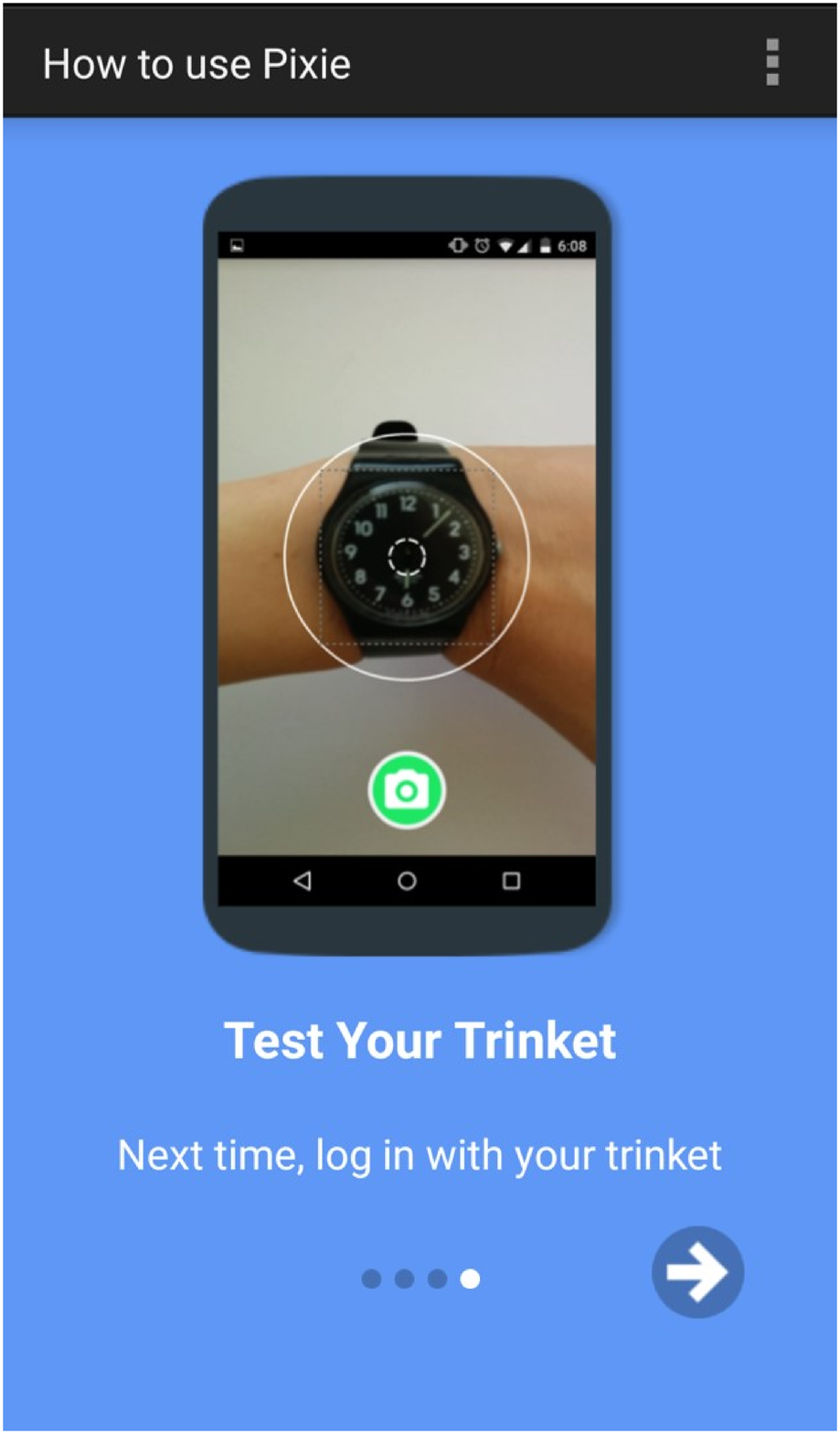}}}
\vspace{-5pt}
\caption{Pixie in-app instructions (best viewed in color) showing how to
(a) setup a trinket,
(b) confirm the trinket,
(c) enter credentials for the MyFIU account the first time the app is used, and
(d) login using the trinket. 
\label{fig:pixie:instruction}}
\vspace{-5pt}
\end{figure*}

\noindent
{\bf The study procedure.}
We have conducted the study in an indoor lab using the existing artificial
lighting. For the authentication device, we have used an HTC One M7 smartphone
(1.7GHz CPU, 2.1 MP camera with f/2.0 aperture, $4.7$ inch display with $1920$
$\times$ $1080$ resolution, and $137.4$ $\times$ $68.2$ $\times$ $9.33$mm
overall size).

The study consisted of 3 sessions, taking place on day 1, day 3 and day 8 of
the experiment. From the total of 42 participants, 31 participants returned for
and completed session 2 (7 female). Due to scheduling constraints, 3
participants returned for session 2 on day 4 or 5. 21 participants returned for
and completed session 3 (4 female). The lab sessions proceeded as follows. 

In the first session, we briefed participants about the purpose of the study: 
To explore the usability and the user interface design of a mobile 
device application. Then we asked them to use
Pixie to login to their MyFIU account, using their credentials (username and
password).  Pixie associates the text credentials with the trinket's reference
images.  During subsequent login sessions, the users only needed to correctly
capture the image of their trinket in order to access their account.
Our goal was to let the participants experience Pixie for
authentication, thus we did not ask them to enter their text password in
subsequent sessions. As a result, the comparison of Pixie with text passwords
is based only on the data collected in session 1.

Subsequently, the first session consisted of 3 steps. In the {\it
discoverability step}, we gave no verbal instructions to participants.
Instead, we asked each participant to try to figure out how to use Pixie, given
only the in-app instructions, that show a watch as a trinket example.
Figure~\ref{fig:pixie:instruction}(a-d) shows snapshots of Pixie app
instructions for setting up a trinket, verifying the trinket, setting up the
MyFIU account when the app is used for the first time and login using trinket.

In the {\it training step}, we explained Pixie's purpose and walked the
participant through the process of setting and testing a trinket using a gum
pack. However, we neither justified why we chose this trinket, nor specified
what other objects can be used as trinkets. We then asked the participants to 
set a trinket for the rest of the study.

In the third, {\it repeatability step} we asked the participant to repeat the
login part of the process. To avoid input based on muscle memory, we distracted
the participant's attention between the second and third step by playing a game 
for 5 minutes.

In session 2 and 3, the participants were asked to login to their MyFIU account
with the trinket they chose in session 1.  At the end of each session, the
participants filled out questionnaires that use Likert scales (ranging from
1-strongly disagree to 5-strongly agree). The questionnaires evaluate Pixie and
compare it against text-based passwords on perceived security, ease of use,
memorability and speed dimensions.

In addition, at the end of session 3, we have used ``emocards''~\cite{DOT01} to
evaluate the emotional responses of users toward Pixie and text password
authentication.  Emocards are 16 cartoon faces, each representing one of 8
distinct recognizable facial expression (1 per gender).  Emocards assist users
to non-verbally express their emotions about products, in terms of pleasantness
(pleasant, neutral, unpleasant) and arousal (calm, average, excited), two
commonly accepted dimensions of emotion responses~\cite{russell1980circumplex}.

\noindent
{\bf Participant dropout}.
The participant drop from session 1 to session 3 is not due to a dislike of
Pixie. To conclude this, we have compared the distributions of the answers of
the 21 participants who dropped and of the 21 participants who stayed until
session 3, on their overall impression of Pixie and their willingness to adopt
it. Both questions were rated on a Likert scale. The Mann-Whitney test shows
that the difference between the two populations is not statistically
significant ($p=0.7532$ for the first question, and $p=0.0701$ for the second
question at $\alpha=0.05$).  The participant drop can be due to the difficulty 
of scheduling 3 sessions across 8 days, at the end of the semester.

\noindent
{\bf Ethical considerations}.
We have worked with our university Institutional Review Board to ensure an
ethical interaction with the participants during the user study.  We have asked
the participants to avoid choosing sensitive trinkets. The entire experiments
took around 40 minutes per participant. 
We compensated each participant with a \$5 gift card.

\subsection{Results}
\label{sec:results}

Pixie is a novel authentication solution. Thus, we first present insights from
its use across the 3 sessions, with a focus on discoverability. We then detail
Pixie's observed memorability and performance, as well as the participant
perception and emotional responses.
%\newmaterial{In addition, we quote some of the qualitative 
%statements the participants expressed during the user study in session 1.}
All the statistical tests performed in
this section used a significance level of $\alpha=0.05$. 

\subsubsection{User Experience}
\label{sec:results:experience}

%\mizhoo{Check if security and usability perceptions are a significant factor in 
%adoption preference for Pixie.}

%\mizhoo {Need to reference Table~\ref{tab:comparison:entrytime} and 
%~\ref{tab:comparison:numtrials} in text.}

We now detail the user experience across the 3 sessions.
%, including the 3 steps of session 1.

\noindent
{\bf Session 1: discoverability}.
Without previous knowledge of Pixie, $86$\% of the participants ($36$) were
able to correctly set up their trinkets. Therefore, Pixie's
Failure to Enroll (FTE) rate is $14\%$. From the $14$\% ($6$) participants 
who failed to enroll, 3 did not notice that the 3 trinket photos had to be 
of the same object, captured from similar angles.  While Pixie provides a tooltip 
on the trinket capture button that guides the user to take another picture of the
trinket when the app is used for the first time (see Figure~\ref{fig:pixie:setup}), 
these 3 participants took random pictures from different objects in
the lab. These participants also did not understand the meaning of several
words, as English was their second language:

\begin{changemargin}{1cm}{1cm} 
[P20]: {\it ``Include one page saying what the trinket is. Like [sic], you can
say that trinket is an object that you will be using to sign in to your account''}.
\end{changemargin}

\begin{changemargin}{1cm}{1cm} 
[P21]: {\it ``I don't understand what {\it plain texture} means''}.
\end{changemargin}

In all 3 cases, the Pixie prefilters identified the issue correctly.  The other
3 unsuccessful participants chose trinkets with a plain texture (e.g., palm of
hand, pencil, objects with plain black surface) that generated errors. They
either dismissed error messages quickly or were not sure what to choose as a
trinket to eliminate the errors.

Subsequently, 3 other participants were unable to perform the trinket
verification step within 3 trials.  This occurred due to
(i) bad lighting conditions around the trinket, (ii) the participant forgetting
the trinket angle, or (iii) a texture-less (plain) trinket. While the
Chi-square test did not identify significant differences in the error rates
caused by any of the aforementioned circumstances ($p>0.05$), this could be
because of the limited number of samples.

%We believe lighting and a choice of trinket with plain texture are two major
%cause of error in Pixie.

\noindent
{\bf Session 1: Training}.
All the participants were able to set up a trinket successfully,
reducing the FTE rate of Pixie from $14\%$ in the discoverability
step  to $0\%$. All the participants then tested their trinkets within $4$
trials (M = $1.29$ trials, Std = $0.6$): $76$\% of the participants were able
to login from the first trial. The other $24$\% had lighting related
difficulties (e.g., the trinket reflected the light, or was in the shadow).
Only one participant required 4 trials. 

\noindent
{\bf Session 1: Repeatability}.
All the participants except one, were able to successfully complete this step within 3 
trails (M = $1.29$trials, Std = $0.6$). One participant required 4 trials.

\noindent
{\bf Sessions 2 and 3}.
In session 2, $84$\% of the participants were able to login from the first trial,
$13$\% logged in within 2-3 trials and only one participant needed 6 trials (M =
$1.35$ trials, Std = $1.02$). 2 participants did not carry their trinkets
%
%(watch and water bottle)
%
and had to reset them. In session 3, $81$\% of the participants were able to 
login from the first trial and all the other participants were able to login
within 2-3 trials (M = $1.20$ trials, Std = $0.40$).

\subsubsection{Participant Performance}

To measure participant performance we use {\it success
rate}~\cite{memorabilitymultiplepasswords}, defined as the number of successful
attempts to the total number of attempts.  In order to compare the success rate
of participants for text-based passwords and Pixie, we analyzed the data from
either of the login recalls of each session. We only consider successful Pixie
authentication sessions within 3 trials (see $\S$~\ref{sec:results:experience}). 
This is similar to MyFIU, where the
participants need to reset their passwords after 3 unsuccessful trials. The
success rate of Pixie improves from session 1 ($82.00$\%) to session 2
($83.33$\%) and session 3 ($84.00$\%).
Throughout all the 3 sessions, the Pixie success rate for
successful authentication sessions is slightly lower than the success rate for
the text-based password in session 1 (88.10\%). This is not surprising, given
the significantly lower number of practice opportunities for Pixie, compared to
the ubiquitous text passwords. However, the Chi-square test 
%(for unpaired results) 
revealed no significant difference between the success
rate for Pixie and text password in session 1 ($\chi^2(1)=0.506, p=0.48$). 
%
%We can also use fisher exact test here. again to significant difference between 
%success rate for different methods  P-Value = 0.380
%
Similarly, the Wilcoxon-Mann-Whitney test found no significant difference in
terms of the number of attempts for a successful login for Pixie within
different sessions, and between Pixie and text-based password in session 1.

\subsubsection{Memorability}
\label{sec:results:memorability}

During session 2, $96$\% of the participants (all except one) were able to
remember their trinkets. 2 participants did not immediately recall the part of
the trinket they used to authenticate, but they figured it out in the 3rd
attempt. These 2 participants were able to login in the first attempt in the
3rd session.  During session 3, all the participants were able to remember
their trinkets. We contrast these results with the memorability of text
passwords: 5 participants did not remember their MyFIU password and had to 
reset it in the first session. This is
consistent with previous findings:  Wiedenbeck et
al.~\cite{wiedenbeck2005passpoints}) report that more than 17\% 
of text-based passwords are forgotten in one week.

\subsubsection{User Entry Time}
\label{sec:results:time}

\begin{figure}[t!]
\centering %width=0.49\textwidth
\includegraphics[height=1.9in, keepaspectratio]{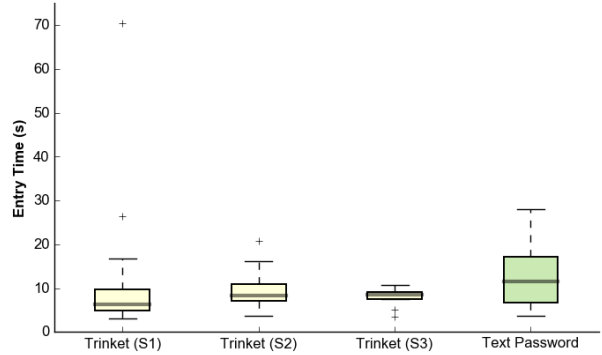}
\caption{Box plot for entry time of Pixie across 3 sessions vs. text password
in session 1. The Wilcoxon-Mann-Whitney test revealed that {\bf Pixie's entry
time in each session was significantly less than the entry time for text
passwords}.  For a single participant, the Pixie entry time
was 70.51s during session 1.}
\label{fig:userstudy:entrytime}
\vspace{-10pt}
\end{figure}

We have measured the {\it user entry time}, the interval from the moment when a
user starts Pixie and when Pixie submits the captured photo to the
authentication module. Figure~\ref{fig:userstudy:entrytime} shows the box plot
of the user entry time for Pixie in different sessions vs. the time for text
passwords, during session 1. The shortest authentication session was $3.01s$
and the longest session was $70.51s$ for Pixie. The average entry time improves
from session 1 (M=$9.71$s, Std=$11.42$s, Mdn=$6.24$s), to session 2 (M=$9.71$s,
Std=$4.66$s, Mdn=$ 8.32$s) and session 3 (M=$7.99$s, Std=$2.26$s, Mdn=$8.51$s).
However, Wilcoxon-Mann-Whitney tests did not
reveal any statistically significant differences between the Pixie user entry
time across the 3 sessions. We expect however that additional practice can
further improve Pixie's entry time.

%A repeated measure ANOVA reveals a significant main effects for authentication
%method on entry time in the first session ($p=0.008$) \bogdan{Mozhgan, explain
%this better}.
Moreover, a Wilcoxon-Mann-Whitney test revealed that the entry time for
Pixie was significantly less than the entry time for text passwords in session
1 ($W=845.0, p=0.000$).  We emphasize that in contrast to text passwords, Pixie
participants did not have the opportunity to practice beyond the steps of the
above procedure. 
%Nevertheless, they were able to enter the trinket faster than
%typing their well rehearsed password.

Table ~\ref{tab:comparison} compares the entry time for Pixie and other
authentication solutions based on biometrics or text and graphical passwords.
Although Pixie's entry time is higher compared to solutions based on face or voice, 
it compares well to several other solutions. For instance, Shay et al.
~\cite{shay2014can} report an entry time of 11.6-16.2 for text passwords.
MyFIU passwords are similar to the comp8 category in~\cite{shay2014can} (at least
8 characters, and include a lowercase English letter, uppercase English letter,
and digit) for which~\cite{shay2014can} report a median entry time of 13.2s.
%
%\cite{SWKW13} reports an entry time of $8$s for 42 bit passwords.
%
The additional safeguards of Boehm et al.'s~\cite{boehm2013safe} face and eyes
based biometric solution result in an entry time of 20-40s. Chiasson et al.
~\cite{memorabilitymultiplepasswords} report an entry time of about 15s for 
Passpoints. Trewin et al.~\cite{Trewin2012} reported an entry time of 
8.1s for gesture (stroke) based biometric.
%
%Pixie's overall authentication time (M=$9.9$s)
%
The eye tracking solution of Liu et al.~\cite{liu2015eyetracking} requires 9.6s
and the audio or haptic based solution of Bianchi et
al.~\cite{bianchi2011spinlock} requires $10.8-20.1$s. 

%\bogdan{Include Spinlock in Table~\ref{tab:comparison}?} \mizhoo{I didn't
%because we they don't report the entry time but the entire authentication
%time. Also some other information for table 1 is missing in their paper.}

%\mizhoo{maybe add the box plot for entry time of different sessions.}
%\mizhoo{What are the shortest and longest authentication session?}

In addition, we evaluated the processing overhead of Pixie: the 
time required to decide if a candidate image matches the reference
set.
%This consists of the
%time required to crop the candidate image to include only the trinket, then to
%extract its keypoints, compute the features needed to match the candidate to
%the reference keypoints, and to perform the classification.
The average processing overhead of Pixie on the HTC One smartphone over $94$
successful authentication trials is $0.5$ seconds.

\subsubsection{Perception}
\label{sec:results:perception}

\begin{figure}[t!]
\centering
\subfigure[]
{\label{fig:pixie:results}{\includegraphics[width=0.49\textwidth]{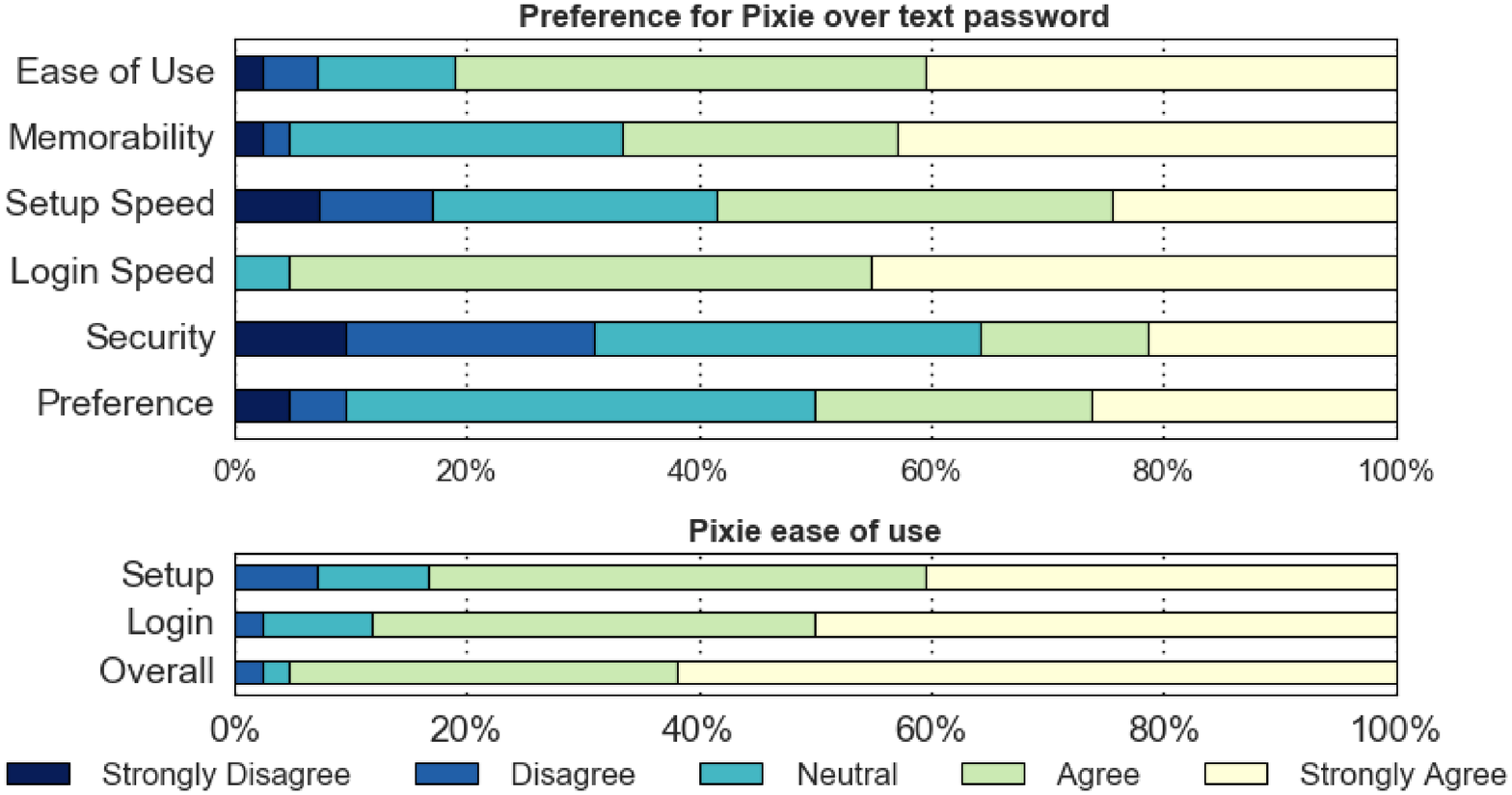}}}
\subfigure[]
{\label{fig:pixie:memorability}{\includegraphics[width=0.49\textwidth]{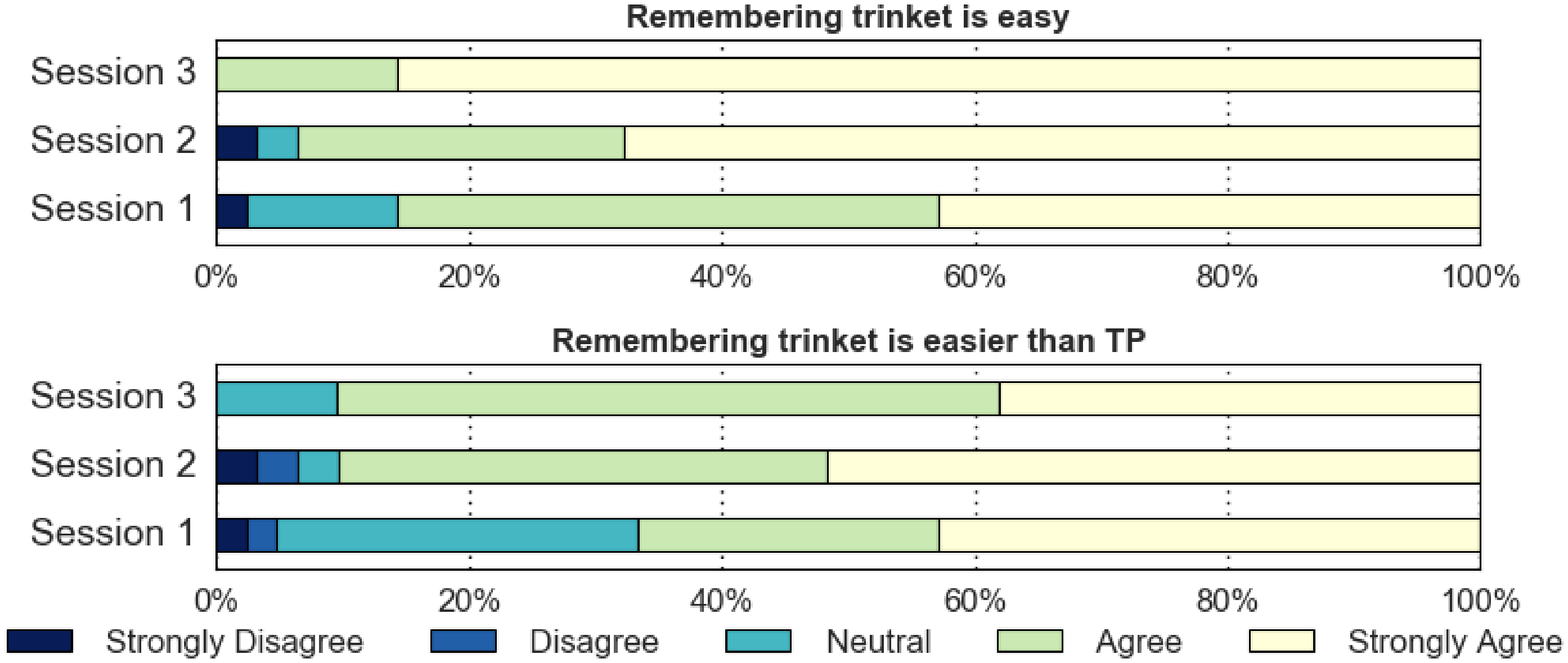}}}
\caption{
(a) Results at the end of session 1.
(a - top) Perceived performance of Pixie compared to text passwords. Pixie
dominates on ease of use, memorability and speed dimensions. 
(b - bottom) Pixie ease of use: $95$\% of
participants agreed that Pixie is easy to use.
(b - top) Pixie perceived memorability. $86$\% of participants agree that the
trinkets are easy to remember after session 1, but reach consensus after
session 3.
(b - bottom) Perceived memorability of Pixie vs. text passwords (TP). No participant
believes text passwords are more memorable after session 3.}
\vspace{-15pt}
\end{figure}

%\begin{figure}[t!]
%\centering
%\includegraphics[width=0.55\textwidth]{graphs/comprehensive.user.study/ColorPlots/remembering.two.row.eps}
%\caption{
%\footnotesize{Memorability over the 3 sessions.
%(top) Pixie perceived memorability. $86$\% of participants agree that the
%trinkets are easy to remember after session 1, but reach consensus after
%session 3.
%(bottom) Perceived memorability of Pixie vs. text passwords (TP). No participant
%believes text passwords are more memorable after session 3.
%}}
%\label{fig:pixie:memorability}
%\vspace{-15pt}
%\end{figure}

We asked the participants to express their perception about Pixie and text
passwords by providing answers to a set of questions in a 5-point Likert scale
(from strongly agree to strongly disagree). In the following we presents the
participants response. 

At the end of session 1, $81$\% of the participants said overall, Pixie
is easier to use than text-based passwords (Figure~\ref{fig:pixie:results}
(top)). $83$\% and $86$\% of the participants agree or strongly agree that the
trinket setup and login steps are easy (Figure~\ref{fig:pixie:results}
(bottom)). $95$\% of participants agree or strongly agree that
overall, Pixie is easy to use.

Furthermore, $86$\% of the participants agree or strongly agree that trinkets
are easy to remember, see Figure~\ref{fig:pixie:memorability} (top). $67$\% of
the participants agree that trinkets are easier to remember than passwords,
while only $5$\% of the participants believe the opposite, see
Figure~\ref{fig:pixie:results} (top) and Figure~\ref{fig:pixie:memorability}
(bottom). 
These results improve in sessions 2 and 3. At the end of session 3,
all the participants agree that trinkets are easy to remember
(Figure~\ref{fig:pixie:memorability} (top)): 12 participants changed their
opinion in favor of Pixie's memorability.  No participants believe that text
passwords are easier to remember than trinkets, see
Figure~\ref{fig:pixie:memorability} (bottom). 
A two-sample proportion test revealed that the proportion of the
participants who think Pixie is memorable, significantly increases from session
1 to session 2 and 3 ($Z=2.36$, $p=0.009$ and  $Z=2.05$, $p=0.020$).

$36$\% of the participants believe that Pixie is more secure than text
passwords, and $31$\% of the participants believe that passwords are more secure
(Figure~\ref{fig:pixie:results} (left)). Several participants felt strongly
about the security of Pixie, e.g.,:

\begin{changemargin}{1cm}{1cm} 
[P27] {\it ``This method is even more secure than text-based passwords,
because even if someone sees me during the password entry, he wouldn't know what
part of the object I have selected as my trinket and cannot easily figure it
out''}.
\end{changemargin}

$68$\% of participants agree or strongly agree that the trinket based login is
fast. $74$\% of participants agree or strongly agree that the trinket setup step
is fast.  $95$\% and $59$\% of the participants agree that Pixie's login and
trinket setup steps are faster compared to the corresponding text password
operations.  (Figure~\ref{fig:pixie:results} (top)).
$50$\% of the participants say that they prefer trinkets over text passwords
(Figure~\ref{fig:pixie:results} (top, bottom bar)).

When asked if they would use trinket based authentication in real life
$26$\% of participants said that they would use Pixie for most of their accounts,
$36$\% would use it for at least some of their accounts, and $36$\% would consider
using it. Only $2$\% of the participants (1) said that they would not use it.
Several participants felt strongly about adopting Pixie:

\vspace{-5pt}

\begin{changemargin}{1cm}{1cm} 
[P18]: {\it ``Why isn't [Pixie] integrated with the original MyFIU mobile
application as another option for signing to my account?''}.
\end{changemargin}

\begin{changemargin}{1cm}{1cm} 
[P40]: {\it ``I always forget my passwords [...] I always store them in my browser.
I would definitely use Pixie if it is available''}.
\end{changemargin}

%[P27]: {\it ``I think this is a good method because I usually forget my 
%passwords for my account''}.
%\end{addmargin}

\begin{table}[t]
\resizebox{0.65\textwidth}{!}{%
\small
\textsf{
\begin{tabular}{l c c c}
\toprule
\textbf{Question} & \textbf{Sample proportion} & \textbf{$95\%$ CI} & \textbf{$p$}\\
%\textbf{Question} & \textbf{proportion} & \textbf{CI} & \textbf{$p$}\\
\midrule
\textbf{Easier to use} & $80.95$ & ($65.88$, $91.39$) & $0.000$*\\
\textbf{More memorable} & $66.67$ & ($50.45$, $80.43$) &$0.044$*\\
\textbf{Faster Login} & $95.24$ & ($83.83$, $99.41$)& $0.000$*\\
\textbf{Faster Setup} & $58.54$ & ($40.96$, $72.27$) & $0.441$\\
\textbf{More secure} & $35.71$ & ($21.55$, $51.97$)& $0.088$\\
%\textbf{Prefer over text password} & $50.00$ & ($34.19$, $65.80$)& $0.999$\\
\bottomrule\\
{* Statistically significant result at $\alpha=0.05$.}
\end{tabular}
}}
\centering
\caption{Confidence interval for the proportion of ``agreement'' answers to
usability and security questions comparing Pixie and text-based authentication.
Pixie is perceived to be easier to use, more memorable and faster than text passwords.
Pixie's perceived advantage in ease of use, memorability, and login speed is
not due to random choice.}
\label{table:perception:ci} 
\vspace{-15pt}
\end{table}

\noindent
{\bf Statistical analysis}.
To differentiate true choice from random chance, we combine the strongly agree
and agree answers into an ``agreement'' answer, and the strongly disagree and
disagree answers into a ``disagreement'' answer.  We then use a one-sample
binomial test with a confidence interval in order to test whether the
proportion of agreement of the participants  with a statement is sufficiently
different from a random choice (50\%). Table ~\ref{table:perception:ci}
presents this result for the proportion of ``agreement'' answers to each
question. Pixie is perceived easier, more memorable and faster than text
passwords for login and the perceived advantage is not due to random choice.
However, the participant do not perceive a significant difference in setup
speed, and security of Pixie over text passwords. 
%A two-sample proportion test
%showed the proportion of participants who voted in favor of Pixie is
%significantly higher than that of text passwords in all dimensions.  

\begin{table}[t]
\centering
\resizebox{0.60\textwidth}{!}{%
\textsf{
\begin{tabular}{l c c}
\toprule
\textbf{Prefer Pixie over text passwords} & \textbf{$\tau_b$} & \textbf{$p$} \\
\midrule
\textbf{Easier to use} & 0.60 & 0.000*\\
\textbf{More memorable} & 0.54 &0.000*\\
\textbf{More secure} & 0.38 & 0.003*\\
\textbf{Faster Setup} & 0.46 & 0.000*\\
\textbf{Faster Login} & 0.48 & 0.000*\\
\midrule
%\textbf{Pixie Ease of Use} & 0.22 & 0.115\\
\textbf{Pixie Memorability} &0.40 & 0.003*\\
\textbf{Willingness to Use Pixie} &0.60 & 0.000*\\
\bottomrule\\
{* Indicates a statistically significant correlation at $\alpha=0.05$.}
\end{tabular}
}}
\caption{Kendall's Tau-b test shows significant positive correlation between
preference of Pixie vs. text passwords, and its preference in terms of ease of
use, memorability, security, faster setup and login time. Preference over text
passwords is also significantly correlated with the overall memorability of the
trinket and willingness to adopt Pixie.
%While all show a positive correlation, ease of use and
%memorability exhibit the highest correlation.
}
\label{table:correlations}
\vspace{-15pt}
\end{table}

\noindent
{\bf Analysis of User Feedback}.
Table~\ref{table:correlations} shows that the general preference of Pixie over
text passwords significantly correlates positively with its preference on ease
of use, memorability and security and speed dimensions.  The preference over text
passwords is also significantly correlated with overall perception of trinket
memorability and willingness to adopt Pixie. 
%However, we did not find a
%significant correlation between preference over text passwords and overall ease
%of use for Pixie. 
Interestingly, we observed a significant 
correlation between preference over text passwords on security and the participant
feeling of owning a unique trinket ($\tau=0.36,p=0.005)$.

The participant willingness to use Pixie also correlates positively with
perceived memorability ($\tau_b=0.29$), perceived ease of use ($\tau_b=0.28$),
general preference over text passwords ($\tau_b=0.32$), preference over text
passwords on security ($\tau_b=0.28$), and preference on ease of use
($\tau_b=0.04$). We observe a negative correlation between the willingness to
use Pixie and the number of login attempts ($\tau_b=-0.16$), highlighting the
impact of unsuccessful logins. However, the correlations are not
statistically significant.

\subsubsection{Emotional Response}
\label{sec:results:emotional}

%\begin{wrapfigure}{r}{0.5\textwidth}
\begin{figure}
\centering
\includegraphics[width=0.55\textwidth]{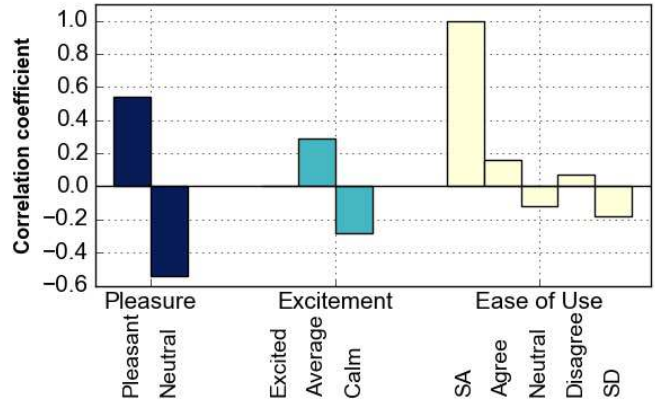}
\caption{Kendall's Tau-b correlations between willingness to use, emotional
responses (pleasure and excitement), and ease of use (SA/SD = Strongly
Agree/Disagree), during session 3. No participant rated Pixie as unpleasant.
Willingness to use correlates positively with pleasant and average levels, as
well as with agreement with ease of use.
\label{fig:pixie:correlations}}
\vspace{-5pt}
\end{figure}

The emocard experiment revealed that Pixie generates only positive emotions:
$81$\% of the participants reported a ``pleasant'', and $19$\% reported a
``neutral'' experience.  In addition, $47$\% of the participants were
``calm'', $34$\% were ``average'' and $19$\% were ``excited''. In contrast to
Pixie, only $5$\% of the participants (1) reported a ``pleasant'' level for text
passwords, while $57$\% reported ``unpleasant'' and $38$\% reported ``neutral''
levels. 
%The emotions toward text passwords are more negative after performing
%the task: 4 participants converted their answers to ``unpleasant'' after the
%task. 
A one-sided test of the difference of proportions revealed that
the proportion of the participants who perceived Pixie as pleasant was
significantly larger than the proportion of the participants who perceived text
passwords as pleasant ($Z = 4.01$, $p = 0.000$).

The Kendall's Tau-b correlations plotted in Figure~\ref{fig:pixie:correlations}
shows that the participant reports of willingness to use Pixie correlate
positively with levels of pleasure and excitement, as well as Pixie's perceived
ease of use. While 4 participants reported excitement for Pixie's
novelty, functionality and performance, we observe no correlation between
``excited'' levels and willingness to use. This is a positive finding,
as authentication solutions should not generate high arousal levels.

\subsubsection{Trinket Choice}
\label{sec:results:choice}

\begin{table}[t]
\centering
\resizebox{0.69\textwidth}{!}{%
\textsf{
\begin{tabular}{l c c c}
\toprule
\textbf{Object type} & \textbf{\# of unique objects} & \textbf{\# of participants} & \textbf{\# of unique trinkets}\\
\midrule
Gum pack & 3 & 16 & 8\\
Watch & 6 & 6 & 6\\
Mug & 3 & 3 & 3\\
Logo & 2 & 2 & 2\\
Keychain & 2 & 2 & 2\\
Car remote control key & 2 & 2 & 2\\
Sunglasses & 2 & 2 & 2\\
A piece of puzzle & 1 & 1 & 1\\
Shoe & 1 & 1 & 1\\
Kohl container & 1 & 1 & 1\\
Backpack pin & 1 & 1 & 1\\
Hair clip & 1 & 1 & 1\\
Cigarette box & 1 & 1 & 1\\
Match box & 1 & 1 & 1\\
Water Bottle & 1 & 1 & 1\\
iphone menu & 1 & 1 & 1\\
University ID card & 1 & 1 & 1\\
Tattoo & 1 & 1 & 1\\
\midrule
\textbf{Total} & \textbf{31} & \textbf{44} & \textbf{36}\\
\bottomrule
\end{tabular}
}}
\caption{Trinket choice: object types chosen by participants, along with the
number of unique objects belonging to each category and number of unique
trinket choice (object + angle) in the study.
%
%The puzzle, logo and mug belong
%to the lab. 2 gum packs are from the lab, the third was owned by a participant.
%The other objects were owned by the participants.
%
The gum pack and watch (used in the training step and on-screen instructions)
are the types most frequently used by the participants. All the captured
watch trinkets are unique.}
\label{table:trinket:choice}
\vspace{-15pt}
\end{table}

We manually analyzed the trinket images captured by the participants in the
first session (42 trinkets) and those captured by the participants who reset
their trinket in session 2 (2 trinkets). We allowed the participants to pick
any nearby object as a trinket. The 42 participants picked a total of 36 unique
trinkets, from 31 unique objects of 18 types, chosen from among participant
owned objects and lab objects. The gum pack and watch were the most frequently
chosen object types. However, all the 6 watch trinkets were different, and the
16 participants who chose a gum pack have captured 8 unique trinket images
(object + angle combination). 

%i.e., watch, shoe, keychain, car remote control key, kohl container, backpack
%pin, hair clip, sunglasses, gum box, cigarette box, match box, water bottle,
%iphone menu, university ID card, body tattoo, and (2) lab objects, i.e., gum
%box, mug, a piece of puzzle, and logo.  The gum box in the lab and watch were
%the most frequently chosen object types.

In the discoverability step, 8 participants used their watches as trinkets.  We
did not observe a significant difference in user choice of trinket between the
discoverability and training steps: 18 participants used the same trinket
in the discoverability and training steps. 8 participant chose their trinket to 
be their watches. The other trinket categories chose by participants that are not 
among those in Table~\ref{table:trinket:choice} include: pen/pencil, 
book and computer mouse.
%moving the discussion about why some users selected random trinket and
%the effect of trinket used in training step on trinket choise 
%to the discussion section.

We have used the images captured by the participants to ``brute force'' the
reference sets of each participant. We removed 8 reference sets as they were
identical (the top view of the same gum pack). This has produced a single
``success'' event, for the two participants who chose the same side of the same
gum pack, with very similar angles.  As we described in
$\S$~\ref{sec:results:perception}, the participant preference of Pixie over
text passwords on security correlates significantly with the participant
feeling of owning a unique trinket. We did not observe a statistically
significant difference between the feeling of owning a unique trinket and
participants gender.% ($p=0.40$). %\chi^2(2)=1.804,

%We have used the images captured by the participants to ``brute force'' the
%reference sets of each participant. We removed the 8 references and corresponding 
%candidate as they were identical (the top view of the same gum pack). 
%%\bogdan{I don't understand this. Why 8. Removed from the set of
%%reference images or from the set of attack images? If you didn't remove, would
%%they generate more successfull events?} \mizhoo{probably. I don't have the data now to test but they were images from the top of the same gum pack}
%This has produced a single ``success''
%event, for the two participants who chose the same side of the same gum pack,
%with very similar angles.  As we described in
%$\S$~\ref{sec:results:perception}, the participant preference of Pixie over
%text passwords on security correlates significantly with the participant
%feeling of owning a unique trinket. We did not observe a statistically
%significant difference between the feeling of owning a unique trinket and
%participants gender ($\chi^2(2)=1.804,p=0.40$).

%As a future work a long term field study of Pixie would provide more
%information about natural user choices for trinket in daily life. 

%\subsection{Researcher Observations}
%\umut{Mozhgan, this section is for you to talk about what you saw.}
%\umut{Examples: see Trewin, Biometric Authentication on a Mobile Device}

\section{Discussion and Limitations}
\label{sec:discussion}

\noindent
{\bf Authentication speed}.
Our user study shows that Pixie's authentication speed in session 1 is $25$\%
faster than well rehearsed text passwords and improves through even mild
repetition.  However, Pixie's entry time is longer than the reported entry time
for face based authentication solutions (see Table~\ref{tab:comparison}). This
may be due to either the novelty of Pixie or the way the images are captured,
i.e.  using the back, not the front camera for capturing trinket images.

\noindent
{\bf Secure image storage and processing}.
The storage and processing of the trinket images needs to be performed
securely. While outside the focus of this paper, we briefly discuss and compare
trinket image storage and processing solutions that are performed on the remote
service vs. the user's authentication device.  A remote server based solution
trivially protects against an adversary that captures the authentication
device, as the device does not store or process sensitive user information.
The image matching process is also faster on a server than on a mobile device
(see $\S$~\ref{sec:evaluation:parameters}).  The drawbacks are the overhead of
transmitting candidate images over the cellular network, and the imposition on
users to register a different reference image set for each remote service.

The authentication device based solution can easily associate the reference
images with the user's authentication credentials (e.g.  OAuth~\cite{rfc6749})
for multiple remote services.  However, since an attacker can capture and thus
access the storage of the mobile device, reference images cannot be stored or
processed in cleartext.
%
%Storing and processing only the keypoints extracted from the
%reference images is also insecure: knowledge of the reference keypoints would
%enable the adversary to fabricate candidate images that contain keypoints that
%match the reference keypoints.
%
The storage and processing of reference images can however be secured through
hardware-level protection, e.g., TrustZone~\cite{TrustZone}, or by using
privacy preserving image feature extraction solutions that work in the
encrypted domain, e.g.,~\cite{WHWR16,QYRCW14,HLP12}.

\noindent
{\bf Deployability}.
Pixie is well suited for OAuth~\cite{rfc6749} authorization to access remote
services from the mobile device: Pixie authenticates the user to the app on the
mobile device, which can then proceed with the OAuth protocol with the remote
server.

\noindent
{\bf Default authentication}.
If the trinket based authentication fails a number of times (due to
e.g., forgotten trinket, poor lighting conditions, unsteady hand), the user is
prompted to use the default authentication solution, e.g., text password.

\noindent
{\bf Strong passwords}.
Popular and ubiquitously available trinkets (e.g., iWatch, Coke can) should not
be chosen as trinkets, as an adversary can easily predict and replicate them.
To address this problem, Pixie can store a dataset of popular trinket images,
then, during the trinket setup process, reject reference sets that match
popular trinkets.

\noindent
{\bf Defense against brute force attacks}.
The brute force attacks of $\S$~\ref{sec:model:adversary} can be made harder to
launch through video ``liveness'' verifications, e.g.,~\cite{RTC16}:
capture both video and accelerometer streams while the user shoots
the trinket, then use video liveness checks to verify the consistency between
the movements extracted from the two streams. The lack of such streams or their
inconsistency can indicate a brute force attack.

%\subsection{Choice of User Study Participants}
%
%\bogdan{Should we move or re-discuss here the discussion on how our
%choice of participants is not unfair, and maybe even how the dropout
%is fine?}
%\mizhoo{I think this is a really a minor point and we don't need to emphasize on that
%again here! The perevious reviewer was absolutely unfair.
%Because everyone does the same thing! }

%\vspace{-7pt}

\noindent
{\bf The user study}.
The study presented in this work was the first attempt to quantify the
usability aspects of an authentication solution based on trinkets.  We
performed the user study in a lab setting. We were able to recruit only 42
participants, of which half did not stay until the last session.  As we wanted
to ensure a consistent and familiar login procedure to remote services for the
participants, we only recruited students from the university who had access to
myFIU, FIU's login portal. While the population of the study is not fully
representative of the users who would use the system, we believe that the
participants had no unfair advantage when compared to other social groups of
similar age in performance: the participants in our study achieved text
password authentication times on par with previously reported results (see
$\S$~\ref{sec:results:time}). 

Pixie works by extracting invariant keypoints from the captured images, using
keypoint extraction algorithms (e.g. SURF~\cite{BTVG06} and ORB~\cite{RRKB11}).
These algorithms are not capable of extracting keypoints from images of object
with constant shade. We attempted to address this issue by providing actionable
feedback to users, and guiding them toward choosing visually complex trinkets.
In addition, to ensure Pixie is able to identify the trinket images even when
captured in slightly different circumstances and to lower the false reject
rate, we required the users to enter 3 trinket images in the registration
phase. This may partially explain why the participants in our study did not
perceive Pixie as significantly faster than text passwords for the registration
phase.

%\umut{Mozhgan please contrast this to the {\it Registration} phase of the 
%Biometric Authentication on iPhone and Android paper; especially the 
%fact that iPhone fingerprint users had difficulty understanding on-screen
%instructions.}

During the discoverability step, we observed that several participants had
difficulties in understanding the in-app instructions on how to use Pixie.
Similar problems have been reported for other authentication mechanisms. For
instance, Bhagavatula et al. ~\cite{2015biometriciphoneandroid} reported that 7
out of 10 participants found understanding on-screen instructions difficult for
iPhone fingerprint authentication.  They recommend to provide clearer
instructions (e.g. through a demo video) on what the users need to do.  We
posit that explaining the meaning of trinkets will help users during the
registration phase and improve the discoverability rate of Pixie. 

%In addition, the in-app instruction can be replaced by showing an expressive demo video on how to use Pixie.

In addition, we observed that similar to face based authentication, Pixie has
problems with insufficient lighting.  A comprehensive study similar to the
studies conducted for biometric based solutions
(e.g.~\cite{2015biometriciphoneandroid}) may help identify other potential
limitations of Pixie in different situations (e.g. authentication while
walking, in public transportation, etc).

The consent form that we read and the participants signed prior to the study,
emphasizes that the focus of the study is on the usability aspects of a new
authentication mechanism. We observed that some of the participants might have
selected trinkets without concerns over security during the study.  We did not
guide the participants towards choosing specific trinkets, as we intended to
observe the personal or lab objects chosen by the participants. Nevertheless,
we observed that the participants preference of Pixie over text passwords on
security correlates significantly with the participants feeling of owning a
unique trinket. This suggests that the participants could corroborate the
relationship between unique trinkets and higher level of security.

%\mizhoo{Does the last sentence make sense?} 

The trinkets used to walk the participants through Pixie (i.e., gum pack) and 
the in-app user guide of Pixie (i.e., watch), appear to influence the participant
trinket selection in the first session: In the discoverability step, 
8 participant chose their watches as trinkets. 
In addition, during session 1, 9 participants
chose the same gum pack as used in the Pixie walk-through without even trying a
different angle, and 5 participants used their watches as trinkets.  
Further studies are required to understand whether other 
means of communicating the goals of Pixie (e.g. using a short video that guides the 
user on how to choose secure and unique trinkets) can reduce this bias.

Further, although 50\% of the participants said they prefer Pixie over text
passwords, 40\% percent of the participants were undecided. This may be due to
the limited experience of the participants with Pixie. 
In addition, 62\% participants said they would use Pixie in real life. 
We did not observe a statistically significant correlation between 
being excited about using Pixie, that could be due to the novelty of the method, 
and willingness to use it. However, future studies are required to understand 
in what scenarios and situations the users are willing to adopt trinket based 
authentication or prefer it over text passwords.
If we were to redo the study, we would split the Likert scale
questions comparing Pixie with text passwords into 2 questions asking the
participants to rate any of them in terms of usability and security. In
addition, we would ask the participants to justify their answers about
perceived usability and security in the form of open ended questions in the
post study interview.

\noindent
{\bf Field study}.
We leave for future work a field study of Pixie to investigate the longer term
effects of using trinket passwords on user entry times, accuracy and
memorability, the factors that impact trinket choice, how users choose and
change their trinkets in real life, as well as the potential improvements 
provided by alternative means of communication of Pixie's goals and
functionality (e.g., through short video instead of text).  
We also leave for
future work the investigation of using mental stories to associate trinkets to
accounts (e.g., use credit card as trinket for bank account) and reducing the
impact of interference~\cite{memorabilitymultiplepasswords,al2015multiple}.

%as well as using ``cues'' to improve the memorability of
%passwords~\cite{al2015impact}.

\section{Conclusions}
\label{sec:conclusions}

We introduced Pixie, a proof of concept implementation of a trinket based
two-factor authentication approach that uses invariant keypoints extracted from
images to perform the matching between the candidate and reference images.
Pixie only requires a camera, thus applies even to simple, traditional mobile devices
as well as resource limited wearable devices such as smartwatch and smartglasses.

We manually captured and collected from public datasets, 40,000 trinket images.
We proposed several attacks against Pixie and have shown that Pixie achieved an
EER of $1.87\%$ and FAR of $0.02\%$ on $122,500$ authentication attempts and an
FAR of less than $0.09\%$ on 14.3 million attack instances generated from the
40,000 images.

We performed an in lab user study to evaluate the usability aspects of Pixie as
a novel authentication solution. Our experiments show that Pixie is
discoverable: without external help and prior training, $86$\% and $78$\% of
the participants were able to correctly set a trinket then authenticate with
it, respectively.  $62\%$ of the participants expressed that they would use
Pixie in real life.  Pixie simplifies the authentication process: The study
shows that trinkets are not only perceived as more memorable than text
passwords, but are also easily remembered 3 and 8 days after being set, without
any inter-session use. In addition, Pixie's authentication speed in session 1
is $25$\% faster than well rehearsed text passwords and improves through even
mild repetition. We believe that Pixie can complement existing authentication
solutions by providing a fast alternative that does not expose sensitive
user information.

%We have introduced Pixie, a two factor trinket based graphical password,
%remote authentication solution for camera equipped mobile devices. The secret
%in Pixie consists of the trinket, its section and angle. We have developed
%robust solutions to match candidate authentication images with reference
%images of the trinket, and to identify low quality images and reference sets.
%We have used datasets of more than 40,000 images to show that Pixie is
%resilient to image based attacks.  We have shown through a user study that
%Pixie's performance, discoverability, speed and memorability make it a viable
%authentication solution for accessing remote services through mobile devices.

A promising approach to improve Pixie is to use more advanced image processing
techniques, e.g. deep neural networks~\cite{szegedy2016rethinking}, for image
feature extraction and processing. Such techniques may improve Pixie's
usability by (i) eliminating the requirement for capturing multiple reference
images of the trinket in the registration phase, (ii) increasing the ability to
extract features even from images of objects with constant shade, and (iii)
further reducing FRRs.

\small
%\bibliographystyle{ACM-Reference-Format}
%\bibliography{authentication,application,biometric,bogdan,entropy,hci,ml,pixie,trustzone,vision,wearable}
\bibliographystyle{ACM}

\end{document}